\newcommand{\LoadPreprintJournalOptions}{\DeclareOption{accountaudit}{%
        \gdef\@journal{accountaudit}
        \gdef\@journalfull{Accounting and Auditing}
        \gdef\@journalshort{Account. Audit.}
        \gdef\@doiabbr{accountaudit}
        \gdef\@ISSN{}
        \gdef\@societyowner{}
}
\DeclareOption{acoustics}{%
	\gdef\@journal{acoustics}
	\gdef\@journalfull{Acoustics}
	\gdef\@journalshort{Acoustics}
	\gdef\@doiabbr{acoustics}
	\gdef\@ISSN{2624-599X}
	\gdef\@societyowner{}
}
\DeclareOption{actuators}{%
	\gdef\@journal{actuators}
	\gdef\@journalfull{Actuators}
	\gdef\@journalshort{Actuators}
	\gdef\@doiabbr{act}
	\gdef\@ISSN{2076-0825}
	\gdef\@societyowner{}
}
\DeclareOption{addictions}{%
	\gdef\@journal{addictions}
	\gdef\@journalfull{Addictions}
	\gdef\@journalshort{Addictions}
	\gdef\@doiabbr{addictions}
	\gdef\@ISSN{0006-0006}
	\gdef\@societyowner{}
}
\DeclareOption{adhesives}{%
	\gdef\@journal{adhesives}
	\gdef\@journalfull{Adhesives}
	\gdef\@journalshort{Adhesives}
	\gdef\@doiabbr{adhesives}
	\gdef\@ISSN{}
	\gdef\@societyowner{}
}
\DeclareOption{admsci}{%
	\gdef\@journal{admsci}
	\gdef\@journalfull{Administrative Sciences}
	\gdef\@journalshort{Adm. Sci.}
	\gdef\@doiabbr{admsci}
	\gdef\@ISSN{2076-3387}
	\gdef\@societyowner{}
}
\DeclareOption{adolescents}{%
	\gdef\@journal{adolescents}
	\gdef\@journalfull{Adolescents}
	\gdef\@journalshort{Adolescents}
	\gdef\@doiabbr{adolescents}
	\gdef\@ISSN{2673-7051}
	\gdef\@societyowner{}
}
\DeclareOption{aerobiology}{%
	\gdef\@journal{aerobiology}
	\gdef\@journalfull{Aerobiology}
	\gdef\@journalshort{Aerobiology}
	\gdef\@doiabbr{aerobiology}
	\gdef\@ISSN{}
	\gdef\@societyowner{}
}
\DeclareOption{aerospace}{%
	\gdef\@journal{aerospace}
	\gdef\@journalfull{Aerospace}
	\gdef\@journalshort{Aerospace}
	\gdef\@doiabbr{aerospace}
	\gdef\@ISSN{2226-4310}
	\gdef\@societyowner{}
}
\DeclareOption{agriculture}{%
	\gdef\@journal{agriculture}
	\gdef\@journalfull{Agriculture}
	\gdef\@journalshort{Agriculture}
	\gdef\@doiabbr{agriculture}
	\gdef\@ISSN{2077-0472}
	\gdef\@societyowner{}
}
\DeclareOption{agriengineering}{%
	\gdef\@journal{agriengineering}
	\gdef\@journalfull{AgriEngineering}
	\gdef\@journalshort{AgriEngineering}
	\gdef\@doiabbr{agriengineering}
	\gdef\@ISSN{2624-7402}
	\gdef\@societyowner{}
}
\DeclareOption{agrochemicals}{%
	\gdef\@journal{agrochemicals}
	\gdef\@journalfull{Agrochemicals}
	\gdef\@journalshort{Agrochemicals}
	\gdef\@doiabbr{agrochemicals}
	\gdef\@ISSN{}
	\gdef\@societyowner{}
}
\DeclareOption{agronomy}{%
	\gdef\@journal{agronomy}
	\gdef\@journalfull{Agronomy}
	\gdef\@journalshort{Agronomy}
	\gdef\@doiabbr{agronomy}
	\gdef\@ISSN{2073-4395}
	\gdef\@societyowner{}
}
\DeclareOption{ai}{%
	\gdef\@journal{ai}
	\gdef\@journalfull{AI}
	\gdef\@journalshort{AI}
	\gdef\@doiabbr{ai}
	\gdef\@ISSN{2673-2688}
	\gdef\@societyowner{}
}
\DeclareOption{aichem}{%
        \gdef\@journal{aichem}
        \gdef\@journalfull{AI Chemistry}
        \gdef\@journalshort{AI Chem.}
        \gdef\@doiabbr{aichem}
        \gdef\@ISSN{3042-6723}
        \gdef\@societyowner{}
}
\DeclareOption{aieduc}{%
        \gdef\@journal{aieduc}
        \gdef\@journalfull{AI in Education}
        \gdef\@journalshort{AI Educ.}
        \gdef\@doiabbr{aieduc}
        \gdef\@ISSN{}
        \gdef\@societyowner{}
}
\DeclareOption{aimed}{%
       \gdef\@journal{aimed}
       \gdef\@journalfull{AI in Medicine}
       \gdef\@journalshort{AI Med.}
       \gdef\@doiabbr{aimed}
       \gdef\@ISSN{3042-6707}
       \gdef\@societyowner{}
}
\DeclareOption{air}{%
	\gdef\@journal{air}
	\gdef\@journalfull{Air}
	\gdef\@journalshort{Air}
	\gdef\@doiabbr{air}
	\gdef\@ISSN{}
	\gdef\@societyowner{}
}
\DeclareOption{aisens}{%
       \gdef\@journal{aisens}
       \gdef\@journalfull{AI Sensors}
       \gdef\@journalshort{AI Sens.}
       \gdef\@doiabbr{aisens}
       \gdef\@ISSN{3042-5999}
       \gdef\@societyowner{}
}
\DeclareOption{algorithms}{%
	\gdef\@journal{algorithms}
	\gdef\@journalfull{Algorithms}
	\gdef\@journalshort{Algorithms}
	\gdef\@doiabbr{a}
	\gdef\@ISSN{1999-4893}
	\gdef\@societyowner{}
}
\DeclareOption{allergies}{%
	\gdef\@journal{allergies}
	\gdef\@journalfull{Allergies}
	\gdef\@journalshort{Allergies}
	\gdef\@doiabbr{allergies}
	\gdef\@ISSN{2313-5786}
	\gdef\@societyowner{}
}
\DeclareOption{alloys}{%
	\gdef\@journal{alloys}
	\gdef\@journalfull{Alloys}
	\gdef\@journalshort{Alloys}
	\gdef\@doiabbr{alloys}
	\gdef\@ISSN{2674-063X}
	\gdef\@societyowner{}
}
\DeclareOption{amh}{%
	\gdef\@journal{amh}
	\gdef\@journalfull{Acta Microbiologica Hellenica}
	\gdef\@journalshort{Acta Microbiol. Hell.}
	\gdef\@doiabbr{amh}
	\gdef\@ISSN{0438-9573}
	\gdef\@societyowner{Hellenic Society for Microbiology}
}
\DeclareOption{analytica}{%
	\gdef\@journal{analytica}
	\gdef\@journalfull{Analytica}
	\gdef\@journalshort{Analytica}
	\gdef\@doiabbr{analytica}
	\gdef\@ISSN{2673-4532}
	\gdef\@societyowner{}
}
\DeclareOption{analytics}{%
	\gdef\@journal{analytics}
	\gdef\@journalfull{Analytics}
	\gdef\@journalshort{Analytics}
	\gdef\@doiabbr{analytics}
	\gdef\@ISSN{}
	\gdef\@societyowner{}
}
\DeclareOption{anatomia}{%
	\gdef\@journal{anatomia}
	\gdef\@journalfull{Anatomia}
	\gdef\@journalshort{Anatomia}
	\gdef\@doiabbr{anatomia}
	\gdef\@ISSN{2813-0545}
	\gdef\@societyowner{}
}
\DeclareOption{anesthres}{%
	\gdef\@journal{anesthres}
	\gdef\@journalfull{Anesthesia Research}
	\gdef\@journalshort{Anesth. Res.}
	\gdef\@doiabbr{anesthres}
	\gdef\@ISSN{}
	\gdef\@societyowner{}
}
\DeclareOption{animals}{%
	\gdef\@journal{animals}
	\gdef\@journalfull{Animals}
	\gdef\@journalshort{Animals}
	\gdef\@doiabbr{ani}
	\gdef\@ISSN{2076-2615}
	\gdef\@societyowner{}
}
\DeclareOption{antibiotics}{%
	\gdef\@journal{antibiotics}
	\gdef\@journalfull{Antibiotics}
	\gdef\@journalshort{Antibiotics}
	\gdef\@doiabbr{antibiotics}
	\gdef\@ISSN{2079-6382}
	\gdef\@societyowner{}
}
\DeclareOption{antibodies}{%
	\gdef\@journal{antibodies}
	\gdef\@journalfull{Antibodies}
	\gdef\@journalshort{Antibodies}
	\gdef\@doiabbr{antib}
	\gdef\@ISSN{2073-4468}
	\gdef\@societyowner{}
}
\DeclareOption{antioxidants}{%
	\gdef\@journal{antioxidants}
	\gdef\@journalfull{Antioxidants}
	\gdef\@journalshort{Antioxidants}
	\gdef\@doiabbr{antiox}
	\gdef\@ISSN{2076-3921}
	\gdef\@societyowner{}
}
\DeclareOption{applbiosci}{%
	\gdef\@journal{applbiosci}
	\gdef\@journalfull{Applied Biosciences}
	\gdef\@journalshort{Appl. Biosci.}
	\gdef\@doiabbr{applbiosci}
	\gdef\@ISSN{2813-0464}
	\gdef\@societyowner{}
}
\DeclareOption{appliedchem}{%
	\gdef\@journal{appliedchem}
	\gdef\@journalfull{AppliedChem}
	\gdef\@journalshort{AppliedChem}
	\gdef\@doiabbr{appliedchem}
	\gdef\@ISSN{2673-9623}
	\gdef\@societyowner{}
}
\DeclareOption{appliedmath}{%
	\gdef\@journal{appliedmath}
	\gdef\@journalfull{AppliedMath}
	\gdef\@journalshort{AppliedMath}
	\gdef\@doiabbr{appliedmath}
	\gdef\@ISSN{2673-9909}
	\gdef\@societyowner{}
}
\DeclareOption{appliedphys}{%
	\gdef\@journal{appliedphys}
	\gdef\@journalfull{AppliedPhys}
	\gdef\@journalshort{AppliedPhys}
	\gdef\@doiabbr{appliedphys}
	\gdef\@ISSN{}
	\gdef\@societyowner{}
}
\DeclareOption{applmech}{%
	\gdef\@journal{applmech}
	\gdef\@journalfull{Applied Mechanics}
	\gdef\@journalshort{Appl. Mech.}
	\gdef\@doiabbr{applmech}
	\gdef\@ISSN{2673-3161}
	\gdef\@societyowner{}
}
\DeclareOption{applmicrobiol}{%
	\gdef\@journal{applmicrobiol}
	\gdef\@journalfull{Applied Microbiology}
	\gdef\@journalshort{Appl. Microbiol.}
	\gdef\@doiabbr{applmicrobiol}
	\gdef\@ISSN{2673-8007}
	\gdef\@societyowner{}
}
\DeclareOption{applnano}{%
	\gdef\@journal{applnano}
	\gdef\@journalfull{Applied Nano}
	\gdef\@journalshort{Appl. Nano}
	\gdef\@doiabbr{applnano}
	\gdef\@ISSN{2673-3501}
	\gdef\@societyowner{}
}
\DeclareOption{applsci}{%
	\gdef\@journal{applsci}
	\gdef\@journalfull{Applied Sciences}
	\gdef\@journalshort{Appl. Sci.}
	\gdef\@doiabbr{app}
	\gdef\@ISSN{2076-3417}
	\gdef\@societyowner{}
}
\DeclareOption{aquacj}{%
	\gdef\@journal{aquacj}
	\gdef\@journalfull{Aquaculture Journal}
	\gdef\@journalshort{Aquac. J.}
	\gdef\@doiabbr{aquacj}
	\gdef\@ISSN{}
	\gdef\@societyowner{}
}
\DeclareOption{architecture}{%
	\gdef\@journal{architecture}
	\gdef\@journalfull{Architecture}
	\gdef\@journalshort{Architecture}
	\gdef\@doiabbr{architecture}
	\gdef\@ISSN{2673-8945}
	\gdef\@societyowner{}
}
\DeclareOption{arm}{%
	\gdef\@journal{arm}
	\gdef\@journalfull{Advances in Respiratory Medicine}
	\gdef\@journalshort{Adv. Respir. Med.}
	\gdef\@doiabbr{arm}
	\gdef\@ISSN{2543-6031}
	\gdef\@societyowner{Polish Respiratory Society}
}
\DeclareOption{arthropoda}{%
	\gdef\@journal{arthropoda}
	\gdef\@journalfull{Arthropoda}
	\gdef\@journalshort{Arthropoda}
	\gdef\@doiabbr{arthropoda}
	\gdef\@ISSN{2813-3323}
	\gdef\@societyowner{}
}
\DeclareOption{arts}{%
	\gdef\@journal{arts}
	\gdef\@journalfull{Arts}
	\gdef\@journalshort{Arts}
	\gdef\@doiabbr{arts}
	\gdef\@ISSN{2076-0752}
	\gdef\@societyowner{}
}
\DeclareOption{asc}{%
	\gdef\@journal{asc}
	\gdef\@journalfull{Automatic Systems and Control}
	\gdef\@journalshort{Autom. Syst. Control}
	\gdef\@doiabbr{asc}
	\gdef\@ISSN{}
	\gdef\@societyowner{}
}
\DeclareOption{asi}{%
	\gdef\@journal{asi}
	\gdef\@journalfull{Applied System Innovation}
	\gdef\@journalshort{Appl. Syst. Innov.}
	\gdef\@doiabbr{asi}
	\gdef\@ISSN{2571-5577}
	\gdef\@societyowner{International Institute of Knowledge Innovation and Invention}
}
\DeclareOption{astronautics}{%
        \gdef\@journal{astronautics}
        \gdef\@journalfull{Astronautics}
        \gdef\@journalshort{Astronautics}
        \gdef\@doiabbr{astronautics}
        \gdef\@ISSN{}
        \gdef\@societyowner{}
}
\DeclareOption{astronomy}{%
	\gdef\@journal{astronomy}
	\gdef\@journalfull{Astronomy}
	\gdef\@journalshort{Astronomy}
	\gdef\@doiabbr{astronomy}
	\gdef\@ISSN{2674-0346}
	\gdef\@societyowner{}
}
\DeclareOption{atmosphere}{%
	\gdef\@journal{atmosphere}
	\gdef\@journalfull{Atmosphere}
	\gdef\@journalshort{Atmosphere}
	\gdef\@doiabbr{atmos}
	\gdef\@ISSN{2073-4433}
	\gdef\@societyowner{}
}
\DeclareOption{atoms}{%
	\gdef\@journal{atoms}
	\gdef\@journalfull{Atoms}
	\gdef\@journalshort{Atoms}
	\gdef\@doiabbr{atoms}
	\gdef\@ISSN{2218-2004}
	\gdef\@societyowner{}
}
\DeclareOption{audiolres}{%
	\gdef\@journal{audiolres}
	\gdef\@journalfull{Audiology Research}
	\gdef\@journalshort{Audiol. Res.}
	\gdef\@doiabbr{audiolres}
	\gdef\@ISSN{2039-4349}
	\gdef\@societyowner{}
}
\DeclareOption{automation}{%
	\gdef\@journal{automation}
	\gdef\@journalfull{Automation}
	\gdef\@journalshort{Automation}
	\gdef\@doiabbr{automation}
	\gdef\@ISSN{2673-4052}
	\gdef\@societyowner{}
}
\DeclareOption{axioms}{%
	\gdef\@journal{axioms}
	\gdef\@journalfull{Axioms}
	\gdef\@journalshort{Axioms}
	\gdef\@doiabbr{axioms}
	\gdef\@ISSN{2075-1680}
	\gdef\@societyowner{}
}
\DeclareOption{bacteria}{%
	\gdef\@journal{bacteria}
	\gdef\@journalfull{Bacteria}
	\gdef\@journalshort{Bacteria}
	\gdef\@doiabbr{bacteria}
	\gdef\@ISSN{2674-1334}
	\gdef\@societyowner{}
}
\DeclareOption{batteries}{%
	\gdef\@journal{batteries}
	\gdef\@journalfull{Batteries}
	\gdef\@journalshort{Batteries}
	\gdef\@doiabbr{batteries}
	\gdef\@ISSN{2313-0105}
	\gdef\@societyowner{}
}
\DeclareOption{bdcc}{%
	\gdef\@journal{bdcc}
	\gdef\@journalfull{Big Data and Cognitive Computing}
	\gdef\@journalshort{Big Data Cogn. Comput.}
	\gdef\@doiabbr{bdcc}
	\gdef\@ISSN{2504-2289}
	\gdef\@societyowner{}
}
\DeclareOption{behavsci}{%
	\gdef\@journal{behavsci}
	\gdef\@journalfull{Behavioral Sciences}
	\gdef\@journalshort{Behav. Sci.}
	\gdef\@doiabbr{bs}
	\gdef\@ISSN{2076-328X}
	\gdef\@societyowner{}
}
\DeclareOption{beverages}{%
	\gdef\@journal{beverages}
	\gdef\@journalfull{Beverages}
	\gdef\@journalshort{Beverages}
	\gdef\@doiabbr{beverages}
	\gdef\@ISSN{2306-5710}
	\gdef\@societyowner{}
}
\DeclareOption{biochem}{%
	\gdef\@journal{biochem}
	\gdef\@journalfull{BioChem}
	\gdef\@journalshort{BioChem}
	\gdef\@doiabbr{biochem}
	\gdef\@ISSN{2673-6411}
	\gdef\@societyowner{}
}
\DeclareOption{bioengineering}{%
	\gdef\@journal{bioengineering}
	\gdef\@journalfull{Bioengineering}
	\gdef\@journalshort{Bioengineering}
	\gdef\@doiabbr{bioengineering}
	\gdef\@ISSN{2306-5354}
	\gdef\@societyowner{}
}
\DeclareOption{biologics}{%
	\gdef\@journal{biologics}
	\gdef\@journalfull{Biologics}
	\gdef\@journalshort{Biologics}
	\gdef\@doiabbr{biologics}
	\gdef\@ISSN{}
	\gdef\@societyowner{}
}
\DeclareOption{biology}{%
	\gdef\@journal{biology}
	\gdef\@journalfull{Biology}
	\gdef\@journalshort{Biology}
	\gdef\@doiabbr{biology}
	\gdef\@ISSN{2079-7737}
	\gdef\@societyowner{}
}
\DeclareOption{biomass}{%
	\gdef\@journal{biomass}
	\gdef\@journalfull{Biomass}
	\gdef\@journalshort{Biomass}
	\gdef\@doiabbr{biomass}
	\gdef\@ISSN{}
	\gdef\@societyowner{}
}
\DeclareOption{biomechanics}{%
	\gdef\@journal{biomechanics}
	\gdef\@journalfull{Biomechanics}
	\gdef\@journalshort{Biomechanics}
	\gdef\@doiabbr{biomechanics}
	\gdef\@ISSN{2673-7078}
	\gdef\@societyowner{}
}
\DeclareOption{biomed}{%
	\gdef\@journal{biomed}
	\gdef\@journalfull{BioMed}
	\gdef\@journalshort{BioMed}
	\gdef\@doiabbr{biomed}
	\gdef\@ISSN{}
	\gdef\@societyowner{}
}
\DeclareOption{biomedicines}{%
	\gdef\@journal{biomedicines}
	\gdef\@journalfull{Biomedicines}
	\gdef\@journalshort{Biomedicines}
	\gdef\@doiabbr{biomedicines}
	\gdef\@ISSN{2227-9059}
	\gdef\@societyowner{}
}
\DeclareOption{biomedinformatics}{%
	\gdef\@journal{biomedinformatics}
	\gdef\@journalfull{BioMedInformatics}
	\gdef\@journalshort{BioMedInformatics}
	\gdef\@doiabbr{biomedinformatics}
	\gdef\@ISSN{2673-7426}
	\gdef\@societyowner{}
}
\DeclareOption{biomimetics}{%
	\gdef\@journal{biomimetics}
	\gdef\@journalfull{Biomimetics}
	\gdef\@journalshort{Biomimetics}
	\gdef\@doiabbr{biomimetics}
	\gdef\@ISSN{2313-7673}
	\gdef\@societyowner{}
}
\DeclareOption{biomolecules}{%
	\gdef\@journal{biomolecules}
	\gdef\@journalfull{Biomolecules}
	\gdef\@journalshort{Biomolecules}
	\gdef\@doiabbr{biom}
	\gdef\@ISSN{2218-273X}
	\gdef\@societyowner{}
}
\DeclareOption{biophysica}{%
	\gdef\@journal{biophysica}
	\gdef\@journalfull{Biophysica}
	\gdef\@journalshort{Biophysica}
	\gdef\@doiabbr{biophysica}
	\gdef\@ISSN{2673-4125}
	\gdef\@societyowner{}
}
\DeclareOption{bioresourbioprod}{%
        \gdef\@journal{bioresourbioprod}
        \gdef\@journalfull{Bioresources and Bioproducts}
        \gdef\@journalshort{Bioresour. Bioprod.}
        \gdef\@doiabbr{bioresourbioprod}
        \gdef\@ISSN{}
        \gdef\@societyowner{}
}
\DeclareOption{biosensors}{%
	\gdef\@journal{biosensors}
	\gdef\@journalfull{Biosensors}
	\gdef\@journalshort{Biosensors}
	\gdef\@doiabbr{bios}
	\gdef\@ISSN{2079-6374}
	\gdef\@societyowner{}
}
\DeclareOption{biosphere}{%
       \gdef\@journal{biosphere}
       \gdef\@journalfull{Biosphere}
       \gdef\@journalshort{Biosphere}
       \gdef\@doiabbr{biosphere}
       \gdef\@ISSN{}
       \gdef\@societyowner{}
}
\DeclareOption{biotech}{%
	\gdef\@journal{biotech}
	\gdef\@journalfull{BioTech}
	\gdef\@journalshort{BioTech}
	\gdef\@doiabbr{biotech}
	\gdef\@ISSN{2673-6284}
	\gdef\@societyowner{}
}
\DeclareOption{birds}{%
	\gdef\@journal{birds}
	\gdef\@journalfull{Birds}
	\gdef\@journalshort{Birds}
	\gdef\@doiabbr{birds}
	\gdef\@ISSN{2673-6004}
	\gdef\@societyowner{}
}
\DeclareOption{blockchains}{%
	\gdef\@journal{blockchains}
	\gdef\@journalfull{Blockchains}
	\gdef\@journalshort{Blockchains}
	\gdef\@doiabbr{blockchains}
	\gdef\@ISSN{2813-5288}
	\gdef\@societyowner{}
}
\DeclareOption{bloods}{%
	\gdef\@journal{bloods}
	\gdef\@journalfull{Bloods}
	\gdef\@journalshort{Bloods}
	\gdef\@doiabbr{bloods}
	\gdef\@ISSN{2673-415X}
	\gdef\@societyowner{}
}
\DeclareOption{blsf}{%
	\gdef\@journal{blsf}
	\gdef\@journalfull{Biology and Life Sciences Forum}
	\gdef\@journalshort{Biol. Life Sci. Forum}
	\gdef\@doiabbr{blsf}
	\gdef\@ISSN{2673-9976}
	\gdef\@societyowner{}
}
\DeclareOption{brainsci}{%
	\gdef\@journal{brainsci}
	\gdef\@journalfull{Brain Sciences}
	\gdef\@journalshort{Brain Sci.}
	\gdef\@doiabbr{brainsci}
	\gdef\@ISSN{2076-3425}
	\gdef\@societyowner{}
}
\DeclareOption{breath}{%
	\gdef\@journal{breath}
	\gdef\@journalfull{Breath}
	\gdef\@journalshort{Breath}
	\gdef\@doiabbr{breath}
	\gdef\@ISSN{}
	\gdef\@societyowner{}
}
\DeclareOption{buildings}{%
	\gdef\@journal{buildings}
	\gdef\@journalfull{Buildings}
	\gdef\@journalshort{Buildings}
	\gdef\@doiabbr{buildings}
	\gdef\@ISSN{2075-5309}
	\gdef\@societyowner{}
}
\DeclareOption{businesses}{%
	\gdef\@journal{businesses}
	\gdef\@journalfull{Businesses}
	\gdef\@journalshort{Businesses}
	\gdef\@doiabbr{businesses}
	\gdef\@ISSN{2673-7116}
	\gdef\@societyowner{}
}
\DeclareOption{cancers}{%
	\gdef\@journal{cancers}
	\gdef\@journalfull{Cancers}
	\gdef\@journalshort{Cancers}
	\gdef\@doiabbr{cancers}
	\gdef\@ISSN{2072-6694}
	\gdef\@societyowner{}
}
\DeclareOption{carbon}{%
	\gdef\@journal{carbon}
	\gdef\@journalfull{C}
	\gdef\@journalshort{C}
	\gdef\@doiabbr{c}
	\gdef\@ISSN{2311-5629}
	\gdef\@societyowner{}
}
\DeclareOption{cardiogenetics}{%
	\gdef\@journal{cardiogenetics}
	\gdef\@journalfull{Cardiogenetics}
	\gdef\@journalshort{Cardiogenetics}
	\gdef\@doiabbr{cardiogenetics}
	\gdef\@ISSN{2035-8148}
	\gdef\@societyowner{}
}
\DeclareOption{catalysts}{%
	\gdef\@journal{catalysts}
	\gdef\@journalfull{Catalysts}
	\gdef\@journalshort{Catalysts}
	\gdef\@doiabbr{catal}
	\gdef\@ISSN{2073-4344}
	\gdef\@societyowner{}
}
\DeclareOption{cells}{%
	\gdef\@journal{cells}
	\gdef\@journalfull{Cells}
	\gdef\@journalshort{Cells}
	\gdef\@doiabbr{cells}
	\gdef\@ISSN{2073-4409}
	\gdef\@societyowner{}
}
\DeclareOption{ceramics}{%
	\gdef\@journal{ceramics}
	\gdef\@journalfull{Ceramics}
	\gdef\@journalshort{Ceramics}
	\gdef\@doiabbr{ceramics}
	\gdef\@ISSN{2571-6131}
	\gdef\@societyowner{}
}
\DeclareOption{challenges}{%
	\gdef\@journal{challenges}
	\gdef\@journalfull{Challenges}
	\gdef\@journalshort{Challenges}
	\gdef\@doiabbr{challe}
	\gdef\@ISSN{2078-1547}
	\gdef\@societyowner{}
}
\DeclareOption{chemengineering}{%
	\gdef\@journal{chemengineering}
	\gdef\@journalfull{ChemEngineering}
	\gdef\@journalshort{ChemEngineering}
	\gdef\@doiabbr{chemengineering}
	\gdef\@ISSN{2305-7084}
	\gdef\@societyowner{}
}
\DeclareOption{chemistry}{%
	\gdef\@journal{chemistry}
	\gdef\@journalfull{Chemistry}
	\gdef\@journalshort{Chemistry}
	\gdef\@doiabbr{chemistry}
	\gdef\@ISSN{2624-8549}
	\gdef\@societyowner{}
}
\DeclareOption{chemosensors}{%
	\gdef\@journal{chemosensors}
	\gdef\@journalfull{Chemosensors}
	\gdef\@journalshort{Chemosensors}
	\gdef\@doiabbr{chemosensors}
	\gdef\@ISSN{2227-9040}
	\gdef\@societyowner{}
}
\DeclareOption{chemproc}{%
	\gdef\@journal{chemproc}
	\gdef\@journalfull{Chemistry Proceedings}
	\gdef\@journalshort{Chem. Proc.}
	\gdef\@doiabbr{chemproc}
	\gdef\@ISSN{2673-4583}
	\gdef\@societyowner{}
}
\DeclareOption{children}{%
	\gdef\@journal{children}
	\gdef\@journalfull{Children}
	\gdef\@journalshort{Children}
	\gdef\@doiabbr{children}
	\gdef\@ISSN{2227-9067}
	\gdef\@societyowner{}
}
\DeclareOption{chips}{%
	\gdef\@journal{chips}
	\gdef\@journalfull{Chips}
	\gdef\@journalshort{Chips}
	\gdef\@doiabbr{chips}
	\gdef\@ISSN{2674-0729}
	\gdef\@societyowner{}
}
\DeclareOption{cimb}{%
	\gdef\@journal{cimb}
	\gdef\@journalfull{Current Issues in Molecular Biology}
	\gdef\@journalshort{Curr. Issues Mol. Biol.}
	\gdef\@doiabbr{cimb}
	\gdef\@ISSN{1467-3045}
	\gdef\@societyowner{}
}
\DeclareOption{civileng}{%
	\gdef\@journal{civileng}
	\gdef\@journalfull{CivilEng}
	\gdef\@journalshort{CivilEng}
	\gdef\@doiabbr{civileng}
	\gdef\@ISSN{2673-4109}
	\gdef\@societyowner{}
}
\DeclareOption{cleantechnol}{%
	\gdef\@journal{cleantechnol}
	\gdef\@journalfull{Clean Technologies}
	\gdef\@journalshort{Clean Technol.}
	\gdef\@doiabbr{cleantechnol}
	\gdef\@ISSN{2571-8797}
	\gdef\@societyowner{}
}
\DeclareOption{climate}{%
	\gdef\@journal{climate}
	\gdef\@journalfull{Climate}
	\gdef\@journalshort{Climate}
	\gdef\@doiabbr{cli}
	\gdef\@ISSN{2225-1154}
	\gdef\@societyowner{}
}
\DeclareOption{clinbioenerg}{%
      \gdef\@journal{clinbioenerg}
      \gdef\@journalfull{Clinical Bioenergetics}
      \gdef\@journalshort{Clin. Bioenerg.}
      \gdef\@doiabbr{clinbioenerg}
      \gdef\@ISSN{3042-5158}
      \gdef\@societyowner{}
}
\DeclareOption{clinpract}{%
	\gdef\@journal{clinpract}
	\gdef\@journalfull{Clinics and Practice}
	\gdef\@journalshort{Clin. Pract.}
	\gdef\@doiabbr{clinpract}
	\gdef\@ISSN{2039-7283}
	\gdef\@societyowner{}
}
\DeclareOption{clockssleep}{%
	\gdef\@journal{clockssleep}
	\gdef\@journalfull{Clocks \& Sleep}
	\gdef\@journalshort{Clocks\&Sleep}
	\gdef\@doiabbr{clockssleep}
	\gdef\@ISSN{2624-5175}
	\gdef\@societyowner{}
}
\DeclareOption{cmd}{%
	\gdef\@journal{cmd}
	\gdef\@journalfull{Corrosion and Materials Degradation}
	\gdef\@journalshort{Corros. Mater. Degrad.}
	\gdef\@doiabbr{cmd}
	\gdef\@ISSN{2624-5558}
	\gdef\@societyowner{}
}
\DeclareOption{cmtr}{%
        \gdef\@journal{cmtr}
        \gdef\@journalfull{Craniomaxillofacial Trauma \& Reconstruction}
        \gdef\@journalshort{Craniomaxillofac. Trauma Reconstr.}
       \gdef\@doiabbr{cmtr}
       \gdef\@ISSN{1943-3883}
       \gdef\@societyowner{AO Foundation}
}
\DeclareOption{coasts}{%
	\gdef\@journal{coasts}
	\gdef\@journalfull{Coasts}
	\gdef\@journalshort{Coasts}
	\gdef\@doiabbr{coasts}
	\gdef\@ISSN{2673-964X}
	\gdef\@societyowner{}
}
\DeclareOption{coatings}{%
	\gdef\@journal{coatings}
	\gdef\@journalfull{Coatings}
	\gdef\@journalshort{Coatings}
	\gdef\@doiabbr{coatings}
	\gdef\@ISSN{2079-6412}
	\gdef\@societyowner{}
}
\DeclareOption{colloids}{%
	\gdef\@journal{colloids}
	\gdef\@journalfull{Colloids and Interfaces}
	\gdef\@journalshort{Colloids Interfaces}
	\gdef\@doiabbr{colloids}
	\gdef\@ISSN{2504-5377}
	\gdef\@societyowner{}
}
\DeclareOption{colorants}{%
	\gdef\@journal{colorants}
	\gdef\@journalfull{Colorants}
	\gdef\@journalshort{Colorants}
	\gdef\@doiabbr{colorants}
	\gdef\@ISSN{2079-6447}
	\gdef\@societyowner{}
}
\DeclareOption{commodities}{%
	\gdef\@journal{commodities}
	\gdef\@journalfull{Commodities}
	\gdef\@journalshort{Commodities}
	\gdef\@doiabbr{commodities}
	\gdef\@ISSN{}
	\gdef\@societyowner{}
}
\DeclareOption{complexities}{%
        \gdef\@journal{complexities}
        \gdef\@journalfull{Complexities}
       \gdef\@journalshort{Complexities}
       \gdef\@doiabbr{complexities}
       \gdef\@ISSN{}
       \gdef\@societyowner{}
}
\DeclareOption{complications}{%
       \gdef\@journal{complications}
       \gdef\@journalfull{Complications}
       \gdef\@journalshort{Complications}
       \gdef\@doiabbr{complications}
       \gdef\@ISSN{}
       \gdef\@societyowner{}
}
\DeclareOption{compounds}{%
	\gdef\@journal{compounds}
	\gdef\@journalfull{Compounds}
	\gdef\@journalshort{Compounds}
	\gdef\@doiabbr{compounds}
	\gdef\@ISSN{2673-6918}
	\gdef\@societyowner{}
}
\DeclareOption{computation}{%
	\gdef\@journal{computation}
	\gdef\@journalfull{Computation}
	\gdef\@journalshort{Computation}
	\gdef\@doiabbr{computation}
	\gdef\@ISSN{2079-3197}
	\gdef\@societyowner{}
}
\DeclareOption{computers}{%
	\gdef\@journal{computers}
	\gdef\@journalfull{Computers}
	\gdef\@journalshort{Computers}
	\gdef\@doiabbr{computers}
	\gdef\@ISSN{2073-431X}
	\gdef\@societyowner{}
}
\DeclareOption{condensedmatter}{%
	\gdef\@journal{condensedmatter}
	\gdef\@journalfull{Condensed Matter}
	\gdef\@journalshort{Condens. Matter}
	\gdef\@doiabbr{condmat}
	\gdef\@ISSN{2410-3896}
	\gdef\@societyowner{}
}
\DeclareOption{conservation}{%
	\gdef\@journal{conservation}
	\gdef\@journalfull{Conservation}
	\gdef\@journalshort{Conservation}
	\gdef\@doiabbr{conservation}
	\gdef\@ISSN{2673-7159}
	\gdef\@societyowner{}
}
\DeclareOption{constrmater}{%
	\gdef\@journal{constrmater}
	\gdef\@journalfull{Construction Materials}
	\gdef\@journalshort{Constr. Mater.}
	\gdef\@doiabbr{constrmater}
	\gdef\@ISSN{2673-7108}
	\gdef\@societyowner{}
}
\DeclareOption{cosmetics}{%
	\gdef\@journal{cosmetics}
	\gdef\@journalfull{Cosmetics}
	\gdef\@journalshort{Cosmetics}
	\gdef\@doiabbr{cosmetics}
	\gdef\@ISSN{2079-9284}
	\gdef\@societyowner{}
}
\DeclareOption{covid}{%
	\gdef\@journal{covid}
	\gdef\@journalfull{COVID}
	\gdef\@journalshort{COVID}
	\gdef\@doiabbr{covid}
	\gdef\@ISSN{2673-8112}
	\gdef\@societyowner{}
}
\DeclareOption{crops}{%
	\gdef\@journal{crops}
	\gdef\@journalfull{Crops}
	\gdef\@journalshort{Crops}
	\gdef\@doiabbr{crops}
	\gdef\@ISSN{2673-7655}
	\gdef\@societyowner{}
}
\DeclareOption{cryo}{%
	\gdef\@journal{cryo}
	\gdef\@journalfull{Cryo}
	\gdef\@journalshort{Cryo}
	\gdef\@doiabbr{cryo}
	\gdef\@ISSN{}
	\gdef\@societyowner{}
}
\DeclareOption{cryptography}{%
	\gdef\@journal{cryptography}
	\gdef\@journalfull{Cryptography}
	\gdef\@journalshort{Cryptography}
	\gdef\@doiabbr{cryptography}
	\gdef\@ISSN{2410-387X}
	\gdef\@societyowner{}
}
\DeclareOption{crystals}{%
	\gdef\@journal{crystals}
	\gdef\@journalfull{Crystals}
	\gdef\@journalshort{Crystals}
	\gdef\@doiabbr{cryst}
	\gdef\@ISSN{2073-4352}
	\gdef\@societyowner{}
}
\DeclareOption{csmf}{%
	\gdef\@journal{csmf}
	\gdef\@journalfull{Computer Sciences \& Mathematics Forum}
	\gdef\@journalshort{Comput. Sci. Math. Forum}
	\gdef\@doiabbr{csmf}
	\gdef\@ISSN{2813-0324}
	\gdef\@societyowner{}
}
\DeclareOption{ctn}{%
	\gdef\@journal{ctn}
	\gdef\@journalfull{Clinical and Translational Neuroscience}
	\gdef\@journalshort{Clin. Transl. Neurosci.}
	\gdef\@doiabbr{ctn}
	\gdef\@ISSN{2514-183X}
	\gdef\@societyowner{Swiss Federation of Clinical Neuro-Societies}
}
\DeclareOption{curroncol}{%
	\gdef\@journal{curroncol}
	\gdef\@journalfull{Current Oncology}
	\gdef\@journalshort{Curr. Oncol.}
	\gdef\@doiabbr{curroncol}
	\gdef\@ISSN{1718-7729}
	\gdef\@societyowner{}
}
\DeclareOption{cyber}{%
	\gdef\@journal{cyber}
	\gdef\@journalfull{Cyber}
	\gdef\@journalshort{Cyber}
	\gdef\@doiabbr{cyber}
	\gdef\@ISSN{2624-800X}
	\gdef\@societyowner{}
}
\DeclareOption{dairy}{%
	\gdef\@journal{dairy}
	\gdef\@journalfull{Dairy}
	\gdef\@journalshort{Dairy}
	\gdef\@doiabbr{dairy}
	\gdef\@ISSN{2624-862X}
	\gdef\@societyowner{}
}
\DeclareOption{data}{%
	\gdef\@journal{data}
	\gdef\@journalfull{Data}
	\gdef\@journalshort{Data}
	\gdef\@doiabbr{data}
	\gdef\@ISSN{2306-5729}
	\gdef\@societyowner{}
}
\DeclareOption{ddc}{%
	\gdef\@journal{ddc}
	\gdef\@journalfull{Drugs and Drug Candidates}
	\gdef\@journalshort{Drugs Drug Candidates}
	\gdef\@doiabbr{ddc}
	\gdef\@ISSN{2813-2998}
	\gdef\@societyowner{}
}
\DeclareOption{dentistry}{%
	\gdef\@journal{dentistry}
	\gdef\@journalfull{Dentistry Journal}
	\gdef\@journalshort{Dent. J.}
	\gdef\@doiabbr{dj}
	\gdef\@ISSN{2304-6767}
	\gdef\@societyowner{}
}
\DeclareOption{dermato}{%
	\gdef\@journal{dermato}
	\gdef\@journalfull{Dermato}
	\gdef\@journalshort{Dermato}
	\gdef\@doiabbr{dermato}
	\gdef\@ISSN{2673-6179}
	\gdef\@societyowner{}
}
\DeclareOption{dermatopathology}{%
	\gdef\@journal{dermatopathology}
	\gdef\@journalfull{Dermatopathology}
	\gdef\@journalshort{Dermatopathology}
	\gdef\@doiabbr{dermatopathology}
	\gdef\@ISSN{2296-3529}
	\gdef\@societyowner{European Society of Dermatopathology}
}
\DeclareOption{designs}{%
	\gdef\@journal{designs}
	\gdef\@journalfull{Designs}
	\gdef\@journalshort{Designs}
	\gdef\@doiabbr{designs}
	\gdef\@ISSN{2411-9660}
	\gdef\@societyowner{}
}
\DeclareOption{devices}{%
	\gdef\@journal{devices}
	\gdef\@journalfull{Devices}
	\gdef\@journalshort{Devices}
	\gdef\@doiabbr{devices}
	\gdef\@ISSN{}
	\gdef\@societyowner{}
}
\DeclareOption{diabetology}{%
	\gdef\@journal{diabetology}
	\gdef\@journalfull{Diabetology}
	\gdef\@journalshort{Diabetology}
	\gdef\@doiabbr{diabetology}
	\gdef\@ISSN{2673-4540}
	\gdef\@societyowner{}
}
\DeclareOption{diagnostics}{%
	\gdef\@journal{diagnostics}
	\gdef\@journalfull{Diagnostics}
	\gdef\@journalshort{Diagnostics}
	\gdef\@doiabbr{diagnostics}
	\gdef\@ISSN{2075-4418}
	\gdef\@societyowner{}
}
\DeclareOption{dietetics}{%
	\gdef\@journal{dietetics}
	\gdef\@journalfull{Dietetics}
	\gdef\@journalshort{Dietetics}
	\gdef\@doiabbr{dietetics}
	\gdef\@ISSN{2674-0311}
	\gdef\@societyowner{}
}
\DeclareOption{digital}{%
	\gdef\@journal{digital}
	\gdef\@journalfull{Digital}
	\gdef\@journalshort{Digital}
	\gdef\@doiabbr{digital}
	\gdef\@ISSN{2673-6470}
	\gdef\@societyowner{}
}
\DeclareOption{disabilities}{%
	\gdef\@journal{disabilities}
	\gdef\@journalfull{Disabilities}
	\gdef\@journalshort{Disabilities}
	\gdef\@doiabbr{disabilities}
	\gdef\@ISSN{2673-7272}
	\gdef\@societyowner{}
}
\DeclareOption{diseases}{%
	\gdef\@journal{diseases}
	\gdef\@journalfull{Diseases}
	\gdef\@journalshort{Diseases}
	\gdef\@doiabbr{diseases}
	\gdef\@ISSN{2079-9721}
	\gdef\@societyowner{}
}
\DeclareOption{diversity}{%
	\gdef\@journal{diversity}
	\gdef\@journalfull{Diversity}
	\gdef\@journalshort{Diversity}
	\gdef\@doiabbr{d}
	\gdef\@ISSN{1424-2818}
	\gdef\@societyowner{}
}
\DeclareOption{dna}{%
	\gdef\@journal{dna}
	\gdef\@journalfull{DNA}
	\gdef\@journalshort{DNA}
	\gdef\@doiabbr{dna}
	\gdef\@ISSN{2673-8856}
	\gdef\@societyowner{}
}
\DeclareOption{drones}{%
	\gdef\@journal{drones}
	\gdef\@journalfull{Drones}
	\gdef\@journalshort{Drones}
	\gdef\@doiabbr{drones}
	\gdef\@ISSN{2504-446X}
	\gdef\@societyowner{}
}
\DeclareOption{dynamics}{%
	\gdef\@journal{dynamics}
	\gdef\@journalfull{Dynamics}
	\gdef\@journalshort{Dynamics}
	\gdef\@doiabbr{dynamics}
	\gdef\@ISSN{2673-8716}
	\gdef\@societyowner{}
}
\DeclareOption{earth}{%
	\gdef\@journal{earth}
	\gdef\@journalfull{Earth}
	\gdef\@journalshort{Earth}
	\gdef\@doiabbr{earth}
	\gdef\@ISSN{2673-4834}
	\gdef\@societyowner{}
}
\DeclareOption{ebj}{%
	\gdef\@journal{ebj}
	\gdef\@journalfull{European Burn Journal}
	\gdef\@journalshort{Eur. Burn J.}
	\gdef\@doiabbr{ebj}
	\gdef\@ISSN{2673-1991}
	\gdef\@societyowner{European Burns Association}
}
\DeclareOption{ecm}{%
	\gdef\@journal{ecm}
	\gdef\@journalfull{Emergency Care and Medicine}
	\gdef\@journalshort{Emerg. Care Med.}
	\gdef\@doiabbr{ecm}
	\gdef\@ISSN{}
	\gdef\@societyowner{}
}
\DeclareOption{ecologies}{%
	\gdef\@journal{ecologies}
	\gdef\@journalfull{Ecologies}
	\gdef\@journalshort{Ecologies}
	\gdef\@doiabbr{ecologies}
	\gdef\@ISSN{2673-4133}
	\gdef\@societyowner{}
}
\DeclareOption{econometrics}{%
	\gdef\@journal{econometrics}
	\gdef\@journalfull{Econometrics}
	\gdef\@journalshort{Econometrics}
	\gdef\@doiabbr{econometrics}
	\gdef\@ISSN{2225-1146}
	\gdef\@societyowner{}
}
\DeclareOption{economies}{%
	\gdef\@journal{economies}
	\gdef\@journalfull{Economies}
	\gdef\@journalshort{Economies}
	\gdef\@doiabbr{economies}
	\gdef\@ISSN{2227-7099}
	\gdef\@societyowner{}
}
\DeclareOption{edm}{%
        \gdef\@journal{edm}
        \gdef\@journalfull{Entropic and Disordered Matter}
        \gdef\@journalshort{Entropic Disord. Matter}
        \gdef\@doiabbr{edm}
        \gdef\@ISSN{3042-7592}
        \gdef\@societyowner{}
}
\DeclareOption{education}{%
	\gdef\@journal{education}
	\gdef\@journalfull{Education Sciences}
	\gdef\@journalshort{Educ. Sci.}
	\gdef\@doiabbr{educsci}
	\gdef\@ISSN{2227-7102}
	\gdef\@societyowner{}
}
\DeclareOption{ejihpe}{%
	\gdef\@journal{ejihpe}
	\gdef\@journalfull{European Journal of Investigation in Health, Psychology and Education}
	\gdef\@journalshort{Eur. J. Investig. Health Psychol. Educ.}
	\gdef\@doiabbr{ejihpe}
	\gdef\@ISSN{2254-9625}
	\gdef\@societyowner{University Association of Education and Psychology}
}
\DeclareOption{electricity}{%
	\gdef\@journal{electricity}
	\gdef\@journalfull{Electricity}
	\gdef\@journalshort{Electricity}
	\gdef\@doiabbr{electricity}
	\gdef\@ISSN{2673-4826}
	\gdef\@societyowner{}
}
\DeclareOption{electrochem}{%
	\gdef\@journal{electrochem}
	\gdef\@journalfull{Electrochem}
	\gdef\@journalshort{Electrochem}
	\gdef\@doiabbr{electrochem}
	\gdef\@ISSN{2673-3293}
	\gdef\@societyowner{}
}
\DeclareOption{electronicmat}{%
	\gdef\@journal{electronicmat}
	\gdef\@journalfull{Electronic Materials}
	\gdef\@journalshort{Electron. Mater.}
	\gdef\@doiabbr{electronicmat}
	\gdef\@ISSN{2673-3978}
	\gdef\@societyowner{}
}
\DeclareOption{electronics}{%
	\gdef\@journal{electronics}
	\gdef\@journalfull{Electronics}
	\gdef\@journalshort{Electronics}
	\gdef\@doiabbr{electronics}
	\gdef\@ISSN{2079-9292}
	\gdef\@societyowner{}
}
\DeclareOption{encyclopedia}{%
	\gdef\@journal{encyclopedia}
	\gdef\@journalfull{Encyclopedia}
	\gdef\@journalshort{Encyclopedia}
	\gdef\@doiabbr{encyclopedia}
	\gdef\@ISSN{2673-8392}
	\gdef\@societyowner{}
}
\DeclareOption{endocrines}{%
	\gdef\@journal{endocrines}
	\gdef\@journalfull{Endocrines}
	\gdef\@journalshort{Endocrines}
	\gdef\@doiabbr{endocrines}
	\gdef\@ISSN{2673-396X}
	\gdef\@societyowner{}
}
\DeclareOption{energies}{%
	\gdef\@journal{energies}
	\gdef\@journalfull{Energies}
	\gdef\@journalshort{Energies}
	\gdef\@doiabbr{en}
	\gdef\@ISSN{1996-1073}
	\gdef\@societyowner{}
}
\DeclareOption{eng}{%
	\gdef\@journal{eng}
	\gdef\@journalfull{Eng}
	\gdef\@journalshort{Eng}
	\gdef\@doiabbr{eng}
	\gdef\@ISSN{2673-4117}
	\gdef\@societyowner{}
}
\DeclareOption{engproc}{%
	\gdef\@journal{engproc}
	\gdef\@journalfull{Engineering Proceedings}
	\gdef\@journalshort{Eng. Proc.}
	\gdef\@doiabbr{engproc}
	\gdef\@ISSN{2673-4591}
	\gdef\@societyowner{}
}
\DeclareOption{entomology}{%
	\gdef\@journal{entomology}
	\gdef\@journalfull{Entomology}
	\gdef\@journalshort{Entomology}
	\gdef\@doiabbr{entomology}
	\gdef\@ISSN{2813-0502}
	\gdef\@societyowner{}
}
\DeclareOption{entropy}{%
	\gdef\@journal{entropy}
	\gdef\@journalfull{Entropy}
	\gdef\@journalshort{Entropy}
	\gdef\@doiabbr{e}
	\gdef\@ISSN{1099-4300}
	\gdef\@societyowner{}
}
\DeclareOption{environments}{%
	\gdef\@journal{environments}
	\gdef\@journalfull{Environments}
	\gdef\@journalshort{Environments}
	\gdef\@doiabbr{environments}
	\gdef\@ISSN{2076-3298}
	\gdef\@societyowner{}
}
\DeclareOption{eesp}{%
	\gdef\@journal{eesp}
	\gdef\@journalfull{Environmental and Earth Sciences Proceedings}
	\gdef\@journalshort{Environ. Earth Sci. Proc.}
	\gdef\@doiabbr{eesp}
	\gdef\@ISSN{3042-5743}
	\gdef\@societyowner{}
}
\DeclareOption{epidemiologia}{%
	\gdef\@journal{epidemiologia}
	\gdef\@journalfull{Epidemiologia}
	\gdef\@journalshort{Epidemiologia}
	\gdef\@doiabbr{epidemiologia}
	\gdef\@ISSN{2673-3986}
	\gdef\@societyowner{}
}
\DeclareOption{epigenomes}{%
	\gdef\@journal{epigenomes}
	\gdef\@journalfull{Epigenomes}
	\gdef\@journalshort{Epigenomes}
	\gdef\@doiabbr{epigenomes}
	\gdef\@ISSN{2075-4655}
	\gdef\@societyowner{}
}
\DeclareOption{esa}{%
	\gdef\@journal{esa}
	\gdef\@journalfull{Energy Storage and Applications}
	\gdef\@journalshort{Energy Storage Appl.}
	\gdef\@doiabbr{esa}
	\gdef\@ISSN{}
	\gdef\@societyowner{}
}
\DeclareOption{est}{%
	\gdef\@journal{est}
	\gdef\@journalfull{Electrochemical Science and Technology}
	\gdef\@journalshort{Electrochem. Sci. Technol.}
	\gdef\@doiabbr{est}
	\gdef\@ISSN{}
	\gdef\@societyowner{}
}
\DeclareOption{famsci}{%
      \gdef\@journal{famsci}
      \gdef\@journalfull{Family Sciences}
      \gdef\@journalshort{Fam. Sci.}
      \gdef\@doiabbr{famsci}
      \gdef\@ISSN{}
      \gdef\@societyowner{}
}
\DeclareOption{fermentation}{%
	\gdef\@journal{fermentation}
	\gdef\@journalfull{Fermentation}
	\gdef\@journalshort{Fermentation}
	\gdef\@doiabbr{fermentation}
	\gdef\@ISSN{2311-5637}
	\gdef\@societyowner{}
}
\DeclareOption{fibers}{%
	\gdef\@journal{fibers}
	\gdef\@journalfull{Fibers}
	\gdef\@journalshort{Fibers}
	\gdef\@doiabbr{fib}
	\gdef\@ISSN{2079-6439}
	\gdef\@societyowner{}
}
\DeclareOption{fintech}{%
	\gdef\@journal{fintech}
	\gdef\@journalfull{FinTech}
	\gdef\@journalshort{FinTech}
	\gdef\@doiabbr{fintech}
	\gdef\@ISSN{2674-1032}
	\gdef\@societyowner{}
}
\DeclareOption{fire}{%
	\gdef\@journal{fire}
	\gdef\@journalfull{Fire}
	\gdef\@journalshort{Fire}
	\gdef\@doiabbr{fire}
	\gdef\@ISSN{2571-6255}
	\gdef\@societyowner{}
}
\DeclareOption{fishes}{%
	\gdef\@journal{fishes}
	\gdef\@journalfull{Fishes}
	\gdef\@journalshort{Fishes}
	\gdef\@doiabbr{fishes}
	\gdef\@ISSN{2410-3888}
	\gdef\@societyowner{}
}
\DeclareOption{fluids}{%
	\gdef\@journal{fluids}
	\gdef\@journalfull{Fluids}
	\gdef\@journalshort{Fluids}
	\gdef\@doiabbr{fluids}
	\gdef\@ISSN{2311-5521}
	\gdef\@societyowner{}
}
\DeclareOption{foods}{%
	\gdef\@journal{foods}
	\gdef\@journalfull{Foods}
	\gdef\@journalshort{Foods}
	\gdef\@doiabbr{foods}
	\gdef\@ISSN{2304-8158}
	\gdef\@societyowner{}
}
\DeclareOption{forecasting}{%
	\gdef\@journal{forecasting}
	\gdef\@journalfull{Forecasting}
	\gdef\@journalshort{Forecasting}
	\gdef\@doiabbr{forecast}
	\gdef\@ISSN{2571-9394}
	\gdef\@societyowner{}
}
\DeclareOption{forensicsci}{%
	\gdef\@journal{forensicsci}
	\gdef\@journalfull{Forensic Sciences}
	\gdef\@journalshort{Forensic Sci.}
	\gdef\@doiabbr{forensicsci}
	\gdef\@ISSN{2673-6756}
	\gdef\@societyowner{}
}
\DeclareOption{forests}{%
	\gdef\@journal{forests}
	\gdef\@journalfull{Forests}
	\gdef\@journalshort{Forests}
	\gdef\@doiabbr{f}
	\gdef\@ISSN{1999-4907}
	\gdef\@societyowner{}
}
\DeclareOption{fossstud}{%
	\gdef\@journal{fossstud}
	\gdef\@journalfull{Fossil Studies}
	\gdef\@journalshort{Foss. Stud.}
	\gdef\@doiabbr{fossils}
	\gdef\@ISSN{2813-6284}
	\gdef\@societyowner{}
}
\DeclareOption{foundations}{%
	\gdef\@journal{foundations}
	\gdef\@journalfull{Foundations}
	\gdef\@journalshort{Foundations}
	\gdef\@doiabbr{foundations}
	\gdef\@ISSN{2673-9321}
	\gdef\@societyowner{}
}
\DeclareOption{fractalfract}{%
	\gdef\@journal{fractalfract}
	\gdef\@journalfull{Fractal and Fractional}
	\gdef\@journalshort{Fractal Fract.}
	\gdef\@doiabbr{fractalfract}
	\gdef\@ISSN{2504-3110}
	\gdef\@societyowner{}
}
\DeclareOption{fuels}{%
	\gdef\@journal{fuels}
	\gdef\@journalfull{Fuels}
	\gdef\@journalshort{Fuels}
	\gdef\@doiabbr{fuels}
	\gdef\@ISSN{2673-3994}
	\gdef\@societyowner{}
}
\DeclareOption{future}{%
	\gdef\@journal{future}
	\gdef\@journalfull{Future}
	\gdef\@journalshort{Future}
	\gdef\@doiabbr{future}
	\gdef\@ISSN{2813-2882}
	\gdef\@societyowner{}
}
\DeclareOption{futureinternet}{%
	\gdef\@journal{futureinternet}
	\gdef\@journalfull{Future Internet}
	\gdef\@journalshort{Future Internet}
	\gdef\@doiabbr{fi}
	\gdef\@ISSN{1999-5903}
	\gdef\@societyowner{}
}
\DeclareOption{futurepharmacol}{%
	\gdef\@journal{futurepharmacol}
	\gdef\@journalfull{Future Pharmacology}
	\gdef\@journalshort{Future Pharmacol.}
	\gdef\@doiabbr{futurepharmacol}
	\gdef\@ISSN{2673-9879}
	\gdef\@societyowner{}
}
\DeclareOption{futurephys}{%
	\gdef\@journal{futurephys}
	\gdef\@journalfull{Future Physics}
	\gdef\@journalshort{Future Phys.}
	\gdef\@doiabbr{futurephys}
	\gdef\@ISSN{2624-6503}
	\gdef\@societyowner{}
}
\DeclareOption{futuretransp}{%
	\gdef\@journal{futuretransp}
	\gdef\@journalfull{Future Transportation}
	\gdef\@journalshort{Future Transp.}
	\gdef\@doiabbr{futuretransp}
	\gdef\@ISSN{}
	\gdef\@societyowner{}
}
\DeclareOption{galaxies}{%
	\gdef\@journal{galaxies}
	\gdef\@journalfull{Galaxies}
	\gdef\@journalshort{Galaxies}
	\gdef\@doiabbr{galaxies}
	\gdef\@ISSN{2075-4434}
	\gdef\@societyowner{}
}
\DeclareOption{games}{%
	\gdef\@journal{games}
	\gdef\@journalfull{Games}
	\gdef\@journalshort{Games}
	\gdef\@doiabbr{g}
	\gdef\@ISSN{2073-4336}
	\gdef\@societyowner{}
}
\DeclareOption{gases}{%
	\gdef\@journal{gases}
	\gdef\@journalfull{Gases}
	\gdef\@journalshort{Gases}
	\gdef\@doiabbr{gases}
	\gdef\@ISSN{2673-5628}
	\gdef\@societyowner{}
}
\DeclareOption{gastroent}{%
	\gdef\@journal{gastroent}
	\gdef\@journalfull{Gastroenterology Insights}
	\gdef\@journalshort{Gastroenterol. Insights}
	\gdef\@doiabbr{gastroent}
	\gdef\@ISSN{2036-7422}
	\gdef\@societyowner{}
}
\DeclareOption{gastrointestdisord}{%
	\gdef\@journal{gastrointestdisord}
	\gdef\@journalfull{Gastrointestinal Disorders}
	\gdef\@journalshort{Gastrointest. Disord.}
	\gdef\@doiabbr{gidisord}
	\gdef\@ISSN{2624-5647}
	\gdef\@societyowner{}
}
\DeclareOption{gastronomy}{%
	\gdef\@journal{gastronomy}
	\gdef\@journalfull{Gastronomy}
	\gdef\@journalshort{Gastronomy}
	\gdef\@doiabbr{gastronomy}
	\gdef\@ISSN{}
	\gdef\@societyowner{}
}
\DeclareOption{gels}{%
	\gdef\@journal{gels}
	\gdef\@journalfull{Gels}
	\gdef\@journalshort{Gels}
	\gdef\@doiabbr{gels}
	\gdef\@ISSN{2310-2861}
	\gdef\@societyowner{}
}
\DeclareOption{genealogy}{%
	\gdef\@journal{genealogy}
	\gdef\@journalfull{Genealogy}
	\gdef\@journalshort{Genealogy}
	\gdef\@doiabbr{genealogy}
	\gdef\@ISSN{2313-5778}
	\gdef\@societyowner{}
}
\DeclareOption{genes}{%
	\gdef\@journal{genes}
	\gdef\@journalfull{Genes}
	\gdef\@journalshort{Genes}
	\gdef\@doiabbr{genes}
	\gdef\@ISSN{2073-4425}
	\gdef\@societyowner{}
}
\DeclareOption{geographies}{%
	\gdef\@journal{geographies}
	\gdef\@journalfull{Geographies}
	\gdef\@journalshort{Geographies}
	\gdef\@doiabbr{geographies}
	\gdef\@ISSN{2673-7086}
	\gdef\@societyowner{}
}
\DeclareOption{geohazards}{%
	\gdef\@journal{geohazards}
	\gdef\@journalfull{GeoHazards}
	\gdef\@journalshort{GeoHazards}
	\gdef\@doiabbr{geohazards}
	\gdef\@ISSN{2624-795X}
	\gdef\@societyowner{}
}
\DeclareOption{geomatics}{%
	\gdef\@journal{geomatics}
	\gdef\@journalfull{Geomatics}
	\gdef\@journalshort{Geomatics}
	\gdef\@doiabbr{geomatics}
	\gdef\@ISSN{2673-7418}
	\gdef\@societyowner{}
}
\DeclareOption{geometry}{%
	\gdef\@journal{geometry}
	\gdef\@journalfull{Geometry}
	\gdef\@journalshort{Geometry}
	\gdef\@doiabbr{geometry}
	\gdef\@ISSN{}
	\gdef\@societyowner{}
}
\DeclareOption{geosciences}{%
	\gdef\@journal{geosciences}
	\gdef\@journalfull{Geosciences}
	\gdef\@journalshort{Geosciences}
	\gdef\@doiabbr{geosciences}
	\gdef\@ISSN{2076-3263}
	\gdef\@societyowner{}
}
\DeclareOption{geotechnics}{%
	\gdef\@journal{geotechnics}
	\gdef\@journalfull{Geotechnics}
	\gdef\@journalshort{Geotechnics}
	\gdef\@doiabbr{geotechnics}
	\gdef\@ISSN{2673-7094}
	\gdef\@societyowner{}
}
\DeclareOption{geriatrics}{%
	\gdef\@journal{geriatrics}
	\gdef\@journalfull{Geriatrics}
	\gdef\@journalshort{Geriatrics}
	\gdef\@doiabbr{geriatrics}
	\gdef\@ISSN{2308-3417}
	\gdef\@societyowner{}
}
\DeclareOption{glacies}{%
	\gdef\@journal{glacies}
	\gdef\@journalfull{Glacies}
	\gdef\@journalshort{Glacies}
	\gdef\@doiabbr{glacies}
	\gdef\@ISSN{}
	\gdef\@societyowner{}
}
\DeclareOption{grasses}{%
	\gdef\@journal{grasses}
	\gdef\@journalfull{Grasses}
	\gdef\@journalshort{Grasses}
	\gdef\@doiabbr{grasses}
	\gdef\@ISSN{}
	\gdef\@societyowner{}
}
\DeclareOption{greenhealth}{%
       \gdef\@journal{greenhealth}
       \gdef\@journalfull{Green Health}
      \gdef\@journalshort{Green Health}
      \gdef\@doiabbr{greenhealth}
      \gdef\@ISSN{}
      \gdef\@societyowner{}
}
\DeclareOption{gucdd}{%
	\gdef\@journal{gucdd}
	\gdef\@journalfull{Gout, Urate, and Crystal Deposition Disease}
	\gdef\@journalshort{Gout Urate Cryst. Depos. Dis.}
	\gdef\@doiabbr{gucdd}
	\gdef\@ISSN{}
	\gdef\@societyowner{Gout, Hyperuricemia and Crystal Associated Disease Network}
}
\DeclareOption{hardware}{%
	\gdef\@journal{hardware}
	\gdef\@journalfull{Hardware}
	\gdef\@journalshort{Hardware}
	\gdef\@doiabbr{hardware}
	\gdef\@ISSN{}
	\gdef\@societyowner{}
}
\DeclareOption{hazardousmatters}{%
	\gdef\@journal{hazardousmatters}
	\gdef\@journalfull{Hazardous Matters}
	\gdef\@journalshort{Hazard. Matters}
	\gdef\@doiabbr{hazardousmatters}
	\gdef\@ISSN{0014-0014}
	\gdef\@societyowner{}
}
\DeclareOption{healthcare}{%
	\gdef\@journal{healthcare}
	\gdef\@journalfull{Healthcare}
	\gdef\@journalshort{Healthcare}
	\gdef\@doiabbr{healthcare}
	\gdef\@ISSN{2227-9032}
	\gdef\@societyowner{}
}
\DeclareOption{hearts}{%
	\gdef\@journal{hearts}
	\gdef\@journalfull{Hearts}
	\gdef\@journalshort{Hearts}
	\gdef\@doiabbr{hearts}
	\gdef\@ISSN{2673-3846}
	\gdef\@societyowner{}
}
\DeclareOption{hemato}{%
	\gdef\@journal{hemato}
	\gdef\@journalfull{Hemato}
	\gdef\@journalshort{Hemato}
	\gdef\@doiabbr{hemato}
	\gdef\@ISSN{2673-6357}
	\gdef\@societyowner{}
}
\DeclareOption{hematolrep}{%
	\gdef\@journal{hematolrep}
	\gdef\@journalfull{Hematology Reports}
	\gdef\@journalshort{Hematol. Rep.}
	\gdef\@doiabbr{hematolrep}
	\gdef\@ISSN{2038-8330}
	\gdef\@societyowner{}
}
\DeclareOption{heritage}{%
	\gdef\@journal{heritage}
	\gdef\@journalfull{Heritage}
	\gdef\@journalshort{Heritage}
	\gdef\@doiabbr{heritage}
	\gdef\@ISSN{2571-9408}
	\gdef\@societyowner{}
}
\DeclareOption{higheredu}{%
	\gdef\@journal{higheredu}
	\gdef\@journalfull{Trends in Higher Education}
	\gdef\@journalshort{Trends High. Educ.}
	\gdef\@doiabbr{higheredu}
	\gdef\@ISSN{}
	\gdef\@societyowner{}
}
\DeclareOption{highthroughput}{%
	\gdef\@journal{highthroughput}
	\gdef\@journalfull{High-Throughput}
	\gdef\@journalshort{High-Throughput}
	\gdef\@doiabbr{ht}
	\gdef\@ISSN{2571-5135}
	\gdef\@societyowner{}
}
\DeclareOption{histories}{%
	\gdef\@journal{histories}
	\gdef\@journalfull{Histories}
	\gdef\@journalshort{Histories}
	\gdef\@doiabbr{histories}
	\gdef\@ISSN{2409-9252}
	\gdef\@societyowner{}
}
\DeclareOption{horticulturae}{%
	\gdef\@journal{horticulturae}
	\gdef\@journalfull{Horticulturae}
	\gdef\@journalshort{Horticulturae}
	\gdef\@doiabbr{horticulturae}
	\gdef\@ISSN{2311-7524}
	\gdef\@societyowner{}
}
\DeclareOption{hospitals}{%
	\gdef\@journal{hospitals}
	\gdef\@journalfull{Hospitals}
	\gdef\@journalshort{Hospitals}
	\gdef\@doiabbr{hospitals}
	\gdef\@ISSN{}
	\gdef\@societyowner{}
}
\DeclareOption{humanities}{%
	\gdef\@journal{humanities}
	\gdef\@journalfull{Humanities}
	\gdef\@journalshort{Humanities}
	\gdef\@doiabbr{h}
	\gdef\@ISSN{2076-0787}
	\gdef\@societyowner{}
}
\DeclareOption{humans}{%
	\gdef\@journal{humans}
	\gdef\@journalfull{Humans}
	\gdef\@journalshort{Humans}
	\gdef\@doiabbr{humans}
	\gdef\@ISSN{}
	\gdef\@societyowner{}
}
\DeclareOption{hydrobiology}{%
	\gdef\@journal{hydrobiology}
	\gdef\@journalfull{Hydrobiology}
	\gdef\@journalshort{Hydrobiology}
	\gdef\@doiabbr{hydrobiology}
	\gdef\@ISSN{2673-9917}
	\gdef\@societyowner{}
}
\DeclareOption{hydrogen}{%
	\gdef\@journal{hydrogen}
	\gdef\@journalfull{Hydrogen}
	\gdef\@journalshort{Hydrogen}
	\gdef\@doiabbr{hydrogen}
	\gdef\@ISSN{2673-4141}
	\gdef\@societyowner{}
}
\DeclareOption{hydrology}{%
	\gdef\@journal{hydrology}
	\gdef\@journalfull{Hydrology}
	\gdef\@journalshort{Hydrology}
	\gdef\@doiabbr{hydrology}
	\gdef\@ISSN{2306-5338}
	\gdef\@societyowner{}
}
\DeclareOption{hydropower}{%
        \gdef\@journal{hydropower}
        \gdef\@journalfull{Hydropower}
        \gdef\@journalshort{Hydropower}
        \gdef\@doiabbr{hydropower}
        \gdef\@ISSN{}
        \gdef\@societyowner{}
}
\DeclareOption{hygiene}{%
	\gdef\@journal{hygiene}
	\gdef\@journalfull{Hygiene}
	\gdef\@journalshort{Hygiene}
	\gdef\@doiabbr{hygiene}
	\gdef\@ISSN{}
	\gdef\@societyowner{}
}
\DeclareOption{idr}{%
	\gdef\@journal{idr}
	\gdef\@journalfull{Infectious Disease Reports}
	\gdef\@journalshort{Infect. Dis. Rep.}
	\gdef\@doiabbr{idr}
	\gdef\@ISSN{2036-7449}
	\gdef\@societyowner{}
}
\DeclareOption{iic}{%
	\gdef\@journal{iic}
	\gdef\@journalfull{Intelligent Infrastructure and Construction}
	\gdef\@journalshort{Intell. Infrastruct. Constr.}
	\gdef\@doiabbr{iic}
	\gdef\@ISSN{3042-4720}
	\gdef\@societyowner{}
}
\DeclareOption{ijcs}{%
       \gdef\@journal{ijcs}
       \gdef\@journalfull{International Journal of Cognitive Sciences}
       \gdef\@journalshort{Int. J. Cogn. Sci.}
       \gdef\@doiabbr{ijcs}
       \gdef\@ISSN{}
       \gdef\@societyowner{}
}
\DeclareOption{ijerph}{%
	\gdef\@journal{ijerph}
	\gdef\@journalfull{International Journal of Environmental Research and Public Health}
	\gdef\@journalshort{Int. J. Environ. Res. Public Health}
	\gdef\@doiabbr{ijerph}
	\gdef\@ISSN{1660-4601}
	\gdef\@societyowner{}
}
\DeclareOption{ijfs}{%
	\gdef\@journal{ijfs}
	\gdef\@journalfull{International Journal of Financial Studies}
	\gdef\@journalshort{Int. J. Financial Stud.}
	\gdef\@doiabbr{ijfs}
	\gdef\@ISSN{2227-7072}
	\gdef\@societyowner{}
}
\DeclareOption{ijgi}{%
	\gdef\@journal{ijgi}
	\gdef\@journalfull{ISPRS International Journal of Geo-Information}
	\gdef\@journalshort{ISPRS Int. J. Geo-Inf.}
	\gdef\@doiabbr{ijgi}
	\gdef\@ISSN{2220-9964}
	\gdef\@societyowner{International Society for Photogrammetry and Remote Sensing}
}
\DeclareOption{ijmd}{%
	\gdef\@journal{ijmd}
	\gdef\@journalfull{International Journal of Medical Devices}
	\gdef\@journalshort{Int. J. Med. Devices}
	\gdef\@doiabbr{ijmd}
	\gdef\@ISSN{}
	\gdef\@societyowner{}
}
\DeclareOption{ijms}{%
	\gdef\@journal{ijms}
	\gdef\@journalfull{International Journal of Molecular Sciences}
	\gdef\@journalshort{Int. J. Mol. Sci.}
	\gdef\@doiabbr{ijms}
	\gdef\@ISSN{1422-0067}
	\gdef\@societyowner{}
}
\DeclareOption{ijns}{%
	\gdef\@journal{ijns}
	\gdef\@journalfull{International Journal of Neonatal Screening}
	\gdef\@journalshort{Int. J. Neonatal Screen.}
	\gdef\@doiabbr{ijns}
	\gdef\@ISSN{2409-515X}
	\gdef\@societyowner{International Society for Neonatal Screening}
}
\DeclareOption{ijom}{%
        \gdef\@journal{ijom}
        \gdef\@journalfull{International Journal of Orofacial Myology and Myofunctional Therapy}
        \gdef\@journalshort{Int. J. Orofac. Myol. Myofunct. Ther.}
        \gdef\@doiabbr{ijom}
        \gdef\@ISSN{}
        \gdef\@societyowner{International Association of Orofacial Myology (IAOM)}
}
\DeclareOption{ijpb}{%
	\gdef\@journal{ijpb}
	\gdef\@journalfull{International Journal of Plant Biology}
	\gdef\@journalshort{Int. J. Plant Biol.}
	\gdef\@doiabbr{ijpb}
	\gdef\@ISSN{2037-0164}
	\gdef\@societyowner{}
}
\DeclareOption{ijt}{%
	\gdef\@journal{ijt}
	\gdef\@journalfull{International Journal of Topology}
	\gdef\@journalshort{Int. J. Topol.}
	\gdef\@doiabbr{ijt}
	\gdef\@ISSN{}
	\gdef\@societyowner{}
}
\DeclareOption{ijtm}{%
	\gdef\@journal{ijtm}
	\gdef\@journalfull{International Journal of Translational Medicine}
	\gdef\@journalshort{Int. J. Transl. Med.}
	\gdef\@doiabbr{ijtm}
	\gdef\@ISSN{2673-8937}
	\gdef\@societyowner{}
}
\DeclareOption{ijtpp}{%
	\gdef\@journal{ijtpp}
	\gdef\@journalfull{International Journal of Turbomachinery, Propulsion and Power}
	\gdef\@journalshort{Int. J. Turbomach. Propuls. Power}
	\gdef\@doiabbr{ijtpp}
	\gdef\@ISSN{2504-186X}
	\gdef\@societyowner{EUROTURBO}
}
\DeclareOption{ime}{%
	\gdef\@journal{ime}
	\gdef\@journalfull{International Medical Education}
	\gdef\@journalshort{Int. Med. Educ.}
	\gdef\@doiabbr{ime}
	\gdef\@ISSN{2813-141X}
	\gdef\@societyowner{Academic Society for International Medical Education}
}
\DeclareOption{immuno}{%
	\gdef\@journal{immuno}
	\gdef\@journalfull{Immuno}
	\gdef\@journalshort{Immuno}
	\gdef\@doiabbr{immuno}
	\gdef\@ISSN{2673-5601}
	\gdef\@societyowner{}
}
\DeclareOption{informatics}{%
	\gdef\@journal{informatics}
	\gdef\@journalfull{Informatics}
	\gdef\@journalshort{Informatics}
	\gdef\@doiabbr{informatics}
	\gdef\@ISSN{2227-9709}
	\gdef\@societyowner{}
}
\DeclareOption{information}{%
	\gdef\@journal{information}
	\gdef\@journalfull{Information}
	\gdef\@journalshort{Information}
	\gdef\@doiabbr{info}
	\gdef\@ISSN{2078-2489}
	\gdef\@societyowner{}
}
\DeclareOption{infrastructures}{%
	\gdef\@journal{infrastructures}
	\gdef\@journalfull{Infrastructures}
	\gdef\@journalshort{Infrastructures}
	\gdef\@doiabbr{infrastructures}
	\gdef\@ISSN{2412-3811}
	\gdef\@societyowner{}
}
\DeclareOption{inorganics}{%
	\gdef\@journal{inorganics}
	\gdef\@journalfull{Inorganics}
	\gdef\@journalshort{Inorganics}
	\gdef\@doiabbr{inorganics}
	\gdef\@ISSN{2304-6740}
	\gdef\@societyowner{}
}
\DeclareOption{insects}{%
	\gdef\@journal{insects}
	\gdef\@journalfull{Insects}
	\gdef\@journalshort{Insects}
	\gdef\@doiabbr{insects}
	\gdef\@ISSN{2075-4450}
	\gdef\@societyowner{}
}
\DeclareOption{instruments}{%
	\gdef\@journal{instruments}
	\gdef\@journalfull{Instruments}
	\gdef\@journalshort{Instruments}
	\gdef\@doiabbr{instruments}
	\gdef\@ISSN{2410-390X}
	\gdef\@societyowner{}
}
\DeclareOption{inventions}{%
	\gdef\@journal{inventions}
	\gdef\@journalfull{Inventions}
	\gdef\@journalshort{Inventions}
	\gdef\@doiabbr{inventions}
	\gdef\@ISSN{2411-5134}
	\gdef\@societyowner{}
}
\DeclareOption{iot}{%
	\gdef\@journal{iot}
	\gdef\@journalfull{IoT}
	\gdef\@journalshort{IoT}
	\gdef\@doiabbr{iot}
	\gdef\@ISSN{2624-831X}
	\gdef\@societyowner{}
}
\DeclareOption{j}{%
	\gdef\@journal{j}
	\gdef\@journalfull{J}
	\gdef\@journalshort{J}
	\gdef\@doiabbr{j}
	\gdef\@ISSN{2571-8800}
	\gdef\@societyowner{}
}
\DeclareOption{jaestheticmed}{%
        \gdef\@journal{jaestheticmed}
        \gdef\@journalfull{Journal of Aesthetic Medicine}
        \gdef\@journalshort{J. Aesthetic Med.}
        \gdef\@doiabbr{jaestheticmed}
        \gdef\@ISSN{}
        \gdef\@societyowner{}
}
\DeclareOption{jal}{%
	\gdef\@journal{jal}
	\gdef\@journalfull{Journal of Ageing and Longevity}
	\gdef\@journalshort{J. Ageing Longev.}
	\gdef\@doiabbr{jal}
	\gdef\@ISSN{2673-9259}
	\gdef\@societyowner{}
}
\DeclareOption{jcdd}{%
	\gdef\@journal{jcdd}
	\gdef\@journalfull{Journal of Cardiovascular Development and Disease}
	\gdef\@journalshort{J. Cardiovasc. Dev. Dis.}
	\gdef\@doiabbr{jcdd}
	\gdef\@ISSN{2308-3425}
	\gdef\@societyowner{}
}
\DeclareOption{jcm}{%
	\gdef\@journal{jcm}
	\gdef\@journalfull{Journal of Clinical Medicine}
	\gdef\@journalshort{J. Clin. Med.}
	\gdef\@doiabbr{jcm}
	\gdef\@ISSN{2077-0383}
	\gdef\@societyowner{}
}
\DeclareOption{jcp}{%
	\gdef\@journal{jcp}
	\gdef\@journalfull{Journal of Cybersecurity and Privacy}
	\gdef\@journalshort{J. Cybersecur. Priv.}
	\gdef\@doiabbr{jcp}
	\gdef\@ISSN{2624-800X}
	\gdef\@societyowner{}
}
\DeclareOption{jcrm}{%
        \gdef\@journal{jcrm}
        \gdef\@journalfull{Journal of CardioRenal Medicine}
        \gdef\@journalshort{J. CardioRenal Med.}
        \gdef\@doiabbr{jcrm}
        \gdef\@ISSN{}
        \gdef\@societyowner{}
}
\DeclareOption{jcs}{%
	\gdef\@journal{jcs}
	\gdef\@journalfull{Journal of Composites Science}
	\gdef\@journalshort{J. Compos. Sci.}
	\gdef\@doiabbr{jcs}
	\gdef\@ISSN{2504-477X}
	\gdef\@societyowner{}
}
\DeclareOption{jcto}{%
	\gdef\@journal{jcto}
	\gdef\@journalfull{Journal of Clinical \& Translational Ophthalmology}
	\gdef\@journalshort{J. Clin. Transl. Ophthalmol.}
	\gdef\@doiabbr{jcto}
	\gdef\@ISSN{2813-1053}
	\gdef\@societyowner{}
}
\DeclareOption{jdad}{%
	\gdef\@journal{jdad}
	\gdef\@journalfull{Journal of Dementia and Alzheimer’s Disease}
	\gdef\@journalshort{J. Dement. Alzheimer's Dis.}
	\gdef\@doiabbr{jdad}
	\gdef\@ISSN{}
	\gdef\@societyowner{}
}
\DeclareOption{jdb}{%
	\gdef\@journal{jdb}
	\gdef\@journalfull{Journal of Developmental Biology}
	\gdef\@journalshort{J. Dev. Biol.}
	\gdef\@doiabbr{jdb}
	\gdef\@ISSN{2221-3759}
	\gdef\@societyowner{}
}
\DeclareOption{jemr}{%
      \gdef\@journal{jemr}
      \gdef\@journalfull{Journal of Eye Movement Research}
      \gdef\@journalshort{J. Eye Mov. Res.}
      \gdef\@doiabbr{jemr}
      \gdef\@ISSN{1995-8692}
      \gdef\@societyowner{}
}
\DeclareOption{jeta}{%
	\gdef\@journal{jeta}
	\gdef\@journalfull{Journal of Experimental and Theoretical Analyses}
	\gdef\@journalshort{J. Exp. Theor. Anal.}
	\gdef\@doiabbr{jeta}
	\gdef\@ISSN{}
	\gdef\@societyowner{}
}
\DeclareOption{jfb}{%
	\gdef\@journal{jfb}
	\gdef\@journalfull{Journal of Functional Biomaterials}
	\gdef\@journalshort{J. Funct. Biomater.}
	\gdef\@doiabbr{jfb}
	\gdef\@ISSN{2079-4983}
	\gdef\@societyowner{}
}
\DeclareOption{jfmk}{%
	\gdef\@journal{jfmk}
	\gdef\@journalfull{Journal of Functional Morphology and Kinesiology}
	\gdef\@journalshort{J. Funct. Morphol. Kinesiol.}
	\gdef\@doiabbr{jfmk}
	\gdef\@ISSN{2411-5142}
	\gdef\@societyowner{}
}
\DeclareOption{jimaging}{%
	\gdef\@journal{jimaging}
	\gdef\@journalfull{Journal of Imaging}
	\gdef\@journalshort{J. Imaging}
	\gdef\@doiabbr{jimaging}
	\gdef\@ISSN{2313-433X}
	\gdef\@societyowner{}
}
\DeclareOption{jintelligence}{%
	\gdef\@journal{jintelligence}
	\gdef\@journalfull{Journal of Intelligence}
	\gdef\@journalshort{J. Intell.}
	\gdef\@doiabbr{jintelligence}
	\gdef\@ISSN{2079-3200}
	\gdef\@societyowner{}
}
\DeclareOption{jlpea}{%
	\gdef\@journal{jlpea}
	\gdef\@journalfull{Journal of Low Power Electronics and Applications}
	\gdef\@journalshort{J. Low Power Electron. Appl.}
	\gdef\@doiabbr{jlpea}
	\gdef\@ISSN{2079-9268}
	\gdef\@societyowner{}
}
\DeclareOption{jmahp}{%
	\gdef\@journal{jmahp}
	\gdef\@journalfull{Journal of Market Access \& Health Policy}
	\gdef\@journalshort{J. Mark. Access Health Policy}
	\gdef\@doiabbr{jmahp}
	\gdef\@ISSN{2001-6689}
	\gdef\@societyowner{Market Access Society}
}
\DeclareOption{jmmp}{%
	\gdef\@journal{jmmp}
	\gdef\@journalfull{Journal of Manufacturing and Materials Processing}
	\gdef\@journalshort{J. Manuf. Mater. Process.}
	\gdef\@doiabbr{jmmp}
	\gdef\@ISSN{2504-4494}
	\gdef\@societyowner{}
}
\DeclareOption{jmms}{%
        \gdef\@journal{jmms}
        \gdef\@journalfull{Journal of Mind and Medical Sciences}
        \gdef\@journalshort{J. Mind Med. Sci.}
        \gdef\@doiabbr{jmms}
        \gdef\@ISSN{2392-7674}
        \gdef\@societyowner{JMMS}
}
\DeclareOption{jmp}{%
	\gdef\@journal{jmp}
	\gdef\@journalfull{Journal of Molecular Pathology}
	\gdef\@journalshort{J. Mol. Pathol.}
	\gdef\@doiabbr{jmp}
	\gdef\@ISSN{2673-5261}
	\gdef\@societyowner{}
}
\DeclareOption{jmse}{%
	\gdef\@journal{jmse}
	\gdef\@journalfull{Journal of Marine Science and Engineering}
	\gdef\@journalshort{J. Mar. Sci. Eng.}
	\gdef\@doiabbr{jmse}
	\gdef\@ISSN{2077-1312}
	\gdef\@societyowner{}
}
\DeclareOption{jne}{%
	\gdef\@journal{jne}
	\gdef\@journalfull{Journal of Nuclear Engineering}
	\gdef\@journalshort{J. Nucl. Eng.}
	\gdef\@doiabbr{jne}
	\gdef\@ISSN{2673-4362}
	\gdef\@societyowner{}
}
\DeclareOption{jnt}{%
	\gdef\@journal{jnt}
	\gdef\@journalfull{Journal of Nanotheranostics}
	\gdef\@journalshort{J. Nanotheranostics}
	\gdef\@doiabbr{jnt}
	\gdef\@ISSN{2624-845X}
	\gdef\@societyowner{}
}
\DeclareOption{jof}{%
	\gdef\@journal{jof}
	\gdef\@journalfull{Journal of Fungi}
	\gdef\@journalshort{J. Fungi}
	\gdef\@doiabbr{jof}
	\gdef\@ISSN{2309-608X}
	\gdef\@societyowner{}
}
\DeclareOption{joitmc}{%
	\gdef\@journal{joitmc}
	\gdef\@journalfull{Journal of Open Innovation: Technology, Market, and Complexity}
	\gdef\@journalshort{J. Open Innov. Technol. Mark. Complex.}
	\gdef\@doiabbr{joitmc}
	\gdef\@ISSN{2199-8531}
	\gdef\@societyowner{}
}
\DeclareOption{joma}{%
	\gdef\@journal{joma}
	\gdef\@journalfull{Journal of the Oman Medical Association}
	\gdef\@journalshort{J. Oman Med. Assoc.}
	\gdef\@doiabbr{joma}
	\gdef\@ISSN{2813-8759}
	\gdef\@societyowner{Oman Medical Association}
}
\DeclareOption{jop}{%
        \gdef\@journal{jop}
        \gdef\@journalfull{Journal of Parks}
        \gdef\@journalshort{J. Parks}
       \gdef\@doiabbr{jop}
       \gdef\@ISSN{}
       \gdef\@societyowner{}
}
\DeclareOption{jor}{%
	\gdef\@journal{jor}
	\gdef\@journalfull{Journal of Respiration}
	\gdef\@journalshort{J. Respir.}
	\gdef\@doiabbr{jor}
	\gdef\@ISSN{2673-527X}
	\gdef\@societyowner{}
}
\DeclareOption{journalmedia}{%
	\gdef\@journal{journalmedia}
	\gdef\@journalfull{Journalism and Media}
	\gdef\@journalshort{Journal. Media}
	\gdef\@doiabbr{journalmedia}
	\gdef\@ISSN{2673-5172}
	\gdef\@societyowner{}
}
\DeclareOption{jox}{%
	\gdef\@journal{jox}
	\gdef\@journalfull{Journal of Xenobiotics}
	\gdef\@journalshort{J. Xenobiot.}
	\gdef\@doiabbr{jox}
	\gdef\@ISSN{2039-4713}
	\gdef\@societyowner{}
}
\DeclareOption{jpbi}{%
	\gdef\@journal{jpbi}
	\gdef\@journalfull{Journal of Pharmaceutical and BioTech Industry}
	\gdef\@journalshort{J. Pharm. BioTech Ind.}
	\gdef\@doiabbr{jpbi}
	\gdef\@ISSN{}
	\gdef\@societyowner{}
}
\DeclareOption{jpm}{%
	\gdef\@journal{jpm}
	\gdef\@journalfull{Journal of Personalized Medicine}
	\gdef\@journalshort{J. Pers. Med.}
	\gdef\@doiabbr{jpm}
	\gdef\@ISSN{2075-4426}
	\gdef\@societyowner{}
}
\DeclareOption{jrfm}{%
	\gdef\@journal{jrfm}
	\gdef\@journalfull{Journal of Risk and Financial Management}
	\gdef\@journalshort{J. Risk Financial Manag.}
	\gdef\@doiabbr{jrfm}
	\gdef\@ISSN{1911-8074}
	\gdef\@societyowner{}
}
\DeclareOption{jsan}{%
	\gdef\@journal{jsan}
	\gdef\@journalfull{Journal of Sensor and Actuator Networks}
	\gdef\@journalshort{J. Sens. Actuator Netw.}
	\gdef\@doiabbr{jsan}
	\gdef\@ISSN{2224-2708}
	\gdef\@societyowner{}
}
\DeclareOption{jtaer}{%
	\gdef\@journal{jtaer}
	\gdef\@journalfull{Journal of Theoretical and Applied Electronic Commerce Research}
	\gdef\@journalshort{J. Theor. Appl. Electron. Commer. Res.}
	\gdef\@doiabbr{jtaer}
	\gdef\@ISSN{0718-1876}
	\gdef\@societyowner{}
}
\DeclareOption{jvd}{%
	\gdef\@journal{jvd}
	\gdef\@journalfull{Journal of Vascular Diseases}
	\gdef\@journalshort{J. Vasc. Dis.}
	\gdef\@doiabbr{jvd}
	\gdef\@ISSN{}
	\gdef\@societyowner{}
}
\DeclareOption{jzbg}{%
	\gdef\@journal{jzbg}
	\gdef\@journalfull{Journal of Zoological and Botanical Gardens}
	\gdef\@journalshort{J. Zool. Bot. Gard.}
	\gdef\@doiabbr{jzbg}
	\gdef\@ISSN{2673-5636}
	\gdef\@societyowner{}
}
\DeclareOption{kidneydial}{%
	\gdef\@journal{kidneydial}
	\gdef\@journalfull{Kidney and Dialysis}
	\gdef\@journalshort{Kidney Dial.}
	\gdef\@doiabbr{kidneydial}
	\gdef\@ISSN{}
	\gdef\@societyowner{}
}
\DeclareOption{kinasesphosphatases}{%
	\gdef\@journal{kinasesphosphatases}
	\gdef\@journalfull{Kinases and Phosphatases}
	\gdef\@journalshort{Kinases Phosphatases}
	\gdef\@doiabbr{kinasesphosphatases}
	\gdef\@ISSN{}
	\gdef\@societyowner{}
}
\DeclareOption{knowledge}{%
	\gdef\@journal{knowledge}
	\gdef\@journalfull{Knowledge}
	\gdef\@journalshort{Knowledge}
	\gdef\@doiabbr{knowledge}
	\gdef\@ISSN{2673-9585}
	\gdef\@societyowner{}
}
\DeclareOption{labmed}{%
	\gdef\@journal{labmed}
	\gdef\@journalfull{LabMed}
	\gdef\@journalshort{LabMed}
	\gdef\@doiabbr{labmed}
	\gdef\@ISSN{}
	\gdef\@societyowner{}
}
\DeclareOption{laboratories}{%
	\gdef\@journal{laboratories}
	\gdef\@journalfull{Laboratories}
	\gdef\@journalshort{Laboratories}
	\gdef\@doiabbr{laboratories}
	\gdef\@ISSN{}
	\gdef\@societyowner{}
}
\DeclareOption{land}{%
	\gdef\@journal{land}
	\gdef\@journalfull{Land}
	\gdef\@journalshort{Land}
	\gdef\@doiabbr{land}
	\gdef\@ISSN{2073-445X}
	\gdef\@societyowner{}
}
\DeclareOption{languages}{%
	\gdef\@journal{languages}
	\gdef\@journalfull{Languages}
	\gdef\@journalshort{Languages}
	\gdef\@doiabbr{languages}
	\gdef\@ISSN{2226-471X}
	\gdef\@societyowner{}
}
\DeclareOption{laws}{%
	\gdef\@journal{laws}
	\gdef\@journalfull{Laws}
	\gdef\@journalshort{Laws}
	\gdef\@doiabbr{laws}
	\gdef\@ISSN{2075-471X}
	\gdef\@societyowner{}
}
\DeclareOption{life}{%
	\gdef\@journal{life}
	\gdef\@journalfull{Life}
	\gdef\@journalshort{Life}
	\gdef\@doiabbr{life}
	\gdef\@ISSN{2075-1729}
	\gdef\@societyowner{}
}
\DeclareOption{lights}{%
	\gdef\@journal{lights}
	\gdef\@journalfull{Lights}
	\gdef\@journalshort{Lights}
	\gdef\@doiabbr{lights}
	\gdef\@ISSN{}
	\gdef\@societyowner{}
}
\DeclareOption{limnolrev}{%
	\gdef\@journal{limnolrev}
	\gdef\@journalfull{Limnological Review}
	\gdef\@journalshort{Limnol. Rev.}
	\gdef\@doiabbr{limnolrev}
	\gdef\@ISSN{2300-7575}
	\gdef\@societyowner{}
}
\DeclareOption{lipidology}{%
	\gdef\@journal{lipidology}
	\gdef\@journalfull{Lipidology}
	\gdef\@journalshort{Lipidology}
	\gdef\@doiabbr{lipidology}
	\gdef\@ISSN{2813-7086}
	\gdef\@societyowner{}
}
\DeclareOption{liquids}{%
	\gdef\@journal{liquids}
	\gdef\@journalfull{Liquids}
	\gdef\@journalshort{Liquids}
	\gdef\@doiabbr{liquids}
	\gdef\@ISSN{2673-8015}
	\gdef\@societyowner{}
}
\DeclareOption{literature}{%
	\gdef\@journal{literature}
	\gdef\@journalfull{Literature}
	\gdef\@journalshort{Literature}
	\gdef\@doiabbr{literature}
	\gdef\@ISSN{2410-9789}
	\gdef\@societyowner{}
}
\DeclareOption{livers}{%
	\gdef\@journal{livers}
	\gdef\@journalfull{Livers}
	\gdef\@journalshort{Livers}
	\gdef\@doiabbr{livers}
	\gdef\@ISSN{2673-4389}
	\gdef\@societyowner{}
}
\DeclareOption{logics}{%
	\gdef\@journal{logics}
	\gdef\@journalfull{Logics}
	\gdef\@journalshort{Logics}
	\gdef\@doiabbr{logics}
	\gdef\@ISSN{2813-0405}
	\gdef\@societyowner{}
}
\DeclareOption{logistics}{%
	\gdef\@journal{logistics}
	\gdef\@journalfull{Logistics}
	\gdef\@journalshort{Logistics}
	\gdef\@doiabbr{logistics}
	\gdef\@ISSN{2305-6290}
	\gdef\@societyowner{}
}
\DeclareOption{lubricants}{%
	\gdef\@journal{lubricants}
	\gdef\@journalfull{Lubricants}
	\gdef\@journalshort{Lubricants}
	\gdef\@doiabbr{lubricants}
	\gdef\@ISSN{2075-4442}
	\gdef\@societyowner{}
}
\DeclareOption{lymphatics}{%
	\gdef\@journal{lymphatics}
	\gdef\@journalfull{Lymphatics}
	\gdef\@journalshort{Lymphatics}
	\gdef\@doiabbr{lymphatics}
	\gdef\@ISSN{}
	\gdef\@societyowner{}
}
\DeclareOption{machines}{%
	\gdef\@journal{machines}
	\gdef\@journalfull{Machines}
	\gdef\@journalshort{Machines}
	\gdef\@doiabbr{machines}
	\gdef\@ISSN{2075-1702}
	\gdef\@societyowner{}
}
\DeclareOption{macromol}{%
	\gdef\@journal{macromol}
	\gdef\@journalfull{Macromol}
	\gdef\@journalshort{Macromol}
	\gdef\@doiabbr{macromol}
	\gdef\@ISSN{2673-6209}
	\gdef\@societyowner{}
}
\DeclareOption{magnetism}{%
	\gdef\@journal{magnetism}
	\gdef\@journalfull{Magnetism}
	\gdef\@journalshort{Magnetism}
	\gdef\@doiabbr{magnetism}
	\gdef\@ISSN{}
	\gdef\@societyowner{}
}
\DeclareOption{magnetochemistry}{%
	\gdef\@journal{magnetochemistry}
	\gdef\@journalfull{Magnetochemistry}
	\gdef\@journalshort{Magnetochemistry}
	\gdef\@doiabbr{magnetochemistry}
	\gdef\@ISSN{2312-7481}
	\gdef\@societyowner{}
}
\DeclareOption{make}{%
	\gdef\@journal{make}
	\gdef\@journalfull{Machine Learning and Knowledge Extraction}
	\gdef\@journalshort{Mach. Learn. Knowl. Extr.}
	\gdef\@doiabbr{make}
	\gdef\@ISSN{2504-4990}
	\gdef\@societyowner{}
}
\DeclareOption{marinedrugs}{%
	\gdef\@journal{marinedrugs}
	\gdef\@journalfull{Marine Drugs}
	\gdef\@journalshort{Mar. Drugs}
	\gdef\@doiabbr{md}
	\gdef\@ISSN{1660-3397}
	\gdef\@societyowner{}
}
\DeclareOption{materials}{%
	\gdef\@journal{materials}
	\gdef\@journalfull{Materials}
	\gdef\@journalshort{Materials}
	\gdef\@doiabbr{ma}
	\gdef\@ISSN{1996-1944}
	\gdef\@societyowner{}
}
\DeclareOption{materproc}{%
	\gdef\@journal{materproc}
	\gdef\@journalfull{Materials Proceedings}
	\gdef\@journalshort{Mater. Proc.}
	\gdef\@doiabbr{materproc}
	\gdef\@ISSN{2673-4605}
	\gdef\@societyowner{}
}
\DeclareOption{mathematics}{%
	\gdef\@journal{mathematics}
	\gdef\@journalfull{Mathematics}
	\gdef\@journalshort{Mathematics}
	\gdef\@doiabbr{math}
	\gdef\@ISSN{2227-7390}
	\gdef\@societyowner{}
}
\DeclareOption{mca}{%
	\gdef\@journal{mca}
	\gdef\@journalfull{Mathematical and Computational Applications}
	\gdef\@journalshort{Math. Comput. Appl.}
	\gdef\@doiabbr{mca}
	\gdef\@ISSN{2297-8747}
	\gdef\@societyowner{}
}
\DeclareOption{measurements}{%
	\gdef\@journal{measurements}
	\gdef\@journalfull{Measurements}
	\gdef\@journalshort{Measurements}
	\gdef\@doiabbr{measurements}
	\gdef\@ISSN{}
	\gdef\@societyowner{}
}
\DeclareOption{medicina}{%
	\gdef\@journal{medicina}
	\gdef\@journalfull{Medicina}
	\gdef\@journalshort{Medicina}
	\gdef\@doiabbr{medicina}
	\gdef\@ISSN{1648-9144}
	\gdef\@societyowner{Lithuanian University of Health Sciences}
}
\DeclareOption{medicines}{%
	\gdef\@journal{medicines}
	\gdef\@journalfull{Medicines}
	\gdef\@journalshort{Medicines}
	\gdef\@doiabbr{medicines}
	\gdef\@ISSN{2305-6320}
	\gdef\@societyowner{}
}
\DeclareOption{medsci}{%
	\gdef\@journal{medsci}
	\gdef\@journalfull{Medical Sciences}
	\gdef\@journalshort{Med. Sci.}
	\gdef\@doiabbr{medsci}
	\gdef\@ISSN{2076-3271}
	\gdef\@societyowner{}
}
\DeclareOption{membranes}{%
	\gdef\@journal{membranes}
	\gdef\@journalfull{Membranes}
	\gdef\@journalshort{Membranes}
	\gdef\@doiabbr{membranes}
	\gdef\@ISSN{2077-0375}
	\gdef\@societyowner{}
}
\DeclareOption{merits}{%
	\gdef\@journal{merits}
	\gdef\@journalfull{Merits}
	\gdef\@journalshort{Merits}
	\gdef\@doiabbr{merits}
	\gdef\@ISSN{2673-8104}
	\gdef\@societyowner{}
}
\DeclareOption{metabolites}{%
	\gdef\@journal{metabolites}
	\gdef\@journalfull{Metabolites}
	\gdef\@journalshort{Metabolites}
	\gdef\@doiabbr{metabo}
	\gdef\@ISSN{2218-1989}
	\gdef\@societyowner{}
}
\DeclareOption{metals}{%
	\gdef\@journal{metals}
	\gdef\@journalfull{Metals}
	\gdef\@journalshort{Metals}
	\gdef\@doiabbr{met}
	\gdef\@ISSN{2075-4701}
	\gdef\@societyowner{}
}
\DeclareOption{meteorology}{%
	\gdef\@journal{meteorology}
	\gdef\@journalfull{Meteorology}
	\gdef\@journalshort{Meteorology}
	\gdef\@doiabbr{meteorology}
	\gdef\@ISSN{2674-0494}
	\gdef\@societyowner{}
}
\DeclareOption{methane}{%
	\gdef\@journal{methane}
	\gdef\@journalfull{Methane}
	\gdef\@journalshort{Methane}
	\gdef\@doiabbr{methane}
	\gdef\@ISSN{}
	\gdef\@societyowner{}
}
\DeclareOption{metrics}{%
	\gdef\@journal{metrics}
	\gdef\@journalfull{Metrics}
	\gdef\@journalshort{Metrics}
	\gdef\@doiabbr{metrics}
	\gdef\@ISSN{}
	\gdef\@societyowner{}
}
\DeclareOption{metrology}{%
	\gdef\@journal{metrology}
	\gdef\@journalfull{Metrology}
	\gdef\@journalshort{Metrology}
	\gdef\@doiabbr{metrology}
	\gdef\@ISSN{}
	\gdef\@societyowner{}
}
\DeclareOption{micro}{%
	\gdef\@journal{micro}
	\gdef\@journalfull{Micro}
	\gdef\@journalshort{Micro}
	\gdef\@doiabbr{micro}
	\gdef\@ISSN{}
	\gdef\@societyowner{}
}
\DeclareOption{microarrays}{%
	\gdef\@journal{microarrays}
	\gdef\@journalfull{Microarrays}
	\gdef\@journalshort{Microarrays}
	\gdef\@doiabbr{microarrays}
	\gdef\@ISSN{2076-3905}
	\gdef\@societyowner{}
}
\DeclareOption{microbiolres}{%
	\gdef\@journal{microbiolres}
	\gdef\@journalfull{Microbiology Research}
	\gdef\@journalshort{Microbiol. Res.}
	\gdef\@doiabbr{microbiolres}
	\gdef\@ISSN{2036-7481}
	\gdef\@societyowner{}
}
\DeclareOption{microelectronics}{%
       \gdef\@journal{microelectronics}
       \gdef\@journalfull{Microelectronics}
       \gdef\@journalshort{Microelectronics}
       \gdef\@doiabbr{microelectronics}
       \gdef\@ISSN{}
       \gdef\@societyowner{}
}
\DeclareOption{micromachines}{%
	\gdef\@journal{micromachines}
	\gdef\@journalfull{Micromachines}
	\gdef\@journalshort{Micromachines}
	\gdef\@doiabbr{mi}
	\gdef\@ISSN{2072-666X}
	\gdef\@societyowner{}
}
\DeclareOption{microorganisms}{%
	\gdef\@journal{microorganisms}
	\gdef\@journalfull{Microorganisms}
	\gdef\@journalshort{Microorganisms}
	\gdef\@doiabbr{microorganisms}
	\gdef\@ISSN{2076-2607}
	\gdef\@societyowner{}
}
\DeclareOption{microplastics}{%
	\gdef\@journal{microplastics}
	\gdef\@journalfull{Microplastics}
	\gdef\@journalshort{Microplastics}
	\gdef\@doiabbr{microplastics}
	\gdef\@ISSN{2673-8929}
	\gdef\@societyowner{}
}
\DeclareOption{microwave}{%
	\gdef\@journal{microwave}
	\gdef\@journalfull{Microwave}
	\gdef\@journalshort{Microwave}
	\gdef\@doiabbr{microwave}
	\gdef\@ISSN{}
	\gdef\@societyowner{}
}
\DeclareOption{minerals}{%
	\gdef\@journal{minerals}
	\gdef\@journalfull{Minerals}
	\gdef\@journalshort{Minerals}
	\gdef\@doiabbr{min}
	\gdef\@ISSN{2075-163X}
	\gdef\@societyowner{}
}
\DeclareOption{mining}{%
	\gdef\@journal{mining}
	\gdef\@journalfull{Mining}
	\gdef\@journalshort{Mining}
	\gdef\@doiabbr{mining}
	\gdef\@ISSN{2673-6489}
	\gdef\@societyowner{}
}
\DeclareOption{mmphys}{%
        \gdef\@journal{mmphys}
        \gdef\@journalfull{Modern Mathematical Physics}
        \gdef\@journalshort{Mod. Math. Phys.}
        \gdef\@doiabbr{mmphys}
        \gdef\@ISSN{3042-5034}
       \gdef\@societyowner{}
}
\DeclareOption{modelling}{%
	\gdef\@journal{modelling}
	\gdef\@journalfull{Modelling}
	\gdef\@journalshort{Modelling}
	\gdef\@doiabbr{modelling}
	\gdef\@ISSN{2673-3951}
	\gdef\@societyowner{}
}
\DeclareOption{molbank}{%
	\gdef\@journal{molbank}
	\gdef\@journalfull{Molbank}
	\gdef\@journalshort{Molbank}
	\gdef\@doiabbr{M}
	\gdef\@ISSN{1422-8599}
	\gdef\@societyowner{}
}
\DeclareOption{molecules}{%
	\gdef\@journal{molecules}
	\gdef\@journalfull{Molecules}
	\gdef\@journalshort{Molecules}
	\gdef\@doiabbr{molecules}
	\gdef\@ISSN{1420-3049}
	\gdef\@societyowner{}
}
\DeclareOption{mps}{%
	\gdef\@journal{mps}
	\gdef\@journalfull{Methods and Protocols}
	\gdef\@journalshort{Methods Protoc.}
	\gdef\@doiabbr{mps}
	\gdef\@ISSN{2409-9279}
	\gdef\@societyowner{}
}
\DeclareOption{msf}{%
	\gdef\@journal{msf}
	\gdef\@journalfull{Medical Sciences Forum}
	\gdef\@journalshort{Med. Sci. Forum}
	\gdef\@doiabbr{msf}
	\gdef\@ISSN{2673-9992}
	\gdef\@societyowner{}
}
\DeclareOption{mti}{%
	\gdef\@journal{mti}
	\gdef\@journalfull{Multimodal Technologies and Interaction}
	\gdef\@journalshort{Multimodal Technol. Interact.}
	\gdef\@doiabbr{mti}
	\gdef\@ISSN{2414-4088}
	\gdef\@societyowner{}
}
\DeclareOption{multimedia}{%
       \gdef\@journal{multimedia}
       \gdef\@journalfull{Multimedia}
       \gdef\@journalshort{Multimedia}
       \gdef\@doiabbr{multimedia}
       \gdef\@ISSN{}
       \gdef\@societyowner{}
}
\DeclareOption{muscles}{%
	\gdef\@journal{muscles}
	\gdef\@journalfull{Muscles}
	\gdef\@journalshort{Muscles}
	\gdef\@doiabbr{muscles}
	\gdef\@ISSN{2813-0413}
	\gdef\@societyowner{}
}
\DeclareOption{nanoenergyadv}{%
	\gdef\@journal{nanoenergyadv}
	\gdef\@journalfull{Nanoenergy Advances}
	\gdef\@journalshort{Nanoenergy Adv.}
	\gdef\@doiabbr{nanoenergyadv}
	\gdef\@ISSN{2673-706X}
	\gdef\@societyowner{}
}
\DeclareOption{nanomanufacturing}{%
	\gdef\@journal{nanomanufacturing}
	\gdef\@journalfull{Nanomanufacturing}
	\gdef\@journalshort{Nanomanufacturing}
	\gdef\@doiabbr{nanomanufacturing}
	\gdef\@ISSN{2673-687X}
	\gdef\@societyowner{}
}
\DeclareOption{nanomaterials}{%
	\gdef\@journal{nanomaterials}
	\gdef\@journalfull{Nanomaterials}
	\gdef\@journalshort{Nanomaterials}
	\gdef\@doiabbr{nano}
	\gdef\@ISSN{2079-4991}
	\gdef\@societyowner{}
}
\DeclareOption{ncrna}{%
	\gdef\@journal{ncrna}
	\gdef\@journalfull{Non-Coding RNA}
	\gdef\@journalshort{Non-Coding RNA}
	\gdef\@doiabbr{ncrna}
	\gdef\@ISSN{2311-553X}
	\gdef\@societyowner{}
}
\DeclareOption{ndt}{%
	\gdef\@journal{ndt}
	\gdef\@journalfull{NDT}
	\gdef\@journalshort{NDT}
	\gdef\@doiabbr{ndt}
	\gdef\@ISSN{}
	\gdef\@societyowner{}
}
\DeclareOption{network}{%
	\gdef\@journal{network}
	\gdef\@journalfull{Network}
	\gdef\@journalshort{Network}
	\gdef\@doiabbr{network}
	\gdef\@ISSN{}
	\gdef\@societyowner{}
}
\DeclareOption{neuroglia}{%
	\gdef\@journal{neuroglia}
	\gdef\@journalfull{Neuroglia}
	\gdef\@journalshort{Neuroglia}
	\gdef\@doiabbr{neuroglia}
	\gdef\@ISSN{2571-6980}
	\gdef\@societyowner{}
}
\DeclareOption{neuroimaging}{%
        \gdef\@journal{neuroimaging}
        \gdef\@journalfull{Neuroimaging}
        \gdef\@journalshort{Neuroimaging}
        \gdef\@doiabbr{neuroimaging}
        \gdef\@ISSN{}
        \gdef\@societyowner{}
}
\DeclareOption{neurolint}{%
	\gdef\@journal{neurolint}
	\gdef\@journalfull{Neurology International}
	\gdef\@journalshort{Neurol. Int.}
	\gdef\@doiabbr{neurolint}
	\gdef\@ISSN{2035-8377}
	\gdef\@societyowner{}
}
\DeclareOption{neurosci}{%
	\gdef\@journal{neurosci}
	\gdef\@journalfull{NeuroSci}
	\gdef\@journalshort{NeuroSci}
	\gdef\@doiabbr{neurosci}
	\gdef\@ISSN{2673-4087}
	\gdef\@societyowner{}
}
\DeclareOption{nitrogen}{%
	\gdef\@journal{nitrogen}
	\gdef\@journalfull{Nitrogen}
	\gdef\@journalshort{Nitrogen}
	\gdef\@doiabbr{nitrogen}
	\gdef\@ISSN{2504-3129}
	\gdef\@societyowner{}
}
\DeclareOption{notspecified}{%
	\gdef\@journal{notspecified}
	\gdef\@journalfull{Journal Not Specified}
	\gdef\@journalshort{Journal Not Specified}
	\gdef\@doiabbr{}
	\gdef\@ISSN{}
	\gdef\@societyowner{}
}
\DeclareOption{nursrep}{%
	\gdef\@journal{nursrep}
	\gdef\@journalfull{Nursing Reports}
	\gdef\@journalshort{Nurs. Rep.}
	\gdef\@doiabbr{nursrep}
	\gdef\@ISSN{2039-4403}
	\gdef\@societyowner{}
}
\DeclareOption{nutraceuticals}{%
	\gdef\@journal{nutraceuticals}
	\gdef\@journalfull{Nutraceuticals}
	\gdef\@journalshort{Nutraceuticals}
	\gdef\@doiabbr{nutraceuticals}
	\gdef\@ISSN{1661-3821}
	\gdef\@societyowner{}
}
\DeclareOption{nutrients}{%
	\gdef\@journal{nutrients}
	\gdef\@journalfull{Nutrients}
	\gdef\@journalshort{Nutrients}
	\gdef\@doiabbr{nu}
	\gdef\@ISSN{2072-6643}
	\gdef\@societyowner{}
}
\DeclareOption{obesities}{%
	\gdef\@journal{obesities}
	\gdef\@journalfull{Obesities}
	\gdef\@journalshort{Obesities}
	\gdef\@doiabbr{obesities}
	\gdef\@ISSN{2673-4168}
	\gdef\@societyowner{}
}
\DeclareOption{oceans}{%
	\gdef\@journal{oceans}
	\gdef\@journalfull{Oceans}
	\gdef\@journalshort{Oceans}
	\gdef\@doiabbr{oceans}
	\gdef\@ISSN{2673-1924}
	\gdef\@societyowner{}
}
\DeclareOption{ohbm}{%
	\gdef\@journal{ohbm}
	\gdef\@journalfull{Journal of Otorhinolaryngology, Hearing and Balance Medicine}
	\gdef\@journalshort{J. Otorhinolaryngol. Hear. Balance Med.}
	\gdef\@doiabbr{ohbm}
	\gdef\@ISSN{2504-463X}
	\gdef\@societyowner{}
}
\DeclareOption{onco}{%
	\gdef\@journal{onco}
	\gdef\@journalfull{Onco}
	\gdef\@journalshort{Onco}
	\gdef\@doiabbr{onco}
	\gdef\@ISSN{2673-7523}
	\gdef\@societyowner{}
}
\DeclareOption{optics}{%
	\gdef\@journal{optics}
	\gdef\@journalfull{Optics}
	\gdef\@journalshort{Optics}
	\gdef\@doiabbr{opt}
	\gdef\@ISSN{2673-3269}
	\gdef\@societyowner{}
}
\DeclareOption{oral}{%
	\gdef\@journal{oral}
	\gdef\@journalfull{Oral}
	\gdef\@journalshort{Oral}
	\gdef\@doiabbr{oral}
	\gdef\@ISSN{2673-6373}
	\gdef\@societyowner{}
}
\DeclareOption{organics}{%
	\gdef\@journal{organics}
	\gdef\@journalfull{Organics}
	\gdef\@journalshort{Organics}
	\gdef\@doiabbr{org}
	\gdef\@ISSN{2673-401X}
	\gdef\@societyowner{}
}
\DeclareOption{organoids}{%
	\gdef\@journal{organoids}
	\gdef\@journalfull{Organoids}
	\gdef\@journalshort{Organoids}
	\gdef\@doiabbr{organoids}
	\gdef\@ISSN{2674-1172}
	\gdef\@societyowner{}
}
\DeclareOption{osteology}{%
	\gdef\@journal{osteology}
	\gdef\@journalfull{Osteology}
	\gdef\@journalshort{Osteology}
	\gdef\@doiabbr{osteology}
	\gdef\@ISSN{2673-4036}
	\gdef\@societyowner{}
}
\DeclareOption{oxygen}{%
	\gdef\@journal{oxygen}
	\gdef\@journalfull{Oxygen}
	\gdef\@journalshort{Oxygen}
	\gdef\@doiabbr{oxygen}
	\gdef\@ISSN{2673-9801}
	\gdef\@societyowner{}
}
\DeclareOption{parasites}{%
	\gdef\@journal{parasites}
	\gdef\@journalfull{Parasites}
	\gdef\@journalshort{Parasites}
	\gdef\@doiabbr{parasites}
	\gdef\@ISSN{2673-5946}
	\gdef\@societyowner{}
}
\DeclareOption{parasitologia}{%
	\gdef\@journal{parasitologia}
	\gdef\@journalfull{Parasitologia}
	\gdef\@journalshort{Parasitologia}
	\gdef\@doiabbr{parasitologia}
	\gdef\@ISSN{2673-6772}
	\gdef\@societyowner{}
}
\DeclareOption{particles}{%
	\gdef\@journal{particles}
	\gdef\@journalfull{Particles}
	\gdef\@journalshort{Particles}
	\gdef\@doiabbr{particles}
	\gdef\@ISSN{2571-712X}
	\gdef\@societyowner{}
}
\DeclareOption{pathogens}{%
	\gdef\@journal{pathogens}
	\gdef\@journalfull{Pathogens}
	\gdef\@journalshort{Pathogens}
	\gdef\@doiabbr{pathogens}
	\gdef\@ISSN{2076-0817}
	\gdef\@societyowner{}
}
\DeclareOption{pathophysiology}{%
	\gdef\@journal{pathophysiology}
	\gdef\@journalfull{Pathophysiology}
	\gdef\@journalshort{Pathophysiology}
	\gdef\@doiabbr{pathophysiology}
	\gdef\@ISSN{0928-4680}
	\gdef\@societyowner{}
}
\DeclareOption{peacestud}{%
       \gdef\@journal{peacestud}
       \gdef\@journalfull{Peace Studies}
       \gdef\@journalshort{Peace Stud.}
       \gdef\@doiabbr{peacestud}
       \gdef\@ISSN{}
       \gdef\@societyowner{}
}
\DeclareOption{pediatrrep}{%
	\gdef\@journal{pediatrrep}
	\gdef\@journalfull{Pediatric Reports}
	\gdef\@journalshort{Pediatr. Rep.}
	\gdef\@doiabbr{pediatric}
	\gdef\@ISSN{2036-7503}
	\gdef\@societyowner{}
}
\DeclareOption{pets}{%
	\gdef\@journal{pets}
	\gdef\@journalfull{Pets}
	\gdef\@journalshort{Pets}
	\gdef\@doiabbr{pets}
	\gdef\@ISSN{}
	\gdef\@societyowner{}
}
\DeclareOption{pharmaceuticals}{%
	\gdef\@journal{pharmaceuticals}
	\gdef\@journalfull{Pharmaceuticals}
	\gdef\@journalshort{Pharmaceuticals}
	\gdef\@doiabbr{ph}
	\gdef\@ISSN{1424-8247}
	\gdef\@societyowner{}
}
\DeclareOption{pharmaceutics}{%
	\gdef\@journal{pharmaceutics}
	\gdef\@journalfull{Pharmaceutics}
	\gdef\@journalshort{Pharmaceutics}
	\gdef\@doiabbr{pharmaceutics}
	\gdef\@ISSN{1999-4923}
	\gdef\@societyowner{}
}
\DeclareOption{pharmacoepidemiology}{%
	\gdef\@journal{pharmacoepidemiology}
	\gdef\@journalfull{Pharmacoepidemiology}
	\gdef\@journalshort{Pharmacoepidemiology}
	\gdef\@doiabbr{pharma}
	\gdef\@ISSN{2813-0618}
	\gdef\@societyowner{}
}
\DeclareOption{pharmacy}{%
	\gdef\@journal{pharmacy}
	\gdef\@journalfull{Pharmacy}
	\gdef\@journalshort{Pharmacy}
	\gdef\@doiabbr{pharmacy}
	\gdef\@ISSN{2226-4787}
	\gdef\@societyowner{}
}
\DeclareOption{philosophies}{%
	\gdef\@journal{philosophies}
	\gdef\@journalfull{Philosophies}
	\gdef\@journalshort{Philosophies}
	\gdef\@doiabbr{philosophies}
	\gdef\@ISSN{2409-9287}
	\gdef\@societyowner{}
}
\DeclareOption{photochem}{%
	\gdef\@journal{photochem}
	\gdef\@journalfull{Photochem}
	\gdef\@journalshort{Photochem}
	\gdef\@doiabbr{photochem}
	\gdef\@ISSN{2673-7256}
	\gdef\@societyowner{}
}
\DeclareOption{photonics}{%
	\gdef\@journal{photonics}
	\gdef\@journalfull{Photonics}
	\gdef\@journalshort{Photonics}
	\gdef\@doiabbr{photonics}
	\gdef\@ISSN{2304-6732}
	\gdef\@societyowner{}
}
\DeclareOption{phycology}{%
	\gdef\@journal{phycology}
	\gdef\@journalfull{Phycology}
	\gdef\@journalshort{Phycology}
	\gdef\@doiabbr{phycology}
	\gdef\@ISSN{2673-9410}
	\gdef\@societyowner{}
}
\DeclareOption{physchem}{%
	\gdef\@journal{physchem}
	\gdef\@journalfull{Physchem}
	\gdef\@journalshort{Physchem}
	\gdef\@doiabbr{physchem}
	\gdef\@ISSN{}
	\gdef\@societyowner{}
}
\DeclareOption{physics}{%
	\gdef\@journal{physics}
	\gdef\@journalfull{Physics}
	\gdef\@journalshort{Physics}
	\gdef\@doiabbr{physics}
	\gdef\@ISSN{2624-8174}
	\gdef\@societyowner{}
}
\DeclareOption{physiologia}{%
	\gdef\@journal{physiologia}
	\gdef\@journalfull{Physiologia}
	\gdef\@journalshort{Physiologia}
	\gdef\@doiabbr{physiologia}
	\gdef\@ISSN{}
	\gdef\@societyowner{}
}
\DeclareOption{plants}{%
	\gdef\@journal{plants}
	\gdef\@journalfull{Plants}
	\gdef\@journalshort{Plants}
	\gdef\@doiabbr{plants}
	\gdef\@ISSN{2223-7747}
	\gdef\@societyowner{}
}
\DeclareOption{plasma}{%
	\gdef\@journal{plasma}
	\gdef\@journalfull{Plasma}
	\gdef\@journalshort{Plasma}
	\gdef\@doiabbr{plasma}
	\gdef\@ISSN{2571-6182}
	\gdef\@societyowner{}
}
\DeclareOption{platforms}{%
	\gdef\@journal{platforms}
	\gdef\@journalfull{Platforms}
	\gdef\@journalshort{Platforms}
	\gdef\@doiabbr{platforms}
	\gdef\@ISSN{}
	\gdef\@societyowner{}
}
\DeclareOption{pollutants}{%
	\gdef\@journal{pollutants}
	\gdef\@journalfull{Pollutants}
	\gdef\@journalshort{Pollutants}
	\gdef\@doiabbr{pollutants}
	\gdef\@ISSN{2673-4672}
	\gdef\@societyowner{}
}
\DeclareOption{polymers}{%
	\gdef\@journal{polymers}
	\gdef\@journalfull{Polymers}
	\gdef\@journalshort{Polymers}
	\gdef\@doiabbr{polym}
	\gdef\@ISSN{2073-4360}
	\gdef\@societyowner{}
}
\DeclareOption{polysaccharides}{%
	\gdef\@journal{polysaccharides}
	\gdef\@journalfull{Polysaccharides}
	\gdef\@journalshort{Polysaccharides}
	\gdef\@doiabbr{polysaccharides}
	\gdef\@ISSN{2673-4176}
	\gdef\@societyowner{}
}
\DeclareOption{populations}{%
	\gdef\@journal{populations}
	\gdef\@journalfull{Populations}
	\gdef\@journalshort{Populations}
	\gdef\@doiabbr{populations}
	\gdef\@ISSN{}
	\gdef\@societyowner{}
}
\DeclareOption{poultry}{%
	\gdef\@journal{poultry}
	\gdef\@journalfull{Poultry}
	\gdef\@journalshort{Poultry}
	\gdef\@doiabbr{poultry}
	\gdef\@ISSN{2674-1164}
	\gdef\@societyowner{}
}
\DeclareOption{powders}{%
	\gdef\@journal{powders}
	\gdef\@journalfull{Powders}
	\gdef\@journalshort{Powders}
	\gdef\@doiabbr{powders}
	\gdef\@ISSN{2674-0516}
	\gdef\@societyowner{}
}
\DeclareOption{precisoncol}{%
        \gdef\@journal{precisoncol}
        \gdef\@journalfull{Precision Oncology}
        \gdef\@journalshort{Precis. Oncol.}
        \gdef\@doiabbr{precisoncol}
        \gdef\@ISSN{3042-7614}
        \gdef\@societyowner{}
}
\DeclareOption{preprints}{%
	\gdef\@journal{preprints}
	\gdef\@journalfull{Preprints}
	\gdef\@journalshort{Preprints}
	\gdef\@doiabbr{}
	\gdef\@ISSN{}
	\gdef\@societyowner{}
}
\DeclareOption{preprintschicago}{%
	\gdef\@journal{preprintschicago}
	\gdef\@journalfull{Preprints}
	\gdef\@journalshort{Preprints}
	\gdef\@doiabbr{}
	\gdef\@ISSN{}
	\gdef\@societyowner{}
}
\DeclareOption{preprintsapa}{%
	\gdef\@journal{preprintsapa}
	\gdef\@journalfull{Preprints}
	\gdef\@journalshort{Preprints}
	\gdef\@doiabbr{}
	\gdef\@ISSN{}
	\gdef\@societyowner{}
}
\DeclareOption{proceedings}{%
	\gdef\@journal{proceedings}
	\gdef\@journalfull{Proceedings}
	\gdef\@journalshort{Proceedings}
	\gdef\@doiabbr{proceedings}
	\gdef\@ISSN{2504-3900}
	\gdef\@societyowner{}
}
\DeclareOption{processes}{%
	\gdef\@journal{processes}
	\gdef\@journalfull{Processes}
	\gdef\@journalshort{Processes}
	\gdef\@doiabbr{pr}
	\gdef\@ISSN{2227-9717}
	\gdef\@societyowner{}
}
\DeclareOption{prosthesis}{%
	\gdef\@journal{prosthesis}
	\gdef\@journalfull{Prosthesis}
	\gdef\@journalshort{Prosthesis}
	\gdef\@doiabbr{prosthesis}
	\gdef\@ISSN{2673-1592}
	\gdef\@societyowner{}
}
\DeclareOption{proteomes}{%
	\gdef\@journal{proteomes}
	\gdef\@journalfull{Proteomes}
	\gdef\@journalshort{Proteomes}
	\gdef\@doiabbr{proteomes}
	\gdef\@ISSN{2227-7382}
	\gdef\@societyowner{}
}
\DeclareOption{psf}{%
	\gdef\@journal{psf}
	\gdef\@journalfull{Physical Sciences Forum}
	\gdef\@journalshort{Phys. Sci. Forum}
	\gdef\@doiabbr{psf}
	\gdef\@ISSN{2673-9984}
	\gdef\@societyowner{}
}
\DeclareOption{psychiatryint}{%
	\gdef\@journal{psychiatryint}
	\gdef\@journalfull{Psychiatry International}
	\gdef\@journalshort{Psychiatry Int.}
	\gdef\@doiabbr{psychiatryint}
	\gdef\@ISSN{2673-5318}
	\gdef\@societyowner{}
}
\DeclareOption{psychoactives}{%
	\gdef\@journal{psychoactives}
	\gdef\@journalfull{Psychoactives}
	\gdef\@journalshort{Psychoactives}
	\gdef\@doiabbr{psychoactives}
	\gdef\@ISSN{2813-1851}
	\gdef\@societyowner{}
}
\DeclareOption{psycholint}{%
	\gdef\@journal{psycholint}
	\gdef\@journalfull{Psychology International}
	\gdef\@journalshort{Psychol. Int.}
	\gdef\@doiabbr{psycholint}
	\gdef\@ISSN{2813-9844}
	\gdef\@societyowner{}
}
\DeclareOption{publications}{%
	\gdef\@journal{publications}
	\gdef\@journalfull{Publications}
	\gdef\@journalshort{Publications}
	\gdef\@doiabbr{publications}
	\gdef\@ISSN{2304-6775}
	\gdef\@societyowner{}
}
\DeclareOption{purification}{%
	\gdef\@journal{purification}
	\gdef\@journalfull{Purification}
	\gdef\@journalshort{Purification}
	\gdef\@doiabbr{purification}
	\gdef\@ISSN{}
	\gdef\@societyowner{}
}
\DeclareOption{quantumrep}{%
	\gdef\@journal{quantumrep}
	\gdef\@journalfull{Quantum Reports}
	\gdef\@journalshort{Quantum Rep.}
	\gdef\@doiabbr{quantum}
	\gdef\@ISSN{2624-960X}
	\gdef\@societyowner{}
}
\DeclareOption{quaternary}{%
	\gdef\@journal{quaternary}
	\gdef\@journalfull{Quaternary}
	\gdef\@journalshort{Quaternary}
	\gdef\@doiabbr{quat}
	\gdef\@ISSN{2571-550X}
	\gdef\@societyowner{}
}
\DeclareOption{qubs}{%
	\gdef\@journal{qubs}
	\gdef\@journalfull{Quantum Beam Science}
	\gdef\@journalshort{Quantum Beam Sci.}
	\gdef\@doiabbr{qubs}
	\gdef\@ISSN{2412-382X}
	\gdef\@societyowner{}
}
\DeclareOption{radiation}{%
	\gdef\@journal{radiation}
	\gdef\@journalfull{Radiation}
	\gdef\@journalshort{Radiation}
	\gdef\@doiabbr{radiation}
	\gdef\@ISSN{2673-592X}
	\gdef\@societyowner{}
}
\DeclareOption{reactions}{%
	\gdef\@journal{reactions}
	\gdef\@journalfull{Reactions}
	\gdef\@journalshort{Reactions}
	\gdef\@doiabbr{reactions}
	\gdef\@ISSN{2624-781X}
	\gdef\@societyowner{}
}
\DeclareOption{realestate}{%
	\gdef\@journal{realestate}
	\gdef\@journalfull{Real Estate}
	\gdef\@journalshort{Real Estate}
	\gdef\@doiabbr{realestate}
	\gdef\@ISSN{2813-8090}
	\gdef\@societyowner{}
}
\DeclareOption{receptors}{%
	\gdef\@journal{receptors}
	\gdef\@journalfull{Receptors}
	\gdef\@journalshort{Receptors}
	\gdef\@doiabbr{receptors}
	\gdef\@ISSN{}
	\gdef\@societyowner{}
}
\DeclareOption{recycling}{%
	\gdef\@journal{recycling}
	\gdef\@journalfull{Recycling}
	\gdef\@journalshort{Recycling}
	\gdef\@doiabbr{recycling}
	\gdef\@ISSN{2313-4321}
	\gdef\@societyowner{}
}
\DeclareOption{regeneration}{%
	\gdef\@journal{regeneration}
	\gdef\@journalfull{Regeneration}
	\gdef\@journalshort{Regeneration}
	\gdef\@doiabbr{regeneration}
	\gdef\@ISSN{}
	\gdef\@societyowner{}
}
\DeclareOption{religions}{%
	\gdef\@journal{religions}
	\gdef\@journalfull{Religions}
	\gdef\@journalshort{Religions}
	\gdef\@doiabbr{rel}
	\gdef\@ISSN{2077-1444}
	\gdef\@societyowner{}
}
\DeclareOption{remotesensing}{%
	\gdef\@journal{remotesensing}
	\gdef\@journalfull{Remote Sensing}
	\gdef\@journalshort{Remote Sens.}
	\gdef\@doiabbr{rs}
	\gdef\@ISSN{2072-4292}
	\gdef\@societyowner{}
}
\DeclareOption{reports}{%
	\gdef\@journal{reports}
	\gdef\@journalfull{Reports}
	\gdef\@journalshort{Reports}
	\gdef\@doiabbr{reports}
	\gdef\@ISSN{2571-841X}
	\gdef\@societyowner{}
}
\DeclareOption{reprodmed}{%
	\gdef\@journal{reprodmed}
	\gdef\@journalfull{Reproductive Medicine}
	\gdef\@journalshort{Reprod. Med.}
	\gdef\@doiabbr{reprodmed}
	\gdef\@ISSN{2673-3897}
	\gdef\@societyowner{}
}
\DeclareOption{resources}{%
	\gdef\@journal{resources}
	\gdef\@journalfull{Resources}
	\gdef\@journalshort{Resources}
	\gdef\@doiabbr{resources}
	\gdef\@ISSN{2079-9276}
	\gdef\@societyowner{}
}
\DeclareOption{rheumato}{%
	\gdef\@journal{rheumato}
	\gdef\@journalfull{Rheumato}
	\gdef\@journalshort{Rheumato}
	\gdef\@doiabbr{rheumato}
	\gdef\@ISSN{2674-0621}
	\gdef\@societyowner{}
}
\DeclareOption{risks}{%
	\gdef\@journal{risks}
	\gdef\@journalfull{Risks}
	\gdef\@journalshort{Risks}
	\gdef\@doiabbr{risks}
	\gdef\@ISSN{2227-9091}
	\gdef\@societyowner{}
}
\DeclareOption{robotics}{%
	\gdef\@journal{robotics}
	\gdef\@journalfull{Robotics}
	\gdef\@journalshort{Robotics}
	\gdef\@doiabbr{robotics}
	\gdef\@ISSN{2218-6581}
	\gdef\@societyowner{}
}
\DeclareOption{rsee}{%
	\gdef\@journal{rsee}
	\gdef\@journalfull{Regional Science and Environmental Economics}
	\gdef\@journalshort{Reg. Sci. Environ. Econ.}
	\gdef\@doiabbr{rsee}
	\gdef\@ISSN{}
	\gdef\@societyowner{}
}
\DeclareOption{ruminants}{%
	\gdef\@journal{ruminants}
	\gdef\@journalfull{Ruminants}
	\gdef\@journalshort{Ruminants}
	\gdef\@doiabbr{ruminants}
	\gdef\@ISSN{2673-933X}
	\gdef\@societyowner{}
}
\DeclareOption{safety}{%
	\gdef\@journal{safety}
	\gdef\@journalfull{Safety}
	\gdef\@journalshort{Safety}
	\gdef\@doiabbr{safety}
	\gdef\@ISSN{2313-576X}
	\gdef\@societyowner{}
}
\DeclareOption{sci}{%
	\gdef\@journal{sci}
	\gdef\@journalfull{Sci}
	\gdef\@journalshort{Sci}
	\gdef\@doiabbr{sci}
	\gdef\@ISSN{2413-4155}
	\gdef\@societyowner{}
}
\DeclareOption{scipharm}{%
	\gdef\@journal{scipharm}
	\gdef\@journalfull{Scientia Pharmaceutica}
	\gdef\@journalshort{Sci. Pharm.}
	\gdef\@doiabbr{scipharm}
	\gdef\@ISSN{2218-0532}
	\gdef\@societyowner{\"{O}sterreichische Pharmazeutische Gesellschaft}
}
\DeclareOption{sclerosis}{%
	\gdef\@journal{sclerosis}
	\gdef\@journalfull{Sclerosis}
	\gdef\@journalshort{Sclerosis}
	\gdef\@doiabbr{sclerosis}
	\gdef\@ISSN{2813-3064}
	\gdef\@societyowner{}
}
\DeclareOption{seeds}{%
	\gdef\@journal{seeds}
	\gdef\@journalfull{Seeds}
	\gdef\@journalshort{Seeds}
	\gdef\@doiabbr{seeds}
	\gdef\@ISSN{2674-1024}
	\gdef\@societyowner{}
}
\DeclareOption{sensors}{%
	\gdef\@journal{sensors}
	\gdef\@journalfull{Sensors}
	\gdef\@journalshort{Sensors}
	\gdef\@doiabbr{s}
	\gdef\@ISSN{1424-8220}
	\gdef\@societyowner{}
}
\DeclareOption{separations}{%
	\gdef\@journal{separations}
	\gdef\@journalfull{Separations}
	\gdef\@journalshort{Separations}
	\gdef\@doiabbr{separations}
	\gdef\@ISSN{2297-8739}
	\gdef\@societyowner{}
}
\DeclareOption{sexes}{%
	\gdef\@journal{sexes}
	\gdef\@journalfull{Sexes}
	\gdef\@journalshort{Sexes}
	\gdef\@doiabbr{sexes}
	\gdef\@ISSN{2411-5118}
	\gdef\@societyowner{}
}
\DeclareOption{signals}{%
	\gdef\@journal{signals}
	\gdef\@journalfull{Signals}
	\gdef\@journalshort{Signals}
	\gdef\@doiabbr{signals}
	\gdef\@ISSN{2624-6120}
	\gdef\@societyowner{}
}
\DeclareOption{sinusitis}{%
	\gdef\@journal{sinusitis}
	\gdef\@journalfull{Sinusitis}
	\gdef\@journalshort{Sinusitis}
	\gdef\@doiabbr{sinusitis}
	\gdef\@ISSN{2309-107X}
	\gdef\@societyowner{}
}
\DeclareOption{siuj}{%
	\gdef\@journal{siuj}
	\gdef\@journalfull{Soci\'{e}t\'{e} Internationale d'Urologie Journal}
	\gdef\@journalshort{Soc. Int. Urol. J.}
	\gdef\@doiabbr{siuj}
	\gdef\@ISSN{2563-6499}
	\gdef\@societyowner{Soci\'{e}t\'{e} Internationale d'Urologie}
}
\DeclareOption{skins}{%
	\gdef\@journal{skins}
	\gdef\@journalfull{Skins}
	\gdef\@journalshort{Skins}
	\gdef\@doiabbr{skins}
	\gdef\@ISSN{}
	\gdef\@societyowner{}
}
\DeclareOption{smartcities}{%
	\gdef\@journal{smartcities}
	\gdef\@journalfull{Smart Cities}
	\gdef\@journalshort{Smart Cities}
	\gdef\@doiabbr{smartcities}
	\gdef\@ISSN{2624-6511}
	\gdef\@societyowner{}
}
\DeclareOption{sna}{%
	\gdef\@journal{sna}
	\gdef\@journalfull{Sinusitis and Asthma}
	\gdef\@journalshort{Sinusitis Asthma}
	\gdef\@doiabbr{sna}
	\gdef\@ISSN{2624-7003}
	\gdef\@societyowner{}
}
\DeclareOption{societies}{%
	\gdef\@journal{societies}
	\gdef\@journalfull{Societies}
	\gdef\@journalshort{Societies}
	\gdef\@doiabbr{soc}
	\gdef\@ISSN{2075-4698}
	\gdef\@societyowner{}
}
\DeclareOption{socsci}{%
	\gdef\@journal{socsci}
	\gdef\@journalfull{Social Sciences}
	\gdef\@journalshort{Soc. Sci.}
	\gdef\@doiabbr{socsci}
	\gdef\@ISSN{2076-0760}
	\gdef\@societyowner{}
}
\DeclareOption{software}{%
	\gdef\@journal{software}
	\gdef\@journalfull{Software}
	\gdef\@journalshort{Software}
	\gdef\@doiabbr{software}
	\gdef\@ISSN{}
	\gdef\@societyowner{}
}
\DeclareOption{soilsystems}{%
	\gdef\@journal{soilsystems}
	\gdef\@journalfull{Soil Systems}
	\gdef\@journalshort{Soil Syst.}
	\gdef\@doiabbr{soilsystems}
	\gdef\@ISSN{2571-8789}
	\gdef\@societyowner{}
}
\DeclareOption{solar}{%
	\gdef\@journal{solar}
	\gdef\@journalfull{Solar}
	\gdef\@journalshort{Solar}
	\gdef\@doiabbr{solar}
	\gdef\@ISSN{2673-9941}
	\gdef\@societyowner{}
}
\DeclareOption{solids}{%
	\gdef\@journal{solids}
	\gdef\@journalfull{Solids}
	\gdef\@journalshort{Solids}
	\gdef\@doiabbr{solids}
	\gdef\@ISSN{2673-6497}
	\gdef\@societyowner{}
}
\DeclareOption{spectroscj}{%
	\gdef\@journal{spectroscj}
	\gdef\@journalfull{Spectroscopy Journal}
	\gdef\@journalshort{Spectrosc. J.}
	\gdef\@doiabbr{spectroscj}
	\gdef\@ISSN{}
	\gdef\@societyowner{}
}
\DeclareOption{sports}{%
	\gdef\@journal{sports}
	\gdef\@journalfull{Sports}
	\gdef\@journalshort{Sports}
	\gdef\@doiabbr{sports}
	\gdef\@ISSN{2075-4663}
	\gdef\@societyowner{}
}
\DeclareOption{standards}{%
	\gdef\@journal{standards}
	\gdef\@journalfull{Standards}
	\gdef\@journalshort{Standards}
	\gdef\@doiabbr{standards}
	\gdef\@ISSN{2305-6703}
	\gdef\@societyowner{}
}
\DeclareOption{stats}{%
	\gdef\@journal{stats}
	\gdef\@journalfull{Stats}
	\gdef\@journalshort{Stats}
	\gdef\@doiabbr{stats}
	\gdef\@ISSN{2571-905X}
	\gdef\@societyowner{}
}
\DeclareOption{std}{%
	\gdef\@journal{std}
	\gdef\@journalfull{Surgical Techniques Development}
	\gdef\@journalshort{Surg. Tech. Dev.}
	\gdef\@doiabbr{std}
	\gdef\@ISSN{2038-9582}
	\gdef\@societyowner{}
}
\DeclareOption{stresses}{%
	\gdef\@journal{stresses}
	\gdef\@journalfull{Stresses}
	\gdef\@journalshort{Stresses}
	\gdef\@doiabbr{stresses}
	\gdef\@ISSN{2673-7140}
	\gdef\@societyowner{}
}
\DeclareOption{surfaces}{%
	\gdef\@journal{surfaces}
	\gdef\@journalfull{Surfaces}
	\gdef\@journalshort{Surfaces}
	\gdef\@doiabbr{surfaces}
	\gdef\@ISSN{2571-9637}
	\gdef\@societyowner{}
}
\DeclareOption{surgeries}{%
	\gdef\@journal{surgeries}
	\gdef\@journalfull{Surgeries}
	\gdef\@journalshort{Surgeries}
	\gdef\@doiabbr{surgeries}
	\gdef\@ISSN{2017-2017}
	\gdef\@societyowner{}
}
\DeclareOption{suschem}{%
	\gdef\@journal{suschem}
	\gdef\@journalfull{Sustainable Chemistry}
	\gdef\@journalshort{Sustain. Chem.}
	\gdef\@doiabbr{suschem}
	\gdef\@ISSN{2673-4079}
	\gdef\@societyowner{}
}
\DeclareOption{sustainability}{%
	\gdef\@journal{sustainability}
	\gdef\@journalfull{Sustainability}
	\gdef\@journalshort{Sustainability}
	\gdef\@doiabbr{su}
	\gdef\@ISSN{2071-1050}
	\gdef\@societyowner{}
}
\DeclareOption{symmetry}{%
	\gdef\@journal{symmetry}
	\gdef\@journalfull{Symmetry}
	\gdef\@journalshort{Symmetry}
	\gdef\@doiabbr{sym}
	\gdef\@ISSN{2073-8994}
	\gdef\@societyowner{}
}
\DeclareOption{synbio}{%
	\gdef\@journal{synbio}
	\gdef\@journalfull{SynBio}
	\gdef\@journalshort{SynBio}
	\gdef\@doiabbr{synbio}
	\gdef\@ISSN{2674-0583}
	\gdef\@societyowner{}
}
\DeclareOption{systems}{%
	\gdef\@journal{systems}
	\gdef\@journalfull{Systems}
	\gdef\@journalshort{Systems}
	\gdef\@doiabbr{systems}
	\gdef\@ISSN{2079-8954}
	\gdef\@societyowner{}
}
\DeclareOption{tae}{%
        \gdef\@journal{tae}
        \gdef\@journalfull{Theoretical and Applied Ergonomics}
        \gdef\@journalshort{Theor. Appl. Ergon.}
        \gdef\@doiabbr{tae}
        \gdef\@ISSN{}
        \gdef\@societyowner{}
 }
\DeclareOption{targets}{%
	\gdef\@journal{targets}
	\gdef\@journalfull{Targets}
	\gdef\@journalshort{Targets}
	\gdef\@doiabbr{targets}
	\gdef\@ISSN{}
	\gdef\@societyowner{}
}
\DeclareOption{taxonomy}{%
	\gdef\@journal{taxonomy}
	\gdef\@journalfull{Taxonomy}
	\gdef\@journalshort{Taxonomy}
	\gdef\@doiabbr{taxonomy}
	\gdef\@ISSN{2673-6500}
	\gdef\@societyowner{}
}
\DeclareOption{technologies}{%
	\gdef\@journal{technologies}
	\gdef\@journalfull{Technologies}
	\gdef\@journalshort{Technologies}
	\gdef\@doiabbr{technologies}
	\gdef\@ISSN{2227-7080}
	\gdef\@societyowner{}
}
\DeclareOption{telecom}{%
	\gdef\@journal{telecom}
	\gdef\@journalfull{Telecom}
	\gdef\@journalshort{Telecom}
	\gdef\@doiabbr{telecom}
	\gdef\@ISSN{2673-4001}
	\gdef\@societyowner{}
}
\DeclareOption{test}{%
	\gdef\@journal{test}
	\gdef\@journalfull{Test}
	\gdef\@journalshort{Test}
	\gdef\@doiabbr{test}
	\gdef\@ISSN{}
	\gdef\@societyowner{}
}
\DeclareOption{textiles}{%
	\gdef\@journal{textiles}
	\gdef\@journalfull{Textiles}
	\gdef\@journalshort{Textiles}
	\gdef\@doiabbr{textiles}
	\gdef\@ISSN{2673-7248}
	\gdef\@societyowner{}
}
\DeclareOption{thalassrep}{%
	\gdef\@journal{thalassrep}
	\gdef\@journalfull{Thalassemia Reports}
	\gdef\@journalshort{Thalass. Rep.}
	\gdef\@doiabbr{thalassrep}
	\gdef\@ISSN{2039-4365}
	\gdef\@societyowner{}
}
\DeclareOption{therapeutics}{%
	\gdef\@journal{therapeutics}
	\gdef\@journalfull{Therapeutics}
	\gdef\@journalshort{Therapeutics}
	\gdef\@doiabbr{therapeutics}
	\gdef\@ISSN{}
	\gdef\@societyowner{}
}
\DeclareOption{thermo}{%
	\gdef\@journal{thermo}
	\gdef\@journalfull{Thermo}
	\gdef\@journalshort{Thermo}
	\gdef\@doiabbr{thermo}
	\gdef\@ISSN{2673-7264}
	\gdef\@societyowner{}
}
\DeclareOption{timespace}{%
	\gdef\@journal{timespace}
	\gdef\@journalfull{Time and Space}
	\gdef\@journalshort{Time Space}
	\gdef\@doiabbr{timespace}
	\gdef\@ISSN{}
	\gdef\@societyowner{}
}
\DeclareOption{tomography}{%
	\gdef\@journal{tomography}
	\gdef\@journalfull{Tomography}
	\gdef\@journalshort{Tomography}
	\gdef\@doiabbr{tomography}
	\gdef\@ISSN{2379-139X}
	\gdef\@societyowner{}
}
\DeclareOption{tourismhosp}{%
	\gdef\@journal{tourismhosp}
	\gdef\@journalfull{Tourism and Hospitality}
	\gdef\@journalshort{Tour. Hosp.}
	\gdef\@doiabbr{tourhosp}
	\gdef\@ISSN{2673-5768}
	\gdef\@societyowner{}
}
\DeclareOption{toxics}{%
	\gdef\@journal{toxics}
	\gdef\@journalfull{Toxics}
	\gdef\@journalshort{Toxics}
	\gdef\@doiabbr{toxics}
	\gdef\@ISSN{2305-6304}
	\gdef\@societyowner{}
}
\DeclareOption{toxins}{%
	\gdef\@journal{toxins}
	\gdef\@journalfull{Toxins}
	\gdef\@journalshort{Toxins}
	\gdef\@doiabbr{toxins}
	\gdef\@ISSN{2072-6651}
	\gdef\@societyowner{}
}
\DeclareOption{transplantology}{%
	\gdef\@journal{transplantology}
	\gdef\@journalfull{Transplantology}
	\gdef\@journalshort{Transplantology}
	\gdef\@doiabbr{transplantology}
	\gdef\@ISSN{2673-3943}
	\gdef\@societyowner{}
}
\DeclareOption{transportation}{%
	\gdef\@journal{transportation}
	\gdef\@journalfull{Transportation}
	\gdef\@journalshort{Transportation}
	\gdef\@doiabbr{transportation}
	\gdef\@ISSN{}
	\gdef\@societyowner{}
}
\DeclareOption{traumacare}{%
	\gdef\@journal{traumacare}
	\gdef\@journalfull{Trauma Care}
	\gdef\@journalshort{Trauma Care}
	\gdef\@doiabbr{traumacare}
	\gdef\@ISSN{2673-866X}
	\gdef\@societyowner{}
}
\DeclareOption{traumas}{%
	\gdef\@journal{traumas}
	\gdef\@journalfull{Traumas}
	\gdef\@journalshort{Traumas}
	\gdef\@doiabbr{traumas}
	\gdef\@ISSN{2673-7477}
	\gdef\@societyowner{}
}
\DeclareOption{tropicalmed}{%
	\gdef\@journal{tropicalmed}
	\gdef\@journalfull{Tropical Medicine and Infectious Disease}
	\gdef\@journalshort{Trop. Med. Infect. Dis.}
	\gdef\@doiabbr{tropicalmed}
	\gdef\@ISSN{2414-6366}
	\gdef\@societyowner{}
}
\DeclareOption{universe}{%
	\gdef\@journal{universe}
	\gdef\@journalfull{Universe}
	\gdef\@journalshort{Universe}
	\gdef\@doiabbr{universe}
	\gdef\@ISSN{2218-1997}
	\gdef\@societyowner{}
}
\DeclareOption{urbansci}{%
	\gdef\@journal{urbansci}
	\gdef\@journalfull{Urban Science}
	\gdef\@journalshort{Urban Sci.}
	\gdef\@doiabbr{urbansci}
	\gdef\@ISSN{2413-8851}
	\gdef\@societyowner{}
}
\DeclareOption{uro}{%
	\gdef\@journal{uro}
	\gdef\@journalfull{Uro}
	\gdef\@journalshort{Uro}
	\gdef\@doiabbr{uro}
	\gdef\@ISSN{2673-4397}
	\gdef\@societyowner{}
}
\DeclareOption{vaccines}{%
	\gdef\@journal{vaccines}
	\gdef\@journalfull{Vaccines}
	\gdef\@journalshort{Vaccines}
	\gdef\@doiabbr{vaccines}
	\gdef\@ISSN{2076-393X}
	\gdef\@societyowner{}
}
\DeclareOption{vehicles}{%
	\gdef\@journal{vehicles}
	\gdef\@journalfull{Vehicles}
	\gdef\@journalshort{Vehicles}
	\gdef\@doiabbr{vehicles}
	\gdef\@ISSN{2624-8921}
	\gdef\@societyowner{}
}
\DeclareOption{venereology}{%
	\gdef\@journal{venereology}
	\gdef\@journalfull{Venereology}
	\gdef\@journalshort{Venereology}
	\gdef\@doiabbr{venereology}
	\gdef\@ISSN{2674-0710}
	\gdef\@societyowner{}
}
\DeclareOption{vetsci}{%
	\gdef\@journal{vetsci}
	\gdef\@journalfull{Veterinary Sciences}
	\gdef\@journalshort{Vet. Sci.}
	\gdef\@doiabbr{vetsci}
	\gdef\@ISSN{2306-7381}
	\gdef\@societyowner{}
}
\DeclareOption{vibration}{%
	\gdef\@journal{vibration}
	\gdef\@journalfull{Vibration}
	\gdef\@journalshort{Vibration}
	\gdef\@doiabbr{vibration}
	\gdef\@ISSN{2571-631X}
	\gdef\@societyowner{}
}
\DeclareOption{virtualworlds}{%
	\gdef\@journal{virtualworlds}
	\gdef\@journalfull{Virtual Worlds}
	\gdef\@journalshort{Virtual Worlds}
	\gdef\@doiabbr{virtualworlds}
	\gdef\@ISSN{}
	\gdef\@societyowner{}
}
\DeclareOption{viruses}{%
	\gdef\@journal{viruses}
	\gdef\@journalfull{Viruses}
	\gdef\@journalshort{Viruses}
	\gdef\@doiabbr{v}
	\gdef\@ISSN{1999-4915}
	\gdef\@societyowner{}
}
\DeclareOption{vision}{%
	\gdef\@journal{vision}
	\gdef\@journalfull{Vision}
	\gdef\@journalshort{Vision}
	\gdef\@doiabbr{vision}
	\gdef\@ISSN{2411-5150}
	\gdef\@societyowner{}
}
\DeclareOption{waste}{%
	\gdef\@journal{waste}
	\gdef\@journalfull{Waste}
	\gdef\@journalshort{Waste}
	\gdef\@doiabbr{waste}
	\gdef\@ISSN{2813-0391}
	\gdef\@societyowner{}
}
\DeclareOption{water}{%
	\gdef\@journal{water}
	\gdef\@journalfull{Water}
	\gdef\@journalshort{Water}
	\gdef\@doiabbr{w}
	\gdef\@ISSN{2073-4441}
	\gdef\@societyowner{}
}
\DeclareOption{wem}{%
	\gdef\@journal{wem}
	\gdef\@journalfull{Wildlife Ecology and Management}
	\gdef\@journalshort{Wildl. Ecol. Manag.}
	\gdef\@doiabbr{wem}
	\gdef\@ISSN{1234-4321}
	\gdef\@societyowner{}
}
\DeclareOption{wevj}{%
	\gdef\@journal{wevj}
	\gdef\@journalfull{World Electric Vehicle Journal}
	\gdef\@journalshort{World Electr. Veh. J.}
	\gdef\@doiabbr{wevj}
	\gdef\@ISSN{2032-6653}
	\gdef\@societyowner{World Electric Vehicle Association}
}
\DeclareOption{wild}{%
	\gdef\@journal{wild}
	\gdef\@journalfull{Wild}
	\gdef\@journalshort{Wild}
	\gdef\@doiabbr{wild}
	\gdef\@ISSN{}
	\gdef\@societyowner{}
}
\DeclareOption{wind}{%
	\gdef\@journal{wind}
	\gdef\@journalfull{Wind}
	\gdef\@journalshort{Wind}
	\gdef\@doiabbr{wind}
	\gdef\@ISSN{2674-032X}
	\gdef\@societyowner{}
}
\DeclareOption{women}{%
	\gdef\@journal{women}
	\gdef\@journalfull{Women}
	\gdef\@journalshort{Women}
	\gdef\@doiabbr{women}
	\gdef\@ISSN{2673-4184}
	\gdef\@societyowner{}
}
\DeclareOption{world}{%
	\gdef\@journal{world}
	\gdef\@journalfull{World}
	\gdef\@journalshort{World}
	\gdef\@doiabbr{world}
	\gdef\@ISSN{2673-4060}
	\gdef\@societyowner{}
}
\DeclareOption{youth}{%
	\gdef\@journal{youth}
	\gdef\@journalfull{Youth}
	\gdef\@journalshort{Youth}
	\gdef\@doiabbr{youth}
	\gdef\@ISSN{2673-995X}
	\gdef\@societyowner{}
}
\DeclareOption{zoonoticdis}{%
	\gdef\@journal{zoonoticdis}
	\gdef\@journalfull{Zoonotic Diseases}
	\gdef\@journalshort{Zoonotic Dis.}
	\gdef\@doiabbr{zoonoticdis}
	\gdef\@ISSN{2813-0227}
	\gdef\@societyowner{}
}
}
\documentclass[preprints,article,accept,pdftex,oneauthor]{Definitions/mdpi}
\makeatletter
\patchcmd{\@maketitle}{\textit{\@arttype}}{}{}
  {\PackageError{arxiv-preprint}{Could not remove the article type from the title}{}\relax}
\makeatother
\AddToHook{cmd/maketitlen/before}{\vspace{-24pt}}
\usepackage{needspace}
\usepackage{xurl}
\NewDocumentCommand{\abstractproofnote}{mm}{\par\smallskip\noindent #1 \url{#2}.}
\AddToHook{package/hyperref/after}{\pdfstringdefDisableCommands{\def\abstractproofnote#1#2{\space#1 #2.}}}
\usepackage{pgfplots}
\pgfplotsset{compat=1.18}
\newtheorem{result}{Theorem}
\newenvironment{theorem}[1][]{\begin{result}[#1]}{\end{result}}
\newtheorem{qlemma}[result]{Lemma}
\newenvironment{lemma}[1][]{\begin{qlemma}[#1]}{\end{qlemma}}
\newtheorem{qproposition}[result]{Proposition}
\newenvironment{proposition}[1][]{\begin{qproposition}[#1]}{\end{qproposition}}
\newtheorem{qcorollary}[result]{Corollary}
\newenvironment{corollary}[1][]{\begin{qcorollary}[#1]}{\end{qcorollary}}
\reftitle{References}
\newcommand{\E}{\mathbb{E}}
\newcommand{\Tr}{\operatorname{Tr}}
\newcommand{\diag}{\operatorname{diag}}
\newcommand{\pg}{p_{\mathrm{guess}}}
\newcommand{\dd}{\,\mathrm{d}}
\Title{Quantum Subentropy: Subadditivity and Sharp Bounds on Discrimination Error}
\TitleCitation{Quantum Subentropy: Subadditivity and Sharp Bounds on Discrimination Error}
\Author{Pratik Patil}
\AuthorNames{Pratik Patil}
\AuthorCitation{Patil, P.}
\address[1]{Department of Statistics and Data Sciences,
University of Texas at Austin, Austin, USA; pratikpatil@utexas.edu}
\corres{Correspondence: pratikpatil@utexas.edu}
\abstract{Subentropy gives a tight spectral lower bound on the accessible information of ensembles of pure states.
We prove its subadditivity for arbitrary bipartite states in finite dimensions and derive a sharp relation between conditional subentropy and the minimum discrimination error.
This relation strengthens a conjectured comparison with conditional min-entropy.
Its inverse gives the strongest universal lower bound on the error determined only by conditional subentropy and the number of labels.
Both main results follow from a comparison of conditional subentropy with the average subentropy of measurement posteriors, valid for mixed signal states and every finite measurement.
We characterize all equality cases in these two results and bound posterior deviations near equality.
We also determine the optimal Lipschitz constant in trace distance for conditional subentropy on classical--quantum states, including the exact constant when the quantum marginal is fixed.
A determinant inequality strengthens subadditivity and proves strong subadditivity for classical distributions, while an exact counterexample shows that strong subadditivity fails for quantum states.%
\abstractproofnote{Lean~4 proofs of all theoretical results are available at}{https://pratikpatil.io/lean/subentropy/}}
\begin{document}
\raggedbottom
\section{Introduction}
\subsection{Subentropy and quantum state discrimination}
Suppose a source prepares a quantum signal $\rho_x$ with prior probability $p_x$.
In minimum-error discrimination, a receiver uses a measurement outcome to guess $x$, with the aim of minimizing the probability of error.
We study bounds on this optimum that use only the spectra of the joint state of label and signal and of the average signal state.
These bounds give guarantees when the relative eigenvectors of the signal states are not known.
We ask how subentropy provides such a bound and how it behaves when quantum systems are combined.

Subentropy was introduced by Jozsa, Robb, and Wootters \cite{jozsa1994lower} to bound accessible information, the largest Shannon mutual information between the label and a measurement outcome.
For ensembles of pure states with a prescribed average state, the subentropy of that state is the greatest universal lower bound on accessible information.
The Scrooge ensemble attains this bound.
Representations through random measurements and Gaussian vectors provide useful tools for studying subentropy \cite{datta2014properties,jozsa2015symmetric,wootters2018classical,mcginley2025scrooge}.

This bound on accessible information does not directly give a lower bound on discrimination error.
Subentropy also appears in bounds on guesswork, the minimum expected number of successive guesses \cite{chen2015minimum}.
The present question concerns the error probability of a single guess.

Datta, Dorlas, Jozsa, and Benatti \cite{datta2014properties} posed two conjectures connecting these structural and operational questions.
Conjecture~1 asserts subadditivity for arbitrary bipartite states; they proved it for product states.
Conjecture~2 bounds conditional subentropy by conditional min-entropy, the negative logarithm of the optimal guessing probability \cite{konig2009operational}.
They proved this comparison for binary ensembles of pure signals.
Zhang, Singh, and Pati \cite{zhang2017average} subsequently highlighted subadditivity and strong subadditivity (SSA) as open questions.
We prove subadditivity and strengthen the conditional min-entropy comparison to a sharp bound on the optimal discrimination error for each fixed number of labels.
We characterize equality in both main results.

While this manuscript was under journal review, we became aware of two related developments.
The Omega Institute's public Lean source repository \cite{omegainstitute2026trureturing} contains formalizations of subadditivity and the linear guessing bound $Q(X|B)\le1-\pg$.
Zhang, Zhang, and Ma \cite{zhang2026subadditivity} establish subadditivity and its equality cases for the Nichols--Wootters family, including subentropy, and prove failure of quantum strong subadditivity.
They also obtain a sharp two-qubit subadditivity deficit bound that strengthens the elementary estimate given here.
Our results and proofs were developed independently of both works.
The present paper additionally establishes the sharp discrimination bound and its equality cases, quantitative posterior stability, and optimal continuity bounds for conditional subentropy on classical--quantum states.
Further results include pointwise determinant inequalities, classical strong subadditivity with equality cases, and bounds that account for a specified ancillary state.

\subsection{Definitions and the main results}
We work with finite-dimensional Hilbert spaces and finite label alphabets.
Logarithms are natural, and entropy is measured in nats.
We write $Q(p)$ for the subentropy of a probability vector and $Q(\rho)$ for the subentropy of the spectrum of a density operator.
If its eigenvalues are $\lambda_1,\ldots,\lambda_d$, then
\begin{equation}\label{eq:definition}
 Q(\rho)=-\sum_{i=1}^d
 \frac{\lambda_i^d\log\lambda_i}
 {\prod_{j\ne i}(\lambda_i-\lambda_j)}.
\end{equation}
The formula is interpreted by continuity when eigenvalues coincide or vanish.
Appending zeros does not change $Q$, and pure states have subentropy zero.

For a bipartite state, $\rho_A=\Tr_B\rho_{AB}$ and $\rho_B=\Tr_A\rho_{AB}$ are its marginal states.
We also write $Q(AB)$ for $Q(\rho_{AB})$, and similarly for other registers.
Following \cite{datta2014properties}, we define conditional subentropy by $Q(A|B)=Q(\rho_{AB})-Q(\rho_B)$.

For matrices, positive means positive semidefinite; positive definiteness is stated explicitly.
Matrix inequalities use the positive semidefinite order.

An ensemble of signal states with $n$ labels is encoded by the classical--quantum (CQ) state $\rho_{XB}=\sum_x|x\rangle\langle x|\otimes A_x$, where $A_x=p_x\rho_x\ge0$ and $\sum_x\Tr A_x=1$.
The classical register $X$ records the label, and $B$ holds the quantum side information.
For this state, the conditional subentropy, average signal state, and optimal guessing probability are
\begin{equation}\label{eq:conditional}
 Q(X|B)=Q(\rho_{XB})-Q(\rho_B),\qquad
 \rho_B=\sum_xA_x,\qquad
 \pg=\max_{\{E_x\}}\sum_x\Tr(E_xA_x).
\end{equation}
The maximum is over positive operator-valued measures (POVMs) with outcomes labeled by $x$: $E_x\ge0$ and $\sum_xE_x=I$.
Each outcome is interpreted as the receiver's guess of the label.

The notation $Q(X|B)$ always denotes the spectral difference in \eqref{eq:conditional}.
Measurement posteriors are the conditional label distributions given the observed outcomes.
Their average subentropy is generally different from conditional subentropy.
Nonnegativity of conditional subentropy for CQ states was established by Datta et al.\ \cite[Proposition 5]{datta2014properties}.
Proposition~\ref{prop:posterior-comparison} recovers this fact and gives an upper bound in terms of the measurement posteriors.

Signals with orthogonal supports can be distinguished perfectly, and their conditional subentropy is zero.
If all signals are the same pure state, no measurement reveals the label: the optimal rule is to guess a label with largest prior probability, and conditional subentropy is the subentropy of the prior.
At a fixed largest prior probability, subentropy is maximized when all remaining probabilities are equal.
This extremal prior determines the curve in the next theorem.

For $n\ge2$, define
\begin{equation}\label{eq:phi}
 \Phi_n(u)=Q\Bigl(1-u,\frac{u}{n-1},\ldots,\frac{u}{n-1}\Bigr),
 \qquad 0\le u\le1.
\end{equation}
The range $0\le u\le1-1/n$ contains all optimal discrimination errors.
Defining the curve on the full interval simplifies the concavity argument.

\begin{theorem}[Sharp guessing bound]\label{thm:guess}
For every classical--quantum ensemble with $n\ge2$ labels, its optimal error $e=1-\pg$ lies in $[0,1-1/n]$ and satisfies
\begin{equation}\label{eq:guess}
 0\le Q(X|B)\le\Phi_n(e)\le e\le-\log\pg.
\end{equation}
The bound $Q(X|B)\le\Phi_n(e)$ is pointwise optimal at fixed $n$: equality is achievable at every admissible error.

For $e>0$, equality holds exactly when, up to label permutation, all signals are the same pure state and $p=(1-e,e/(n-1),\ldots,e/(n-1))$.
Moreover, $Q(X|B)=0$ if and only if $e=0$, equivalently when the signals with positive prior probability have pairwise orthogonal supports.

For $e>0$, $\Phi_n(e)<e$; the coefficient $1$ in the linear bound $Q(X|B)\le e$ is nevertheless optimal uniformly over ensembles.
For $n=1$, both the error and conditional subentropy are zero.
\end{theorem}
Since $H_{\min}(X|B)=-\log\pg$ \cite{konig2009operational}, Theorem~\ref{thm:guess} proves the conditional min-entropy comparison proposed in \cite{datta2014properties}.
The sharper bound can be inverted to give a lower bound on discrimination error.
On $[0,1-1/n]$, $\Phi_n$ is strictly increasing, so
\begin{equation}\label{eq:inverse}
 e\ge\Phi_n^{-1}\bigl(Q(X|B)\bigr).
\end{equation}
The right-hand side depends only on the joint and average spectra and requires no optimization over measurements.
At the upper endpoint, $\Phi_n(1-1/n)=\log n-H_n+1$, where $H_n=\sum_{k=1}^n k^{-1}$.

The equality cases in Theorem~\ref{thm:guess} show that \eqref{eq:inverse} is the strongest universal lower bound on the error determined by $Q(X|B)$ and $n$ alone.
For a particular ensemble, the inequality may be strict.

When the full operators $A_x$ are known, the optimization in \eqref{eq:conditional} can be solved numerically as a semidefinite program \cite[Section I]{montanaro2008lower}.
Spectral data can also be obtained without reconstructing those operators: unitarily invariant collective measurements on repeated preparations can estimate the signal spectra and the spectrum of the average state without estimating eigenvectors \cite{keyl2001estimating}.
The known priors and signal spectra determine the joint spectrum because $\rho_{XB}$ is block diagonal.
Thus \eqref{eq:inverse} gives a guarantee from these data even when the relative eigenvectors required to specify the semidefinite program are unavailable.
Section~\ref{sec:spectral-bounds} gives two commuting ensembles with the same priors, individual signal spectra, and average state, but different optimal errors.

\Needspace{8\baselineskip}
Our second main result establishes subadditivity and characterizes equality for arbitrary bipartite states.
\begin{theorem}[Subadditivity]\label{thm:sa}
Every finite-dimensional bipartite state satisfies
\begin{equation}\label{eq:sa}
 Q(\rho_{AB})\le Q(\rho_A)+Q(\rho_B).
\end{equation}
Equality holds exactly when at least one marginal is pure.
\end{theorem}

The equality cases of the guessing bound raise a stability question: for an optimal guessing measurement, does a small deficit in the sharp bound force the posterior distributions to be close to the maximizing vector?
Proposition~\ref{prop:posterior-stability} shows that their average squared Euclidean distance is at most six times the deficit.
Each posterior is compared with the vector in \eqref{eq:phi} at the optimal error, with the largest coordinate placed at the guessed label.
The estimate concerns measurement posteriors.
As Section~\ref{sec:ensemble-stability} explains, a vanishing absolute deficit alone need not imply that the signal ensemble is close to an ensemble attaining equality at positive error.

To control the effect of state perturbations on the discrimination bound, we use the trace distance $D(\rho,\sigma)=\|\rho-\sigma\|_1/2$, where $\|\cdot\|_1$ denotes the trace norm.
For states diagonal in a common basis, this is the total variation distance of their diagonal distributions.
When there are at least two labels, $Q(X|B)$ is Lipschitz continuous on CQ states with optimal constant one (Theorem~\ref{thm:cq-lipschitz}).
Corollary~\ref{cor:fixed-marginal} identifies the exact constant when the marginal state of $B$ is fixed.
These estimates control the change in conditional subentropy before the inverse bound is applied.

\subsection{Proof strategy and relation to earlier results}
A measurement replaces quantum side information $B$ by a classical outcome.
The average subentropy of the resulting posteriors generally differs from conditional subentropy, which is a difference of two spectral quantities.
Proposition~\ref{prop:posterior-comparison} relates them: for every finite POVM with outcome probabilities $q_y$ and posteriors $b^y$,
\[
 Q(X|B)\le\sum_yq_yQ(b^y).
\]
The guessing bound then follows from a classical maximization.
At a fixed largest probability, subentropy is maximized when the remaining probabilities are equal.
Applying this fact to each posterior and using concavity gives a bound in terms of the average guessing error.
Choosing an optimal guessing measurement gives Theorem~\ref{thm:guess}.
The same comparison also proves subadditivity.
Removing off-diagonal blocks in an eigenbasis of one marginal gives a CQ state with the same marginals.
For a measurement with a single outcome, the posterior is the spectrum of the classical marginal.

To prove the comparison, we construct a Gram matrix with the same nonzero spectrum as the average signal state.
Its block diagonal has the same nonzero spectrum as the joint state.
The construction for mixed signals and the refinement of measurements follow Montanaro \cite[Section III]{montanaro2008lower}.
A related refinement of pretty good measurements for mixed states appears in Barnum and Knill \cite[Section V]{barnum2002reversing}.
The key property is convexity of the increase in subentropy under pinching, the operation that removes off-diagonal blocks.
A Gaussian representation expresses this increase as a classical relative entropy.
Joint convexity of relative entropy then bounds the increase for the Gram matrix by its average over the pure components induced by the refined measurement.
Concavity combines the refined outcomes to give the displayed comparison.

The maximizing spectrum in \eqref{eq:phi} was identified for subentropy in \cite[Proposition 4]{datta2014properties}.
The bound on posterior averages is a specialization of Rioul's Fano framework \cite[Theorems 4 and 5]{rioul2023interplay}; see also \cite{scarlett2021introductory}.
These classical results concern posterior averages; the measurement comparison relates them to conditional subentropy.
A spectrum with all but one eigenvalue equal also arises in optimization at fixed purity \cite{dallarno2019tradeoff}.
Here the constraint is discrimination error.

Strong concavity makes the posterior argument quantitative and gives the stability estimate.
The optimal modulus of continuity for subentropy itself, including its spectral equality cases, follows from the general theorem of Hanson and Datta \cite{hanson2022entropies}.
Our proof uses maximal coupling to relate this known modulus to the guessing bound.
For conditional subentropy, the shared CQ block structure yields the continuity bounds.

A separate proof of subadditivity uses the reciprocal determinant representation of subentropy.
Merging rows and columns of a classical probability table gives a determinant inequality at each value of the integration parameter.
Hadamard's determinant inequality transfers the resulting bipartite comparison to quantum states.
The merging inequalities also yield classical strong subadditivity.
An exact counterexample shows that strong subadditivity fails for general quantum states.

The theoretical results established in this paper have been formalized in Lean~4.
Parts of the Lean formalization adapt code from the Omega Institute's development~\cite{omegainstitute2026trureturing}.
The companion proof index at \mbox{\url{https://pratikpatil.io/lean/subentropy/}} links the statements and proofs and includes certificates for the numerical bounds in Tables~\ref{tab:binary} and~\ref{tab:cyclic}.

The remainder of the paper is organized as follows.
Section~\ref{sec:representations} gives the representations used in the proofs.
Sections~\ref{sec:guessing}--\ref{sec:composition} develop the measurement comparison, stability and continuity estimates, and determinant inequalities.
Section~\ref{sec:examples} compares spectral lower bounds on the error and studies independent ancillary states.
The appendices give polynomial and asymptotic calculations, details of the continuity equality cases, and an example with three labels.

\section{Representations of subentropy}\label{sec:representations}
The spectral definition is convenient for evaluation.
Our proofs use two equivalent representations: a Gaussian expectation and an integral of reciprocal determinants.
The Gaussian representation relates concavity of subentropy to scalar convexity and leads to the measurement comparison.
The determinant representation reduces composition inequalities to determinant inequalities at each value of the integration parameter.

Throughout the paper, the norms $\|\cdot\|_1$, $\|\cdot\|_2=\|\cdot\|_{\mathrm{HS}}$, and $\|\cdot\|_\infty$ are the trace, Hilbert--Schmidt, and operator norms, respectively.
On vectors, the subscript two denotes the Euclidean norm.

Let $H_d=\sum_{k=1}^d k^{-1}$ and let $\gamma$ be Euler's constant.
The random measurement representation \cite{datta2014properties} gives
\begin{equation}\label{eq:simplex}
 Q(\lambda)=-d\E[(\lambda\cdot X)\log(\lambda\cdot X)]-(H_d-1),
\end{equation}
where $X$ is uniform on the simplex of probability vectors with the same number of entries as the spectrum.

Gaussian constructions for the Scrooge ensemble are discussed by Wootters \cite{wootters2018classical}; McGinley and Schuster \cite{mcginley2025scrooge} state the Gaussian subentropy identity explicitly.
The determinant formula is equivalent, under a reciprocal change of variable, to the integral representation of Jozsa and Mitchison \cite{jozsa2015symmetric}.
We include a derivation that also covers singular states.
\begin{lemma}[Gaussian and integral forms]\label{lem:forms}
Let $g$ be a standard circular complex Gaussian vector, with independent coordinates satisfying $\E|g_i|^2=1$.
Then
\begin{align}
 Q(\rho)&=1-\gamma-F(\rho),&
 F(\rho)&=\E[(g^*\rho g)\log(g^*\rho g)],\label{eq:gaussian}\\
 Q(\rho)&=\int_0^\infty
 \Bigl[\frac1{1+t}-\frac1{\det(I+t\rho)}\Bigr]\frac{\dd t}{t^2}.
 \label{eq:integral}
\end{align}
Both formulas remain valid for singular states.
\end{lemma}
\begin{proof}
Unitary invariance of the Gaussian law allows us to work in an eigenbasis of the state.
The squared moduli of the Gaussian coordinates are independent exponential variables with mean one.
Normalizing by their sum gives a vector that is uniform on the probability simplex and independent of that sum.
Thus we may write $|g_i|^2=RX_i$, where the sum $R$ has a $\operatorname{Gamma}(d,1)$ distribution and is independent of $X$.
Since $\E R=d$ and $\E[R\log R]=d(H_d-\gamma)$, comparison with \eqref{eq:simplex} gives \eqref{eq:gaussian}.

For the determinant formula, we express the logarithm as an integral of exponentials.
Set $T=g^*\rho g$.
Independence of the Gaussian coordinates in an eigenbasis of the state gives its Laplace transform, $L_\rho(t)=\E e^{-tT}=\det(I+t\rho)^{-1}$.
For $y>0$, the scalar identity
\[
 y\log y=\int_0^\infty(ye^{-t}-ye^{-ty})\frac{\dd t}{t}
\]
has an integrand whose absolute value integrates to $y|\log y|$.
Since $0\le T\le\|g\|^2$ and Gaussian moments are finite, the positive part of $T\log T$ is integrable.
The negative part is bounded by $1/\exp(1)$.
Thus the integral of the absolute value has finite expectation, and Fubini's theorem allows us to interchange expectation and integration.
For the exponential distribution with mean one, the Laplace transform is $L_0(t)=(1+t)^{-1}$ and the corresponding value of $\E[T\log T]$ is $1-\gamma$.
Subtracting the two expectation formulas cancels the terms involving their common first moment and gives $F(\rho)-(1-\gamma)=\int_0^\infty[L_\rho'(t)-L_0'(t)]\dd t/t$.
Integrating by parts on a truncated interval and then taking limits gives
\[
 F(\rho)-(1-\gamma)
 =\int_0^\infty [L_\rho(t)-L_0(t)]\frac{\dd t}{t^2}.
\]
The boundary terms vanish at both endpoints.
The two distributions have the same mean and finite second moments, so the difference of their Laplace transforms is $O(t^2)$ at zero.
At infinity, both transforms are bounded, so their difference divided by $t$ vanishes.
This proves \eqref{eq:integral} for every density operator, including singular states.

Finally, the Gaussian formula recovers concavity of $Q$ \cite{datta2014properties} and shows that it is strict.
Indeed, $y\log y$ is strictly convex on $[0,\infty)$, and for distinct Hermitian matrices $R,S$, the quadratic form $g^*(R-S)g$ is nonzero on a set of positive Gaussian measure.
Thus $F$ is strictly convex on density operators, and subentropy is strictly concave.
\end{proof}

\section{Measurement comparison, guessing, and subadditivity}\label{sec:guessing}
A classical Fano argument bounds the subentropy of each posterior distribution in terms of its largest probability, then uses concavity to average over measurement outcomes.
This bounds the average posterior subentropy.
Conditional subentropy in \eqref{eq:conditional} is a difference of spectral quantities and need not equal that average.
We first bound conditional subentropy by the posterior average for every finite measurement.
Choosing an optimal guessing measurement then gives the sharp error bound.

The joint and average states act on different spaces.
We use a Gram matrix with the same nonzero spectrum as the average state.
The block diagonal of this matrix has the spectrum of the joint state.
Conditional subentropy is therefore the increase under block pinching.
Convexity bounds this increase by its average over a decomposition into pure states.
This property does not follow from concavity of subentropy alone; a Gaussian relative entropy identity establishes the required convexity.

We use Montanaro's construction for mixed signals \cite[Section III]{montanaro2008lower}, which factors the signals weighted by their priors.
Refining a measurement then gives a pure decomposition of the Gram matrix, extending the Gram method for pure signals.
Barnum and Knill describe related refinement for mixed signals and pretty good measurements \cite[Section V, Equations (35)--(37)]{barnum2002reversing}.
Gram constructions also connect coherence and discrimination \cite{xiong2018geometric}.

Related coherence results concern averages over bases or properties of dephasing channels.
The Haar average of the relative entropy of coherence can be expressed in terms of $S-Q$, where $S$ is the von Neumann entropy \cite{cheng2015complementarity}.
Subentropies of stochastic columns also enter expressions for the coherence-generating power of dephasing channels \cite{styliaris2018coherence}.
The functional defined next measures the increase in subentropy of a fixed state under a specified pinching.

\subsection{A convex functional for block pinching}
Fix mutually orthogonal projections $P_1,\ldots,P_n$ on a finite-dimensional space, with $\sum_x P_x=I$.
The associated pinching map removes the off-diagonal blocks.
Define this map and the resulting increase in subentropy by
\begin{equation}\label{eq:pinching-gain}
 \Delta(R)=\sum_xP_xRP_x,\qquad C_Q(R)=Q(\Delta(R))-Q(R).
\end{equation}
Pinching is an average over unitaries that change the relative phases of the blocks.
Together with the Gaussian representation, this averaging identifies the increase in \eqref{eq:pinching-gain} with classical relative entropy between two probability measures.
Their densities are Gaussian quadratic forms and depend linearly on the state, so joint convexity of relative entropy gives convexity of the increase.

\begin{lemma}[Convexity of the increase in subentropy under pinching]\label{lem:convex}
The functional $C_Q$ is nonnegative and convex on density operators.
For a normalized vector $\psi$, put $b_x=\|P_x\psi\|^2$.
Then
\begin{equation}\label{eq:pinching-pure}
 C_Q(|\psi\rangle\langle\psi|)=Q(b_1,\ldots,b_n).
\end{equation}
\end{lemma}
\begin{proof}
In the Gaussian representation, the quadratic form multiplying the logarithm defines a probability density relative to Gaussian measure.
Let $\Gamma$ be standard complex Gaussian probability measure and define
\[
 a_R(g)=g^*Rg,\qquad b_R(g)=g^*\Delta(R)g,\qquad
 \dd\mu_R=a_R\dd\Gamma,\quad
 \dd\mu_{\Delta R}=b_R\dd\Gamma.
\]
Both measures have mass one because $\Tr R=\Tr\Delta(R)=1$.
To identify the cross term in relative entropy, average over the relative phases of the blocks.
Put $\omega=\exp(2\pi i/n)$ and $U_k=\sum_x\omega^{kx}P_x$, $0\le k<n$.
Then $n^{-1}\sum_kU_k^*RU_k=\Delta(R)$, and $b_R(U_kg)=b_R(g)$.
Gaussian invariance and this finite average yield
\[
 \int a_R\log b_R\,\dd\Gamma
 =\frac1n\sum_k\int a_R(U_kg)\log b_R(g)\,\dd\Gamma
 =\int b_R\log b_R\,\dd\Gamma.
\]
The identities remain valid for singular states.
To justify the integrals, Cauchy--Schwarz gives
\[
 a_R(g)=\Bigl\|\sum_x R^{1/2}P_xg\Bigr\|^2
 \le n\sum_x\|R^{1/2}P_xg\|^2=n b_R(g).
\]
Thus $b_R=0$ implies $a_R=0$, and $a_R|\log b_R|\le n b_R|\log b_R|$ off this zero set.
The right-hand side is integrable by the Gaussian moment bounds used in Lemma~\ref{lem:forms}.
The positive and negative parts of $a_R\log(a_R/b_R)$ are bounded by $a_R\log n$ and $b_R/\exp(1)$, respectively.
Integrands on the common zero set are defined to be zero.

With $D$ denoting classical relative entropy, the phase average and \eqref{eq:gaussian} give
\begin{equation}\label{eq:pinching-relative-entropy}
 D(\mu_R\Vert\mu_{\Delta R})
 =\int a_R\log(a_R/b_R)\,\dd\Gamma
 =F(R)-F(\Delta(R))=C_Q(R).
\end{equation}
Relative entropy is nonnegative and jointly convex.
Since both densities depend linearly on $R$, \eqref{eq:pinching-relative-entropy} proves nonnegativity and convexity of $C_Q$.

For a pure state, each nonzero block $P_x|\psi\rangle\langle\psi|P_x$ has the single eigenvalue $b_x$.
The blocks have orthogonal supports, so the nonzero eigenvalues of the pinched state are these block probabilities.
The original pure state has subentropy zero, and adding zero eigenvalues does not change $Q$.
This proves \eqref{eq:pinching-pure}.
\end{proof}

Gaussian density tilting is also used in the subentropy analysis of \cite[Lemma 3]{mcginley2025scrooge}.
The phase average above supplies the cross term for the specified pinching map.

Equality in the measurement comparison depends on equality in the convexity statement above.
For pure decompositions, the relative entropy identity makes this condition explicit.
\begin{lemma}[Equality for pure decompositions]\label{lem:pinching-equality}
Let $R=\sum_{\ell=1}^N w_\ell|v_\ell\rangle\langle v_\ell|$, where $w_\ell>0$, $\sum_\ell w_\ell=1$, and $\|v_\ell\|=1$.
Then
\[
 C_Q(R)=\sum_\ell w_\ell C_Q(|v_\ell\rangle\langle v_\ell|)
\]
if and only if either each $v_\ell$ lies in a single pinching block, or all $v_\ell$ are collinear.
In particular, equality with $C_Q(R)>0$ forces $R$ to have rank one.
\end{lemma}
\begin{proof}
Write $a_v(g)=|v^*g|^2$ and $b_v(g)=\sum_x|(P_xv)^*g|^2$.
For each nonzero $v$, $b_v>0$ almost everywhere under the Gaussian measure.
At such a $g$, the ratio for the mixture is a weighted average of the component ratios.
The relative entropy inequality for mixtures is therefore Jensen's inequality for the strictly convex function $u\log u$, with weights $w_\ell b_{v_\ell}/b_R$.
The Jensen gap is nonnegative, and the weights are strictly positive.
Integrating the gap against the mixture density shows that equality holds exactly when all ratios $a_{v_\ell}/b_{v_\ell}$ agree almost everywhere.
The integrability established in Lemma~\ref{lem:convex} justifies this equality argument.
Clearing denominators gives, for every pair $v,w$ in the decomposition,
\begin{equation}\label{eq:ratio-polynomial}
 a_v(g)b_w(g)=a_w(g)b_v(g).
\end{equation}
The identity initially holds almost everywhere.
Both sides are polynomials in the real and imaginary parts of $g$, so continuity and full support of the Gaussian measure make it valid everywhere.
In particular, we may evaluate it on a hyperplane, even though that hyperplane has Gaussian measure zero.

Suppose one vector $v$ has nonzero components in at least two blocks.
The complex hyperplane $H=\{g:v^*g=0\}$ has codimension one.
The zero set $\ker b_v$ has codimension at least two, since the nonzero vectors $P_xv$ are orthogonal.
Hence $b_v>0$ on a nonempty relatively open subset of $H$.
Equation~\eqref{eq:ratio-polynomial} implies $w^*g=0$ on that subset for every vector $w$ in the decomposition.
A linear form vanishing on a relatively open subset of a linear space vanishes throughout that space.
Thus $H\subseteq\ker w^*$, which makes $w$ collinear with $v$.
If no such $v$ exists, every vector in the decomposition lies in a single block.

Conversely, collinear vectors represent the same pure state, so the convexity inequality is an equality.
If each vector lies in a single block, $R=\Delta(R)$ and all subentropy increases vanish.
Thus both alternatives give equality.
Equality with a positive increase forces the collinear case, and hence a state of rank one.
\end{proof}

\subsection{Comparison with measurement posteriors}
Each outcome of a measurement on the signal determines a posterior distribution of the label.
Conditional subentropy is bounded above by the average subentropy of these posteriors, as the following proposition shows.
The measurement outcomes need not be guesses of the signal label.

For a finite POVM $\{F_y\}$ on system $B$, write
\begin{equation}\label{eq:measurement-posteriors}
 q_y=\Tr(F_y\rho_B),\qquad
 b_x^y=\frac{\Tr(F_yA_x)}{q_y}\quad(q_y>0),\qquad
 \rho_B=\sum_xA_x.
\end{equation}
Thus $b^y$ is the posterior distribution of $X$ given outcome $y$.

\begin{proposition}[Comparison with measurement posteriors]
\label{prop:posterior-comparison}
For every finite POVM,
\begin{equation}\label{eq:posterior-comparison}
 0\le Q(X|B)\le\sum_{y:q_y>0}q_yQ(b^y).
\end{equation}
If $Q(X|B)>0$, upper equality holds if and only if all signals with positive prior probability are identical and pure; in that case it holds for every finite POVM.
If $Q(X|B)=0$, upper equality holds if and only if every posterior associated with an outcome of positive probability is a point mass.
\end{proposition}
\begin{proof}
The joint state and the average state act on different spaces.
The Gram construction expresses their subentropy difference in terms of a single state with diagonal blocks $A_x$.
On $\mathcal H=\bigoplus_x\mathcal H_B$, form
\[
 M=[\sqrt{A_1}\ \cdots\ \sqrt{A_n}]:\mathcal H\to\mathcal H_B,
 \qquad G=M^*M,
\]
and use the projections onto these $n$ summands.
Then $G$ is a density operator, $\Tr G=1$, $MM^*=\rho_B$, and $\Delta(G)=\bigoplus_xA_x$.
The nonzero spectra of $M^*M$ and $MM^*$ coincide, and $Q$ is unchanged by adding zero eigenvalues.
Consequently $Q(X|B)=C_Q(G)$, which is nonnegative by Lemma~\ref{lem:convex}.

To use the formula for pure states, refine each measurement effect into rank-one terms, $F_y=\sum_k|f_{yk}\rangle\langle f_{yk}|$, using eigenvectors scaled by the square roots of the nonzero eigenvalues.
Set $v_{yk}=M^*f_{yk}$ and $w_{yk}=\|v_{yk}\|^2$.
Completeness of the POVM gives
\[
 G=\sum_{y,k}|v_{yk}\rangle\langle v_{yk}|,
 \qquad \sum_{y,k}w_{yk}=1,\qquad \sum_kw_{yk}=q_y.
\]
The weights are the probabilities of the refined outcomes.
Discard terms with $w_{yk}=0$.
The normalized vector $v_{yk}/\sqrt{w_{yk}}$ has block probabilities
\[
 b_x^{yk}=\frac{\|P_xv_{yk}\|^2}{w_{yk}}
          =\frac{\langle f_{yk},A_xf_{yk}\rangle}{w_{yk}}.
\]
Thus the block probabilities are the posterior probabilities for the refined outcome.
Convexity of the increase under pinching and \eqref{eq:pinching-pure} therefore yield
\begin{equation}\label{eq:posterior-refinement}
 Q(X|B)\le\sum_{y,k:w_{yk}>0}w_{yk}Q(b^{yk}).
\end{equation}
For $q_y>0$, the posterior for the original outcome is the mixture $b^y=\sum_{k:w_{yk}>0}(w_{yk}/q_y)b^{yk}$.
Concavity of $Q$, established in Lemma~\ref{lem:forms}, then gives
\begin{equation}\label{eq:posterior-coarsening}
 \sum_{k:w_{yk}>0}w_{yk}Q(b^{yk})\le q_yQ(b^y).
\end{equation}
If $q_y=0$, every $w_{yk}$ is zero and that outcome contributes nothing.
Combining \eqref{eq:posterior-refinement} and \eqref{eq:posterior-coarsening}, and summing over the original outcomes, proves the bound.
No assumption about ranks or optimality is needed.

For equality with positive conditional subentropy, both the refinement and the coarsening inequalities must be equalities.
Lemma~\ref{lem:pinching-equality} applied to the finite decomposition into pure states above implies that $G$ has rank one.
Its nonzero spectrum is that of $\rho_B$, so $\rho_B$ is pure.
Positivity of $A_x$ and $\sum_x A_x=\rho_B$ imply $A_x=p_x\rho_B$: all signals with positive prior probability are that same pure state.
Conversely, for identical pure signals every posterior equals $p$, and $Q(X|B)=Q(p)$, giving equality for every POVM.

When $Q(X|B)=0$, equality means that the posterior average vanishes.
Each term is nonnegative, so this happens exactly when every posterior associated with an outcome of positive probability has zero subentropy, that is, is a point mass.
\end{proof}

Zero conditional subentropy has a direct operational interpretation: it is equivalent to perfect discrimination.
\begin{corollary}[Zero conditional subentropy]\label{cor:zero-conditional}
The following are equivalent: $Q(X|B)=0$; $\pg=1$; and the supports of the signals with positive priors are pairwise orthogonal.
\end{corollary}
\begin{proof}
The Gram construction identifies $Q(X|B)$ with $D(\mu_G\Vert\mu_{\Delta G})$.
This vanishes if and only if $a_G=b_G$ almost everywhere.
The quadratic forms are continuous and Gaussian measure has full support, so they agree everywhere.
This implies equality of the operators: $G=\Delta(G)$.
Its off-diagonal blocks are $\sqrt{A_x}\sqrt{A_z}$.
These vanish exactly when the supports of the corresponding signals with positive priors are orthogonal.
In that case their support projections distinguish the signals perfectly.
Conversely, a perfectly distinguishing POVM has posteriors that are point masses; Proposition~\ref{prop:posterior-comparison} then gives $Q(X|B)=0$.
Signals with zero priors impose no constraint.
\end{proof}

\subsection{The classical extremal spectrum and the sharp bound}
At a fixed largest probability, subentropy is maximized by equalizing the remaining probabilities.
The spectrum in \eqref{eq:phi} follows by Schur concavity and was used in \cite[Proposition 4 and Equation (12)]{datta2014properties}.
The following lemma specializes Rioul's classical Fano bounds \cite[Theorems 4 and 5]{rioul2023interplay} to subentropy.
Rioul's conditional quantity is the posterior average; Proposition~\ref{prop:posterior-comparison} connects it to spectral conditional subentropy.

\begin{lemma}[Classical error curve]\label{lem:classical-fano-q}
For a probability vector $b$ with $n\ge2$ entries, set $\varepsilon(b)=1-\max_xb_x$.
Then $Q(b)\le\Phi_n(\varepsilon(b))$, and $\Phi_n$ is concave on $[0,1]$.
For posteriors $b^y$ with probabilities $q_y$, their maximum a posteriori (MAP) error $\varepsilon_{\mathrm{MAP}}=\sum_yq_y\varepsilon(b^y)$ satisfies
\begin{equation}\label{eq:posterior-fano-q}
 \sum_yq_yQ(b^y)\le\Phi_n(\varepsilon_{\mathrm{MAP}}).
\end{equation}
Zero-probability outcomes are omitted.
\end{lemma}
\begin{proof}
Choose a largest coordinate $b_{x_*}=1-\varepsilon(b)$ and average $b$ over permutations of its other $n-1$ coordinates.
The averaged vector is $(1-\varepsilon(b),\varepsilon(b)/(n-1),\ldots)$, up to permutation.
Concavity and permutation invariance of $Q$ give the first bound.
The vector in \eqref{eq:phi} depends affinely on its argument, so its subentropy is concave on $[0,1]$.
Applying Jensen's inequality to the posterior errors gives \eqref{eq:posterior-fano-q}.
\end{proof}

Concavity bounds subentropy by its tangent at a point mass \cite[Proposition 4]{datta2014properties}:
\begin{lemma}[Comparison with the tangent at a pure state]\label{lem:purebound}
For a probability vector $b$,
\begin{equation}\label{eq:purebound}
 Q(b)\le1-\max_x b_x,
\end{equation}
with strict inequality unless $b$ is a point mass.
For $n\ge2$, $\Phi_n'(0)=1$.
\end{lemma}
\begin{proof}
Relabel a largest coordinate as $b_1$ and write $h=b-(1,0,\ldots,0)$.
Let $F(p)=\E[(p\cdot Z)\log(p\cdot Z)]$, where the independent exponential random variables $Z_i$ have mean one.
To justify boundary differentiation, put $S=\sum_iZ_i$.
Along any affine segment in the simplex from the pure spectrum, $Z_1/2\le p\cdot Z\le S$ sufficiently near that endpoint.
The derivative integrand has an integrable majorant $C S(1+S+|\log Z_1|)$, so dominated convergence applies.
The derivatives at the pure spectrum are
\[
 \partial_1F=\E[Z_1(1+\log Z_1)]=2-\gamma,\qquad
 \partial_jF=1-\gamma\quad(j>1).
\]
Since $\sum_jh_j=0$ and $Q=1-\gamma-F$, the directional derivative of $Q$ is $-h_1=1-b_1$.
Concavity gives \eqref{eq:purebound}; strict concavity from Lemma~\ref{lem:forms} makes the bound strict unless $b$ is a point mass.
Along the affine curve defining $\Phi_n$, its direction has first coordinate $-1$, giving $\Phi_n'(0)=1$.
\end{proof}
The binary expansion shows that the linear bound is asymptotically attained:
\begin{equation}\label{eq:sharp}
 Q(1-\varepsilon,\varepsilon)
 =\varepsilon+\varepsilon^2(\log\varepsilon+\tfrac12)
 +O(\varepsilon^3|\log\varepsilon|).
\end{equation}

\begin{proof}[Proof of Theorem~\ref{thm:guess}]
Choose an optimal guessing POVM $\{E_x\}_{x=1}^n$; the optimum is attained because the set of such POVMs is compact in finite dimension.
For every outcome $x$ with positive probability, the reported label $x$ must maximize its posterior probability.
Otherwise some label $z$ has $\Tr(E_xA_z)>\Tr(E_xA_x)$; replacing $E_z$ by $E_z+E_x$ and $E_x$ by zero gives a valid POVM with a larger success probability.
Ties may be resolved arbitrarily.
Thus the classical MAP rule has exactly the optimal quantum error:
\[
 \varepsilon_{\mathrm{MAP}}
 =1-\sum_x\Tr(E_xA_x)=1-\pg=e.
\]
Applying the measurement comparison in Proposition~\ref{prop:posterior-comparison} and the classical bound in Lemma~\ref{lem:classical-fano-q} gives
\[
 0\le Q(X|B)\le\sum_{x:q_x>0}q_xQ(b^x)\le\Phi_n(e).
\]
Always guessing a label with largest prior probability shows $\pg\ge1/n$, hence $e\le1-1/n$.
On this interval the largest entry of the spectrum in \eqref{eq:phi} is $1-e$.
Equation~\eqref{eq:purebound} gives $\Phi_n(e)\le e$, strictly for $e>0$.
The elementary inequality $1-u\le-\log u$ gives the final comparison in \eqref{eq:guess}.

For $n=1$, the joint and average states have the same nonzero spectrum and the label is always guessed correctly.

For exact attainment, take $\rho_x=|\varphi\rangle\langle\varphi|$ for every label, with priors
\[
 p_1=1-e,\qquad p_x=\frac{e}{n-1}\quad(2\le x\le n).
\]
The average state is pure and the joint spectrum is the prior vector padded with zeros.
Measurement outcomes are independent of the label, so the optimal success probability is $\max_xp_x=1-e$.
Therefore $Q(X|B)=\Phi_n(e)$ for every admissible $e$, including both endpoints.

For necessity at $e>0$, $\Phi_n(e)>0$.
Equality in the guessing bound forces equality in Proposition~\ref{prop:posterior-comparison} for an optimal guessing POVM.
Thus all signals with positive prior probability are identical and pure, so $e=1-\max_xp_x$ and $Q(X|B)=Q(p)$.
Strict concavity in the classical permutation average forces every coordinate other than a selected largest one to equal $e/(n-1)$.
Thus all $n$ priors are positive and have the stated form.
At $e=0$, Corollary~\ref{cor:zero-conditional} gives the complete equality condition, including orthogonal mixed signals.

Finally, two identical pure signals with priors $(1-\varepsilon,\varepsilon)$ have error $\varepsilon$ and conditional subentropy $Q(1-\varepsilon,\varepsilon)$.
Equation~\eqref{eq:sharp} proves that the coefficient $1$ in the linear relaxation cannot be reduced uniformly.
\end{proof}

Strict concavity of $Q$ from Lemma~\ref{lem:forms} makes $\Phi_n$ strictly concave.
Its unique maximum occurs at $u=1-1/n$, where the spectrum is uniform; permutation averaging proves this maximality.
Strict concavity therefore makes $\Phi_n$ strictly increasing on $[0,1-1/n]$, which justifies the inverse in \eqref{eq:inverse}.
Removing labels with zero prior probability allows the theorem to use the number of positive priors.

\subsection{Subadditivity as a consequence of the comparison}\label{sec:subadditivity-proof}
The main proof of Theorem~\ref{thm:sa} uses Proposition~\ref{prop:posterior-comparison}.
Pinching in a marginal eigenbasis preserves both marginals and permits the posterior comparison.
\begin{proof}[Proof of Theorem~\ref{thm:sa}]
Pinch $A$ in any eigenbasis of $\rho_A$, and set $\sigma_{AB}=\Delta_A(\rho_{AB})$.
As pinching averages unitary conjugations, concavity and unitary invariance of $Q$ give $Q(\rho_{AB})\le Q(\sigma_{AB})$.
The pinched state is CQ, with $B$ marginal $\rho_B$.
For the POVM with one outcome $\{I_B\}$, the posterior is the spectrum of $\rho_A$.
Proposition~\ref{prop:posterior-comparison} therefore gives
\[
 Q(\rho_{AB})\le Q(\sigma_{AB})\le Q(\rho_B)+Q(\rho_A).
\]
If $\rho_A$ is mixed, $Q(\rho_A)>0$.
Equality in the second inequality would make the pinched state's conditional subentropy positive.
The proposition then forces the signals on $B$ with positive priors to be identical and pure, so $\rho_B$ is pure.
Subadditivity is therefore strict when both marginals are mixed.
Conversely, a pure marginal forces a product state with the other marginal's nonzero spectrum, giving equality.
Zero eigenvalues and degeneracy of the marginal eigenbasis do not affect the argument.
\end{proof}

\begin{corollary}[Triangle inequality]\label{cor:triangle}
Every bipartite state satisfies $|Q(\rho_A)-Q(\rho_B)|\le Q(\rho_{AB})$.
Equality holds exactly when at least one of the states $\rho_A$, $\rho_B$, and $\rho_{AB}$ is pure.
\end{corollary}
\begin{proof}
Purify $\rho_{AB}$ on $ABC$.
Complementary reduced states have identical nonzero spectra, so
\[
 Q(BC)=Q(A),\qquad Q(C)=Q(AB).
\]
Apply Theorem~\ref{thm:sa} to $BC$ and use these identities; interchange $A,B$ for the other direction.

Equality in $Q(A)-Q(B)\le Q(AB)$ holds, by Theorem~\ref{thm:sa}, exactly when $\rho_B$ or $\rho_C$ is pure.
Interchanging $A,B$ gives the condition that $\rho_A$ or $\rho_C$ is pure for the other bound.
Equality in the triangle inequality holds if and only if one of these two bounds is an equality.
Since $\rho_C$ and $\rho_{AB}$ have the same nonzero spectrum, this proves the stated condition.
\end{proof}

\subsection{Quantitative posterior stability}\label{sec:ensemble-stability}
The sharp guessing curve is attained by a specific prior and identical pure signals.
Near equality, quantitative concavity controls the average squared deviation of measurement posteriors from the maximizing distribution.
The relevant Jensen gap is the subentropy of the average minus the average subentropy.
The Gaussian Hessian bounds this gap below by a multiple of the variance, with a coefficient bounded away from zero in every dimension.
\begin{lemma}[An estimate from strong concavity]\label{lem:strong-concavity}
For density operators $\rho_j$ on a common $d$-dimensional space, $d\ge2$, weights $w_j\ge0$ summing to one, and $\bar\rho=\sum_jw_j\rho_j$,
\begin{equation}\label{eq:strong-concavity}
 Q(\bar\rho)-\sum_jw_jQ(\rho_j)
 \ge\frac{\mu_d}{2}\sum_jw_j\|\rho_j-\bar\rho\|_2^2,
 \qquad \mu_d=\frac{d}{3d-2}.
\end{equation}
Here $\|H\|_2^2=\Tr(H^*H)$ is the squared Hilbert--Schmidt norm.
The strong concavity parameter satisfies $\mu_d\ge1/3$.
The coefficient of the variance in \eqref{eq:strong-concavity} is $\mu_d/2$, which is at least $1/6$; its optimality is not asserted.
In dimension one the variance vanishes.
\end{lemma}
\begin{proof}
At a positive definite state $\rho$, take a nonzero traceless Hermitian direction $H$ and put $X=g^*Hg$, $T=g^*\rho g$, and $a=\Tr H^2$.
To justify differentiation under the expectation, write $\lambda=\lambda_{\min}(\rho)>0$ and $R=\|g\|^2$.
For $|s|\le\lambda/(2\|H\|_\infty)$, the state $\rho+sH$ has trace one and is bounded below by $(\lambda/2)I$.
Thus $T_s=g^*(\rho+sH)g$ satisfies $(\lambda/2)R\le T_s\le R$.
The first two derivatives of $T_s\log T_s$ obey, almost surely,
\[
 \begin{aligned}
 |X(1+\log T_s)|&\le\|H\|_\infty R\bigl(1+|\log R|+\log(2/\lambda)\bigr),\\
 \frac{X^2}{T_s}&\le\frac{2\|H\|_\infty^2}{\lambda}R.
 \end{aligned}
\]
Both bounds are integrable under the Gaussian law, since $R$ and $R|\log R|$ have finite expectations.
Dominated convergence therefore permits both differentiations.
Differentiation of \eqref{eq:gaussian} and weighted Cauchy--Schwarz give
\[
 D^2F_\rho[H,H]=\E\frac{X^2}{T}
 \ge\frac{(\E X^2)^2}{\E[X^2T]}
 =\frac{a^2}{a+2\Tr(\rho H^2)}.
\]
The last identity uses the complex Gaussian second and third moments.
In the third moment, the six pairings in Wick's formula leave $\Tr H^2\Tr\rho+2\Tr(\rho H^2)$ because $\Tr H=0$.
Replacing $\rho$ by $I$ would remove the state dependence from the denominator.
If $h$ is an eigenvalue of $H$ of largest absolute value, the remaining eigenvalues sum to $-h$.
Their squared sum is at least $h^2/(d-1)$, so $\|H\|_\infty^2\le(d-1)a/d$.
Consequently $\Tr(\rho H^2)\le(d-1)a/d$ and the Hessian is at least $\mu_d a$.
Thus $F(\rho)-\mu_d\|\rho\|_2^2/2$ is convex on positive definite states.
Its Jensen inequality gives \eqref{eq:strong-concavity}.
Mixing every state with the same $I/d$ and letting the mixing weight tend to zero extends the bound to boundary states by continuity.
\end{proof}
An earlier dimension-independent estimate follows from the information bound for uniform measurements of Matthews, Wehner, and Winter \cite[Section 5]{matthews2009distinguishability}.
Combining that bound with the identity for $Q$ using random measurements gives a Jensen gap of at least $1/18$ times the Hilbert--Schmidt variance, in nats.
Equation~\eqref{eq:strong-concavity} gives a larger explicit coefficient by the short argument using third moments above.

The permutation average relates a small gap in the guessing bound to posterior deviations.
The variance term controls both the spread of posterior errors and the imbalance among labels other than the MAP guess.
\begin{proposition}[Quantitative posterior stability]\label{prop:posterior-stability}
Let a finite POVM on a CQ ensemble with $n\ge2$ labels have outcome probabilities $q_y$ and posteriors $b^y$.
Omit outcomes with $q_y=0$, choose any MAP label $j(y)$, and put $e_M=\sum_yq_y(1-b^y_{j(y)})$.
Let $v_j(e_M)$ have entry $1-e_M$ at $j$ and $e_M/(n-1)$ elsewhere.
Then, with $\Delta_M=\Phi_n(e_M)-Q(X|B)$,
\begin{equation}\label{eq:posterior-stability}
 \sum_yq_y\|b^y-v_{j(y)}(e_M)\|_2^2
 \le\Bigl(6-\frac4n\Bigr)\Delta_M\le6\Delta_M.
\end{equation}
In particular, for every optimal guessing POVM this holds with $e_M=e$ and $\Delta_M=\Phi_n(e)-Q(X|B)$.
\end{proposition}
\begin{proof}
Move a chosen MAP coordinate of each $b^y$ to the first position and average uniformly over permutations of the other coordinates.
The mean of the resulting random vector is $v_1(e_M)$.
Subentropy is invariant under these permutations, and their squared distances to $v_1(e_M)$ are $\|b^y-v_{j(y)}(e_M)\|_2^2$.
Apply Lemma~\ref{lem:strong-concavity} to these diagonal states in dimension $n$ to obtain
\[
 \Phi_n(e_M)-\sum_yq_yQ(b^y)
 \ge\frac{n}{2(3n-2)}
       \sum_yq_y\|b^y-v_{j(y)}(e_M)\|_2^2.
\]
Proposition~\ref{prop:posterior-comparison} bounds the left side above by $\Delta_M$.
The proof of Theorem~\ref{thm:guess} above shows that reporting each outcome of an optimal guessing POVM is a MAP decision, so its posterior error is $e$.
\end{proof}
For $\varepsilon_y=1-b^y_{j(y)}$, the quantity controlled here is
\begin{equation}\label{eq:posterior-variance}
 \sum_yq_y\|b^y-v_{j(y)}(e_M)\|_2^2
 =\sum_yq_y\sum_{x\ne j(y)}
       \Bigl(b^y_x-\frac{\varepsilon_y}{n-1}\Bigr)^2
   +\frac{n}{n-1}\operatorname{Var}_q(\varepsilon_y).
\end{equation}
The deviations among labels other than the guess sum to zero.
They are therefore orthogonal to the change caused by varying the total error, which proves the identity.
Thus a small deficit bounds both the variation in posterior errors and the departure from uniformity among labels other than the guess.
The estimate applies to a specified measurement without computing the optimal discrimination error.

The stability bound concerns measurement posteriors.
Near zero error, a small absolute deficit need not make the CQ state close to an ensemble attaining equality at positive error.
To see this, take identical qubit signals $I_2/2$ with priors $(1-e,e)$, $0<e<1/2$.
Their error is $e$ and $0<\Phi_2(e)-Q(X|B)\le e\to0$.
Every ensemble attaining the guessing bound with positive error has a pure marginal state $P$ on $B$.
Contraction under partial trace therefore puts the constructed ensemble at distance at least $D(I_2/2,P)=1/2$ from any such equality case.
Identical pure signals with the same priors attain this distance.
For fixed dimensions and errors in a closed interval away from zero, compactness and the equality classification imply closeness to an equality case as the deficit tends to zero.
This argument gives neither an explicit rate nor a bound uniform in the dimension.

\section{Continuity of subentropy and conditional subentropy}\label{sec:continuity}
Continuity estimates control perturbations of the joint and average subentropies in the discrimination bound.
For subentropy, coupling the spectra connects continuity to the guessing bound.
The shared CQ block structure gives the optimal Lipschitz bound for conditional subentropy.

\subsection{Continuity of subentropy}
Maximal coupling represents two spectra as the marginals of a joint classical distribution whose mismatch probability is their total variation distance.
The guessing theorem then bounds the subentropy difference, which explains the shared maximizing spectrum.
For $d\ge2$, define
\[
 \omega_d(\delta)=\Phi_d(\min\{\delta,1-1/d\}),\qquad 0\le\delta\le1;
 \qquad \omega_1(\delta)=0.
\]
The following theorem gives the optimal modulus of continuity: the largest subentropy difference under a bound on trace distance.
The modulus and its spectral equality cases are consequences of Hanson and Datta's concave-type entropy theorem \cite[arXiv version 3, Corollary 6.2, Equations (22)--(23)]{hanson2022entropies}.
Earlier subentropy estimates were of Fannes and Audenaert type \cite[Equations (8)--(9)]{datta2014properties}.
We give a new proof using maximal coupling and Theorem~\ref{thm:guess}.
The equality argument also determines the alignment of the eigenvectors.

\begin{theorem}[Optimal modulus of continuity; after Hanson and Datta~\cite{hanson2022entropies}]\label{thm:modulus}
For states on a common $d$-dimensional space,
\begin{equation}\label{eq:modulus}
 \sup_{D(\rho,\sigma)\le\delta}|Q(\rho)-Q(\sigma)|=\omega_d(\delta).
\end{equation}
For $\delta\le1-1/d$ in dimension at least two, a pure state and a commuting state with spectrum $(1-\delta,\delta/(d-1),\ldots)$ attain the supremum at distance $\delta$ when the first eigenvector represents the pure state.
For larger allowed distances, the pair consisting of a pure state and the maximally mixed state attains the plateau $\log d-H_d+1$ at distance $1-1/d$.

\label{cor:modulus-equality}
For $d\ge2$ and $0<\delta<1-1/d$, the pairs attaining \eqref{eq:modulus} are, up to exchange,
\begin{equation}\label{eq:modulus-attainers}
 \rho=|v\rangle\langle v|,\qquad
 \sigma=(1-\delta)|v\rangle\langle v|
       +\frac{\delta}{d-1}(I-|v\rangle\langle v|),\quad\|v\|=1.
\end{equation}
For $\delta\ge1-1/d$, they are exactly pairs consisting of a pure state and the maximally mixed state.
For $\delta=0$, they are identical pairs; dimension one is trivial.

\label{prop:lip}
On every common finite-dimensional space,
\begin{equation}\label{eq:lip}
 |Q(\rho)-Q(\sigma)|\le D(\rho,\sigma).
\end{equation}
The inequality is strict for distinct states, and its dimension-independent coefficient $1$ is optimal.

\end{theorem}
\begin{proof}[New proof by maximal coupling]
The case $d=1$ is immediate.
For $d\ge2$, let $p,q$ be the ordered eigenvalue vectors of $\rho,\sigma$.
The eigenvalue perturbation inequality \cite{bhatia1997matrix} gives $\tau=\|p-q\|_1/2\le D(\rho,\sigma)$.
We adapt the proof of Shannon continuity using maximal coupling \cite[Theorem 3]{sason2013entropy}.
If $\tau>0$, put $s_i=\min(p_i,q_i)$, $p_i'=p_i-s_i$, $q_i'=q_i-s_i$ and define
\[
 P_{ij}=s_i\mathbf1_{i=j}+p_i'q_j'/\tau.
\]
This has marginals $p,q$.
The residual supports are disjoint and each has mass $\tau$, so $\Pr[X\ne Y]=\tau$.
For $\tau=0$ use the diagonal coupling.
Assume $Q(p)\ge Q(q)$.
Nonnegativity of conditional subentropy for this classical coupling gives $Q(P)-Q(p)\ge0$, hence
\[
 Q(p)-Q(q)=[Q(P)-Q(q)]-[Q(P)-Q(p)]\le Q(X|Y).
\]
Apply Theorem~\ref{thm:guess} to this classical coupling with $d$ labels for $X$.
Guessing $X=Y$ has error $\tau$, so the optimal error $e$ satisfies $e\le\min\{\tau,1-1/d\}$.
Monotonicity of $\Phi_d$ gives
\[
 |Q(\rho)-Q(\sigma)|\le\Phi_d(e)
 \le\omega_d(\tau)\le\omega_d(\delta).
\]
Choose the eigenvector corresponding to the first entry in \eqref{eq:phi} to represent the pure state.
The resulting pair has trace distance $\delta$ and subentropy difference $\Phi_d(\delta)$ up to the uniform endpoint.
Beyond that endpoint, a pure state and the maximally mixed state have distance at most $\delta$ and difference $\Phi_d(1-1/d)$.
Both pairs attain the bound.

Appendix~\ref{app:ordinary-continuity-equality} proves the equality cases, including the alignment of the eigenvectors and the plateau.
The strict linear bound follows from $\Phi_d(u)<u$ for $0<u\le1-1/d$; on the plateau the modulus stays below the uniform endpoint distance.
The binary perturbation in \eqref{eq:sharp} proves optimality of the coefficient.
\end{proof}

The determinant representation makes the connection to Hanson and Datta's theorem explicit.
For $t>0$ the kernel in \eqref{eq:integral} has the form
\[
 q_t(\rho)=h_t(\Tr\varphi_t(\rho)),\qquad
 \varphi_t(x)=\log(1+tx),\qquad h_t(y)=\frac1{1+t}-e^{-y}.
\]
The function $\varphi_t$ is strictly concave and vanishes at zero; $h_t$ is increasing and concave, with $h_t(\varphi_t(1))=0$.
Their theorem bounds the change in each kernel by its value at the spectrum in \eqref{eq:phi}, with the same cutoff at the uniform spectrum.
The maximizing spectra are the same for every $t>0$.
Integration with the positive weight $\dd t/t^2$ gives the modulus above and its spectral equality cases.

\subsection{Optimal Lipschitz constant for conditional subentropy}
Separate continuity bounds for the joint and marginal subentropies give a Lipschitz constant of two.
The shared CQ block structure reduces it to one: the common scalar in the block gradients cancels, and the remaining terms share an upper bound in the positive semidefinite order.
The matrix giving this bound is the population complex spatial sign covariance matrix \cite[Theorem 1]{raninen2022bias}.
For a nonzero positive semidefinite matrix $M$, define
\begin{equation}\label{eq:K-matrix}
 K(M)=\int_0^\infty
       \frac{M(I+tM)^{-1}}{\det(I+tM)}\dd t
 =\E\frac{M^{1/2}gg^*M^{1/2}}{g^*Mg},
 \qquad \kappa(M)=\|K(M)\|_\infty.
\end{equation}
The expectation uses a standard complex Gaussian vector of the same dimension as $M$.
The normalized outer product is a rank-one state, so $\Tr K(M)=1$ and $0\le K(M)\le I$.
Its expectation has the same support as $M$ and is positive definite there.
Both $K$ and $\kappa$ are continuous on the cone of nonzero positive semidefinite matrices; see Appendix~\ref{app:conditional-gradient}.

\begin{theorem}[Optimal Lipschitz bound for conditional subentropy on CQ states]\label{thm:cq-lipschitz}
Let $\rho$ and $\sigma$ be CQ states on the same label alphabet and quantum system $B$, and set $\rho_s=(1-s)\rho+s\sigma$.
Then
\begin{equation}\label{eq:cq-path-bound}
 |Q(X|B)_\rho-Q(X|B)_\sigma|
 \le D(\rho,\sigma)\int_0^1\kappa((\rho_s)_B)\dd s
 \le D(\rho,\sigma).
\end{equation}
The bound with coefficient one is strict for distinct states.
For every $n\ge2$ and every dimension $m\ge1$ of system $B$, its coefficient one is optimal.
For $n=1$, the functional is identically zero.
\end{theorem}
\begin{proof}
For positive definite CQ blocks $A_x(s)$, write their sum as $A(s)$ and their direct sum as $R(s)$.
The Gaussian representation gives $Q(X|B)_{\rho_s}=F(A(s))-F(R(s))$.
Appendix~\ref{app:conditional-gradient} proves that the derivative with respect to block $A_x$ is $cI+B_x$, where $c$ is the same for every block and
\begin{equation}\label{eq:block-gradient-order}
 0\le B_x=K(A)-K_x(R)\le K(A)\le\kappa(A)I,
 \quad
 K_x(R)=\int_0^\infty
   \frac{A_x(I+tA_x)^{-1}}{\det(I+tR)}\dd t.
\end{equation}
For $H=\sigma-\rho$, the common scalar cancels because $\Tr H=0$.
The directional derivative is therefore $\Tr[(\bigoplus_xB_x)H]$.
The factor one half follows from the trace constraint: if $0\le B\le\kappa I$ and $H$ is trace zero, its positive and negative parts each have trace $\|H\|_1/2$, giving $|\Tr BH|\le\kappa\|H\|_1/2$.
Integrating this estimate along the segment proves \eqref{eq:cq-path-bound}.
For arbitrary CQ states, mix both endpoints with the same $I_{nm}/(nm)$.
Their distance scales by one minus the mixing weight.
Continuity of $Q$ and $K$, with $\kappa\le1$, allows this weight to tend to zero.
Dominated convergence justifies the limit under the path integral.

Unless both endpoint marginals are the same pure state, every interior marginal on $B$ is mixed.
The corresponding $K$ has trace one and at least two positive eigenvalues, so $\kappa<1$ and the path integral is strictly below one.
Otherwise all blocks are scalar multiples of that pure state, so conditional subentropy is $Q$ of the priors.
Theorem~\ref{prop:lip} gives strictness.
For sharpness, compare ensembles of identical pure signals with priors $(1,0)$ and $(1-\varepsilon,\varepsilon)$, adding zero labels if needed.
Their distance is $\varepsilon$ and their difference in conditional subentropy is $\Phi_2(\varepsilon)$, whose ratio to $\varepsilon$ tends to one by \eqref{eq:sharp}.
For a single label, the joint and marginal states have the same nonzero spectrum, so conditional subentropy vanishes.
\end{proof}

\begin{corollary}[Optimal coefficient with a fixed marginal state]\label{cor:fixed-marginal}
If $\rho_B=\sigma_B=\tau$, then
\begin{equation}\label{eq:fixed-marginal}
 |Q(X|B)_\rho-Q(X|B)_\sigma|
 \le\kappa(\tau)D(\rho,\sigma).
\end{equation}
For $n\ge2$, $\kappa(\tau)$ is the optimal coefficient among CQ states with marginal $\tau$.
In particular, it is $1/m$ for $\tau=I_m/m$, one for pure $\tau$, and strictly smaller than one for mixed $\tau$.
\end{corollary}
\begin{proof}
The marginal is constant along the segment, giving the bound.
For sharpness, restrict to the support of $\tau$ and choose a rank-one projection $P$ onto a common eigenvector of $\tau$ and $K(\tau)$ for the largest eigenvalue of $K(\tau)$.
Starting from blocks $(\tau,0,\ldots)$, replace the first two by $(\tau-\varepsilon P,\varepsilon P)$.
For sufficiently small $\varepsilon>0$ they are positive and their trace distance is $\varepsilon$.
The boundary derivative of their conditional subentropy at zero is $\kappa(\tau)$, as computed in Appendix~\ref{app:conditional-gradient}.
This proves optimality.
Unitary covariance and trace one give $K(I_m/m)=I_m/m$.
The pure and mixed cases follow from support and positivity.
\end{proof}

\begin{corollary}[Continuity with an independent maximally mixed ancilla]\label{cor:ancilla}
Let $u_k=I_k/k$ be the maximally mixed state of an independent ancillary register $C$.
For any two CQ states on the original common spaces,
\begin{equation}\label{eq:ancilla-continuity}
 |Q(X|BC)_{\rho\otimes u_k}-Q(X|BC)_{\sigma\otimes u_k}|
 \le\frac{D(\rho,\sigma)}{k}.
\end{equation}
\end{corollary}
\begin{proof}
By unitary covariance, $K(\tau\otimes u_k)$ is invariant under every unitary $I\otimes U$.
It has the form $T\otimes I_k$ with $T\ge0$ and $k\Tr T=1$, hence $\kappa(\tau\otimes u_k)\le1/k$.
Tensoring with $u_k$ preserves trace distance, so \eqref{eq:cq-path-bound} on the augmented segment gives the bound.
\end{proof}
The sharp guessing curve for a maximally mixed ancilla has slope $1/k$ at zero error (Section~\ref{sec:ancilla-calibration}).
Equation~\eqref{eq:calibrated-robust-certificate} combines this curve with Corollary~\ref{cor:ancilla} to bound the discrimination error under perturbations.

The restriction to CQ states in Theorem~\ref{thm:cq-lipschitz} is essential.
For two qubits, let $|\psi\rangle=(2|00\rangle+|11\rangle)/\sqrt5$, $\rho_0=|\psi\rangle\langle\psi|$, and $\rho_t=(1-t)\rho_0+t|10\rangle\langle10|$.
The pure components are orthogonal, so $D(\rho_0,\rho_t)=t$.
The marginal spectra give
\begin{equation}\label{eq:quantum-conditional-continuity}
 Q(A|B)_{\rho_t}-Q(A|B)_{\rho_0}
 =\Phi_2(t)+\Phi_2(1/5)-\Phi_2((1-t)/5).
\end{equation}
Dividing by $t$ and letting $t\downarrow0$ gives $1+\Phi_2'(1/5)/5>1$.
At $t=1/100$, rational bounds on the logarithms place the change between $0.01045371079$ and $0.01045371080$, strictly above the distance $0.01$.
Both state matrices have rational entries at this parameter value.
The example does not determine the optimal coefficient for general quantum states.

\section{Determinant inequalities and strong subadditivity}\label{sec:composition}
The reciprocal determinant representation admits a pointwise form of subadditivity at each value of the integration parameter.
For classical tables, merging rows or columns cannot increase the deficit, which vanishes for a single row or column.
The same comparison applies to the unnormalized slices needed for classical strong subadditivity.
Hadamard's inequality transfers the bipartite comparison to quantum states in marginal eigenbases.

\begin{proposition}[Pointwise determinant comparison]\label{prop:pointwise-determinant}
For every bipartite state and every $t\ge0$,
\begin{equation}\label{eq:pointwise}
 \frac1{\det(I+t\rho_{AB})}+\frac1{1+t}
 \ge \frac1{\det(I+t\rho_A)}+\frac1{\det(I+t\rho_B)}.
\end{equation}
For each $t>0$, equality holds exactly when at least one marginal is pure.
In particular, both mixed marginals give strict inequality at every positive parameter.
\end{proposition}

\subsection{Coarse graining of classical distributions}
For a nonnegative vector $v$, which need not sum to one, write $L_t(v)=\prod_i(1+tv_i)^{-1}$.

\begin{lemma}[Four cells]\label{lem:four}
For $a,b,c,d\ge0$ with total mass $M$ and every $t\ge0$,
\[
 L_t(a,b,c,d)+(1+tM)^{-1}
 \ge L_t(a+b,c+d)+L_t(a+c,b+d).
\]
For $t>0$ and $M>0$, the inequality is strict exactly when both binary marginal sum vectors have two positive entries.
\end{lemma}
\begin{proof}
Zero mass or $t=0$ gives equality.
Otherwise divide the entries by $M$ and replace $t$ by $tM$ to obtain a table of total mass one.
Clearing the positive denominators of the four terms leaves a polynomial with nonnegative coefficients; the full certificate is given in Appendix~\ref{app:four_cell_certificate}.
Its coefficient at degree two is $ad+bc$.
This quantity is positive exactly when both marginal sum vectors have two positive entries.
If it vanishes, the table is supported in one row or one column and the two sides agree by cancellation.
\end{proof}

For $X,Y\ge0$, set
\begin{equation}\label{eq:merge-gain}
 G(X,Y)=\frac1{1+X+Y}-\frac1{(1+X)(1+Y)}\ge0.
\end{equation}
This difference is the increase in the reciprocal product when two masses are merged.
The next lemma bounds the corresponding increase when these masses are distributed among several rows.

\begin{lemma}[Binary splitting]\label{lem:split}
For nonnegative $x_i,y_i$, $X=\sum_i x_i$, $Y=\sum_i y_i$,
\begin{equation}\label{eq:split}
 \prod_i\frac1{1+x_i+y_i}-\prod_i\frac1{(1+x_i)(1+y_i)}
 \le\frac1{1+X+Y}-\frac1{(1+X)(1+Y)}.
\end{equation}
\end{lemma}
\begin{proof}
With one index, the two sides are equal.
For the induction step, let $A,B$ be the first and second products over the preceding indices and $X,Y$ their partial sums.
For the last index put $a=(1+x+y)^{-1}$, $b=[(1+x)(1+y)]^{-1}$.
Using the induction hypothesis and the bounds $a-b\ge0$ and $B\le D=[(1+X)(1+Y)]^{-1}$ gives
\[
 aA-bB=a(A-B)+(a-b)B\le aG(X,Y)+(a-b)D
 =\frac a{1+X+Y}-\frac b{(1+X)(1+Y)}.
\]
Apply Lemma~\ref{lem:four} to the unnormalized table $\bigl(\begin{smallmatrix}X&Y\\x&y\end{smallmatrix}\bigr)$ at $t=1$.
The final expression is the reciprocal product of the row sums minus that of the cells.
Rearranging this inequality gives the bound $G(X+x,Y+y)$.
\end{proof}

\begin{proposition}[Classical merging and subadditivity]
\label{prop:classical-merging}
Let $p=(p_{ij})$ be a finite nonnegative table of total mass $M$, with row sums $r=(r_i)$ and column sums $c=(c_j)$.
For $t\ge0$, define
\[
 D_t(p)=(1+tM)^{-1}+L_t(p)-L_t(r)-L_t(c).
\]
If $p'$ is obtained by merging rows or columns, then $D_t(p)\ge D_t(p')\ge0$.
In particular,
\begin{equation}\label{eq:classical-pointwise}
 L_t(p)+(1+tM)^{-1}\ge L_t(r)+L_t(c).
\end{equation}
For $t>0$, the deficit is strictly positive exactly when both marginal sum vectors have at least two positive entries.
\end{proposition}
\begin{proof}
Merge two columns $x_i,y_i$ to obtain $p'$, with column sum vector $c'$.
The total mass and row sums are unchanged, so compare the reciprocal products over cells and column sums.
Write $\mathcal J$ for the indices of the other columns and put
\[
 U=\prod_{j\in\mathcal J}\prod_i(1+tp_{ij})^{-1},\qquad
 V=\prod_{j\in\mathcal J}(1+tc_j)^{-1},
\]
\[
 H=\prod_i(1+t(x_i+y_i))^{-1},\qquad
 K=\prod_i[(1+tx_i)(1+ty_i)]^{-1}.
\]
Empty products equal one.
For each remaining column, $\prod_i(1+tp_{ij})\ge1+tc_j$, so $U\le V$.
Set $X=\sum_i x_i$ and $Y=\sum_i y_i$.
Lemma~\ref{lem:split}, applied to $tx_i,ty_i$, gives
\[
 L_t(p')-L_t(p)=U(H-K)
 \le U G(tX,tY)\le V G(tX,tY)
 =L_t(c')-L_t(c).
\]
Thus $D_t(p)\ge D_t(p')$.
Repeated merging leaves one column, where $D_t=0$.
Hence $D_t\ge0$ for every table.
Transposition gives the same monotonicity for row merging.
If both marginal sum vectors have at least two positive entries, partition the rows into two groups of positive total mass and do the same for the columns.
The resulting table has two positive row sums and two positive column sums.
Lemma~\ref{lem:four} makes its deficit strictly positive for $t>0$; merging monotonicity gives the same conclusion for $p$.
If a marginal sum vector has at most one positive entry, the table is supported in a single row or column, or has zero mass.
Direct cancellation in $D_t$ then gives equality.
\end{proof}

The comparison \eqref{eq:classical-pointwise} keeps the same parameter $t$ throughout the merging argument.
Related determinant inequalities for partial traces use positive powers and dimension factors \cite{li2022extensions}; applied directly to $I+t\rho_{AB}$, they introduce dimension multiples of the identity in the marginal terms.
The classical comparison retains the identity matrices in \eqref{eq:pointwise}.

\begin{proof}[Proof of Proposition~\ref{prop:pointwise-determinant}]
Choose eigenbases of $\rho_A$ and $\rho_B$.
Set $p_{ij}=\langle i,j|\rho_{AB}|i,j\rangle$ in their product basis.
Its row and column sums are the marginal eigenvalues.
Apply Proposition~\ref{prop:classical-merging}, then use Hadamard's determinant inequality:
\[
 \det(I+t\rho_{AB})\le\prod_{i,j}(1+tp_{ij}),
 \qquad\det(I+t\rho_{AB})^{-1}\ge L_t(p).
\]
The quantum deficit in \eqref{eq:pointwise} is at least $D_t(p)$.
If both marginals are mixed, this lower bound is positive for every $t>0$ by Proposition~\ref{prop:classical-merging}.
A pure marginal forces $\rho_{AB}$ to be its tensor product with the other marginal; the nonzero joint spectrum then equals the other marginal spectrum, giving pointwise cancellation.
\end{proof}
\phantomsection\label{par:determinant-proof} Integrating \eqref{eq:pointwise} and using \eqref{eq:integral} gives a second proof of Theorem~\ref{thm:sa}, including its equality condition.
This determinant argument is independent of Proposition~\ref{prop:posterior-comparison}.
It uses the integral representation, the classical merging comparison in Proposition~\ref{prop:classical-merging}, and Hadamard's inequality.

For two qubits, the polynomial estimate also gives an explicit lower bound in terms of the marginal determinants.
\begin{corollary}[Subadditivity deficit for two qubits]\label{cor:qubit-deficit}
Every state $\rho_{AB}$ on $\mathbb C^2\otimes\mathbb C^2$ satisfies
\begin{equation}\label{eq:qubit-deficit}
 Q(\rho_A)+Q(\rho_B)-Q(\rho_{AB})
 \ge\frac13\det(\rho_A)\det(\rho_B).
\end{equation}
The right side is positive when both marginals are mixed.
If either marginal is pure, both sides vanish.
The coefficient $1/3$ is not claimed to be optimal.
\end{corollary}
\begin{proof}
Choose marginal eigenbases and let $a,b,c,d$ be the four diagonal entries of $\rho_{AB}$ in their product basis, arranged as a table $p=\bigl(\begin{smallmatrix}a&b\\c&d\end{smallmatrix}\bigr)$.
Then $a+b+c+d=1$ and
\[
 u=(a+b)(c+d)=\det(\rho_A),\qquad
 v=(a+c)(b+d)=\det(\rho_B).
\]
Let $R(t)$ be the left side minus the right side of \eqref{eq:pointwise}.
Hadamard's inequality gives $R(t)\ge D_t(p)$.
For this normalized table, let $P_A,P_B,P$ and $N(t)$ be as in Appendix~\ref{app:four_cell_certificate}.
Writing $E=ad+bc$ and $x=abc+abd+acd+bcd=ad(b+c)+bc(a+d)$ gives $x\le E$.
The coefficients of $N$ are nonnegative, and its terms of degrees two and three therefore give
\[
 N(t)\ge Et^2+(2E-x)t^3\ge Et^2(1+t).
\]
Consequently
\[
 D_t(p)=\frac{N(t)}{(1+t)P_AP_BP}
 \ge\frac{Et^2}{P_AP_BP}\ge Et^2e^{-3t}.
\]
For the last inequality, each of $P_A,P_B,P$ is a product of factors $1+t\lambda_j$ with $\lambda_j\ge0$ and $\sum_j\lambda_j=1$.
The bound $1+y\le e^y$ for $y\ge0$ thus bounds each product by $e^t$.
Also, using $a+b+c+d=1$,
\[
 E-uv=ad(a+d-ad)+bc(b+c-bc)+2abcd\ge0,
\]
since all four entries lie in $[0,1]$.
Integrating with \eqref{eq:integral} now yields
\[
 Q(\rho_A)+Q(\rho_B)-Q(\rho_{AB})
 =\int_0^\infty\frac{R(t)}{t^2}\,\dd t
 \ge E\int_0^\infty e^{-3t}\,\dd t
 =\frac E3\ge\frac{uv}{3}.\qedhere
\]
\end{proof}

\subsection{Classical strong subadditivity and its quantum violation}\label{sec:ssa}
For three classical variables, the reciprocal determinant products factor over the middle variable.
We combine the inequalities for unnormalized slices using an auxiliary probability construction.

\subsubsection{Classical distributions}
\begin{theorem}[Classical strong subadditivity]\label{thm:classical_ssa}
For every finite classical joint distribution $p_{ijk}$,
\begin{equation}\label{eq:classical_ssa}
 Q(p_{AB})+Q(p_{BC})\ge Q(p_B)+Q(p_{ABC}).
\end{equation}
More strongly, for every $t\ge0$,
\[
 L_t(p_B)+L_t(p_{ABC})\ge L_t(p_{AB})+L_t(p_{BC}).
\]
The pointwise statement holds also for unnormalized nonnegative tables.
For a probability distribution, equality in \eqref{eq:classical_ssa} holds exactly when $A=f(B)$ almost surely or $C=g(B)$ almost surely.
For each fixed $t>0$, the pointwise equality condition is the same.
\end{theorem}
\begin{proof}
Fix $t>0$.
For each middle label $j$ put
\[
 d_j=(1+tp_B(j))^{-1},\quad
 a_j=\prod_{i,k}(1+tp_{ijk})^{-1},
\]
\[
 x_j=\prod_i(1+tp_{AB}(i,j))^{-1},\quad
 y_j=\prod_k(1+tp_{BC}(j,k))^{-1}.
\]
Refinement gives $0<a_j\le x_j,y_j\le d_j$.
Proposition~\ref{prop:classical-merging}, applied to the unnormalized slice at $j$, gives $a_j+d_j\ge x_j+y_j$.
Set $A_j=a_j/d_j$, $X_j=x_j/d_j$, and $Y_j=y_j/d_j$.
The four numbers
\[
 A_j,\quad X_j-A_j,\quad Y_j-A_j,\quad 1-X_j-Y_j+A_j
\]
are nonnegative and sum to one.
Use them as probabilities that two auxiliary events $E_j,F_j$ both occur, only the first occurs, only the second occurs, or neither occurs.
The marginal probabilities are $X_j,Y_j$, and the intersection has probability $A_j$.
Construct these event pairs independently across $j$; they need not belong to the original probability space.
Inclusion--exclusion gives
\[
 \prod_j X_j+\prod_j Y_j-\prod_j A_j
 =\Pr\Bigl(\bigcap_j E_j\ \cup\ \bigcap_j F_j\Bigr)\le1.
\]
Multiplying by $\prod_jd_j$ proves the pointwise statement.
Zero slices give $a_j=x_j=y_j=d_j=1$; $t=0$ is immediate.

For equality at $t>0$, put $u_j=X_j-A_j$, $v_j=Y_j-A_j$ and $w_j=1-X_j-Y_j+A_j$.
The global deficit divided by $\prod_jd_j$ is the probability that both global intersections fail.
If $w_j>0$ at any slice, choose neither event there and both events in every other slice.
Because every $A_\ell>0$, this gives a strictly positive global deficit.
Equality thus forces $w_j=0$ throughout; by Proposition~\ref{prop:classical-merging}, each positive slice has one positive row or one positive column.

In a slice with a single row, $X_j=1$, $A_j=Y_j$, and $v_j=0$; if $C$ varies there, strict refinement gives $u_j>0$.
In a slice with a single column, $Y_j=1$, $A_j=X_j$, and $u_j=0$; if $A$ varies there, $v_j>0$.
If neither outer variable is determined by $B$, there is a slice where $A$ varies and a different slice where $C$ varies.
Choose only $F$ in the first, only $E$ in the second, and both elsewhere.
This event has positive probability and makes both global intersections fail.
Thus equality requires the same outer variable to be deterministic in every positive middle slice.

Conversely, if $A=f(B)$, the nonzero entries of $p_{AB}$ are those of $p_B$, and those of $p_{ABC}$ are those of $p_{BC}$; both kernel and subentropy deficits vanish.
The case $C=g(B)$ is symmetric.
Otherwise the pointwise deficit is strictly positive at every $t>0$.
Integrating using \eqref{eq:integral} therefore proves \eqref{eq:classical_ssa} and its complete equality condition.
\end{proof}

The same outer variable must be deterministic across all positive middle slices.
For example, let $\Pr[B=0]=\Pr[B=1]=1/2$ and take the conditional $(A,C)$ tables
\[
 p_{AC|B=0}=\begin{pmatrix}1/2&1/2\\0&0\end{pmatrix},\qquad
 p_{AC|B=1}=\begin{pmatrix}1/2&0\\1/2&0\end{pmatrix}.
\]
Each slice has a deterministic outer variable, so $A\perp C\mid B$.
Nevertheless, the full distribution has pointwise deficit
\[
 \bigl[(1+t/2)^{-1}-(1+t/4)^{-2}\bigr]^2>0\quad(t>0),
\]
and its subentropy deficit is $25/12-3\log2>0$.
Conditional independence therefore does not characterize subentropy SSA equality as it does for Shannon entropy.

\subsubsection{An exact violation for three qubits}
A Bell pair on $AB$ with an independent maximally mixed qubit $C$ already violates strong subadditivity, as recorded in \cite{omegainstitute2026trureturing}.
Its deficit is $Q(I_4/4)-2Q(I_2/2)=-1/12$: the $AB$ state is pure, the $BC$ state is maximally mixed, and the nonzero joint spectrum is that of $I_2/2$.
The following example has $A$ classical and shows failure even when discarding a subsystem implements an optimal discrimination measurement.
Its state matrix is rational, and its deficit has a strictly negative determinant integrand at every positive parameter.

\begin{proposition}[Quantum strong subadditivity fails]\label{prop:quantum_ssa}
Let $A$ be a classical binary register and $B,C$ qubits.
Define
\[
 |\phi_0\rangle=\frac{2|00\rangle+|11\rangle}{\sqrt5},\qquad
 |\phi_1\rangle=\frac{|00\rangle+2|11\rangle}{\sqrt5},
\]
\[
 \rho_{ABC}=\tfrac12|0\rangle\langle0|\otimes|\phi_0\rangle\langle\phi_0|
 +\tfrac12|1\rangle\langle1|\otimes|\phi_1\rangle\langle\phi_1|.
\]
Then
\begin{align}\label{eq:ssa_deficit}
 &Q(AB)+Q(BC)-Q(B)-Q(ABC)\notag\\
 &\qquad=2\log5-\frac{256}{135}\log2-\frac{81}{40}\log3+\frac5{18}
 \simeq-0.04244871319<0.
\end{align}
Its determinant integrand is strictly negative for every $t>0$.
\end{proposition}
\begin{proof}
All matrix entries are rational.
The two nonzero blocks, on the basis pairs $(000,011)$ and $(100,111)$ respectively, are
\[
 \frac1{10}\begin{pmatrix}4&2\\2&1\end{pmatrix},\qquad
 \frac1{10}\begin{pmatrix}1&2\\2&4\end{pmatrix}.
\]
The block structure and partial traces give the nonzero spectra
\begin{align*}
 \operatorname{spec}_{+}(\rho_{ABC})&=\operatorname{spec}_{+}(\rho_B)
 =(1/2,1/2),\\
 \operatorname{spec}_{+}(\rho_{AB})&=(2/5,2/5,1/10,1/10),\\
 \operatorname{spec}_{+}(\rho_{BC})&=(9/10,1/10).
\end{align*}
Put $D_0=(1+t/2)^2$, $D_1=(1+2t/5)^2(1+t/10)^2$, and $D_2=(1+9t/10)(1+t/10)$.
Exact expansion gives
\begin{equation}\label{eq:ssa_kernel}
\frac2{D_0}-\frac1{D_1}-\frac1{D_2}
 =-\frac{t^2(5000+2500t+250t^2+75t^3+7t^4)}{62500D_0D_1D_2}<0\quad(t>0).
\end{equation}
This is the reciprocal determinant combination for the SSA deficit.
The weight in \eqref{eq:integral} is positive, so the subentropy deficit is strictly negative.
Its exact value follows from \eqref{eq:definition}, which gives
\begin{align*}
 Q(AB)&=\log5-\frac{121}{135}\log2-\frac{13}{18},\\
 Q(BC)&=\log10-\frac{81}{40}\log3,\qquad
 Q(B)=Q(ABC)=\log2-\frac12.
\end{align*}
For the repeated spectrum, with $a=2/5$, $b=1/10$, and $f(z)=z^4\log z$, the confluent divided difference is
\[
 Q(a,a,b,b)=-\frac{f'(a)+f'(b)}{(a-b)^2}
             +\frac{2[f(a)-f(b)]}{(a-b)^3}.
\]
Substituting these values yields the stated subentropies and \eqref{eq:ssa_deficit}.
\end{proof}

The same state shows that an optimal discrimination measurement can decrease conditional subentropy.
For the equally likely input signals $(2|0\rangle+|1\rangle)/\sqrt5$ and $(|0\rangle+2|1\rangle)/\sqrt5$, the isometry $|0\rangle\mapsto|00\rangle$, $|1\rangle\mapsto|11\rangle$ gives the counterexample above.
Discarding $C$ implements dephasing in the computational basis, with output signals $\diag(4/5,1/5)$ and $\diag(1/5,4/5)$.
The difference of the input signal states weighted by their priors is $\diag(3/10,-3/10)$, so this measurement is optimal and preserves $\pg=4/5$.
Nevertheless, the negative deficit in \eqref{eq:ssa_deficit} means $Q(A|B)<Q(A|BC)$.
Thus conditional subentropy does not satisfy the usual data processing inequality, even when the label is classical.
The comparison with measurement posteriors remains valid: it bounds conditional subentropy by an average of posterior subentropies.

\section{Spectral lower bounds on the error and independent ancillary states}\label{sec:examples}
\subsection{Two lower bounds from the same spectra}\label{sec:spectral-bounds}
Both conditional subentropy and a difference of purities are determined by the joint and average spectra.
Purity is the trace of the squared density operator; the difference below gives a complementary lower bound on the discrimination error.
For $n\ge2$ put
\begin{equation}\label{eq:quadratic-statistic}
 c(\rho)=\sum_{x<z}\Tr(A_xA_z)
 =\frac12\bigl(\Tr\rho_B^2-\Tr\rho_{XB}^2\bigr),\qquad
 g_n(u)=u-\frac{n}{2(n-1)}u^2.
\end{equation}
The curve $g_n$ is half the classical Fano upper bound for the noncollision probability (the probability that two independent draws from the same distribution differ) \cite[Example 16, Equations (75)--(76)]{rioul2023interplay}.
The same Gram decomposition used for subentropy gives a quantum bound in terms of the difference of purities.

\begin{proposition}[A complementary quadratic lower bound]\label{prop:quadratic-certificate}
For every $n$-label CQ ensemble with optimal error $e$,
\begin{equation}\label{eq:quadratic-certificate}
 0\le c(\rho)\le g_n(e),\qquad
 e\ge g_n^{-1}(c(\rho))=
 \frac{n-1}{n}\biggl(1-\sqrt{1-\frac{2n}{n-1}c(\rho)}\biggr).
\end{equation}
The curve bound is pointwise sharp at fixed $n$ and $c$.
\end{proposition}
\begin{proof}
For the Gram matrix $G=M^*M$ of Proposition~\ref{prop:posterior-comparison}, define $J(G)=\|G-\Delta G\|_{\mathrm{HS}}^2/2$.
Pinching is a Hilbert--Schmidt orthogonal projection, so
\[
 J(G)=\frac12\bigl(\Tr G^2-\Tr(\Delta G)^2\bigr)=c(\rho).
\]
The functional is convex because it is the squared norm of a linear map.
On a pure state with block weights $b$, it equals $j(b)=(1-\sum_xb_x^2)/2$.
Apply convexity to the pure decomposition obtained by refining the measurement in \eqref{eq:posterior-refinement}, then use concavity of $j$ to combine the refined outcomes belonging to each original outcome.
This gives $c(\rho)\le\sum_yq_yj(b^y)$ for every finite POVM.
If $u=1-\max_xb_x$, Cauchy--Schwarz on the other $n-1$ coordinates gives $\sum_xb_x^2\ge(1-u)^2+u^2/(n-1)$, hence $j(b)\le g_n(u)$.
Concavity of the quadratic curve and Jensen's inequality applied to an optimal guessing POVM give $c\le g_n(e)$.
Since the curve is strictly increasing on $[0,1-1/n]$, the quadratic formula gives the inverse bound.
Identical pure signals with priors $(1-e,e/(n-1),\ldots)$ have $c=j(p)=g_n(e)$ and attain every point.
\end{proof}
Montanaro's lower bound \cite[Theorem 1]{montanaro2008lower} is
\[
 e\ge\sum_{x<z}p_xp_zF(\rho_x,\rho_z),\qquad
 F(\rho,\sigma)=\|\sqrt\rho\sqrt\sigma\|_1^2.
\]
We use squared fidelity.
Since $F(\rho,\sigma)\ge\Tr(\rho\sigma)$, this already implies $e\ge c(\rho)$.
Proposition~\ref{prop:quadratic-certificate} improves this scalar bound to $g_n^{-1}(c)>c$ for $c>0$, since $g_n(u)<u$ for $u>0$.
The quadratic bound and Montanaro's full fidelity bound do not dominate each other.
For two equiprobable identical signals, the quadratic and Montanaro bounds are $1/2$ and $1/4$ when the common state is pure, and $(1-1/\sqrt2)/2$ and $1/4$ when it is $I_2/2$.

Qiu's trace norm bound \cite[Theorem 1]{qiu2008minimum} gives
\[
 e\ge\frac12\biggl(1-\frac1{n-1}
          \sum_{x<z}\|A_x-A_z\|_1\biggr).
\]
Audenaert and Mosonyi \cite[Equation (6)]{audenaert2014upper} restate this as a sum of weighted binary errors divided by $n-1$.
For $n=2$ it is the exact Helstrom error and hence is at least either spectral lower bound.
In general it requires pairwise operator information beyond the two spectra.

Barnum and Knill \cite[Sections IV--V]{barnum2002reversing} compare optimal discrimination with the pretty good measurement (PGM).
Write
\[
 M_x^{\mathrm{PGM}}=\rho_B^{-1/2}A_x\rho_B^{-1/2},\qquad
 e_{\mathrm{PGM}}=1-\sum_x\Tr(A_xM_x^{\mathrm{PGM}}).
\]
The inverse is taken on the support of $\rho_B$, and the POVM is completed arbitrarily outside that support.
Their guarantee $(1-e)^2\le1-e_{\mathrm{PGM}}$, together with the fidelity upper bound of Audenaert and Mosonyi \cite[Theorem 3.1]{audenaert2014upper}, yields in our fidelity convention
\[
 1-\sqrt{1-e_{\mathrm{PGM}}}\le e\le e_{\mathrm{PGM}}
 \le\sum_{x<z}\sqrt{p_xp_zF(\rho_x,\rho_z)}.
\]
The last inequality improves the original Barnum--Knill fidelity upper bound by a factor of two.
These guarantees use the ensemble operators or pairwise fidelities; the lower bounds in \eqref{eq:quadratic-certificate} and Theorem~\ref{thm:guess} use the joint and average spectra.

The two lower bounds can be used together:
\[
 e\ge\max\{\Phi_n^{-1}(Q(X|B)),\,g_n^{-1}(c(\rho))\}.
\]
Each bound is optimal when its statistic is fixed separately; optimality with both statistics fixed does not follow.
Even fixing the priors, individual signal spectra, and average state does not in general determine the optimal error.
Consider three commuting signals on $\mathbb C^2$, represented by the joint probability tables
\begin{equation}\label{eq:equal-spectra-ensembles}
 T=\frac1{21}\begin{pmatrix}6&2\\1&5\\7&0\end{pmatrix},\qquad
 \widetilde T=\frac1{21}\begin{pmatrix}2&6\\5&1\\7&0\end{pmatrix}.
\end{equation}
Row $x$ gives $A_x=\diag(T_{x1},T_{x2})$, and similarly for $\widetilde T$.
The row sums give the same priors $p=(8,6,7)/21$, and each row has the same pair of entries in both tables.
Thus each signal has the same spectrum in the two ensembles.
Both average states are $\rho_B=\diag(14,7)/21$, and both joint spectra are $(0,1,2,5,6,7)/21$.
Dephasing a POVM in the common eigenbasis leaves its success probability unchanged.
Optimizing its diagonal entries independently in each column therefore gives
\[
 \pg(T)=\sum_{y=1}^2\max_{1\le x\le3}T_{xy}=\frac{12}{21},\qquad
 \pg(\widetilde T)=\frac{13}{21}.
\]
The optimal errors are $3/7$ and $8/21$, respectively, although each spectral lower bound has the same value on both ensembles.

The following examples show that neither spectral bound dominates the other.

\subsection{Comparison of the bounds}\label{sec:comparison-bounds}
For two pure signals with priors $p,1-p$ and overlap $s$,
\begin{equation}\label{eq:binary-quadratic-exact}
 c=p(1-p)s^2,\qquad
 g_2^{-1}(c)=\frac{1-\sqrt{1-4p(1-p)s^2}}2=e.
\end{equation}
Here $P_i=|\psi_i\rangle\langle\psi_i|$ are the signal states and $s=|\langle\psi_0|\psi_1\rangle|$ is their overlap.
Indeed, $pP_0-(1-p)P_1$ has trace $2p-1$ and determinant $-p(1-p)(1-s^2)$ on the two-dimensional signal span when $s<1$.
Its trace norm is $\sqrt{1-4p(1-p)s^2}$, giving the Helstrom minimum error.
The cases of identical signals or zero priors follow directly or by continuity.
Thus the quadratic bound equals the optimal error for binary pure signals with arbitrary priors.

\begin{figure}[tb]
\centering
\begin{minipage}[t]{0.48\textwidth}
\centering
\begin{tikzpicture}
\begin{axis}[width=\linewidth,height=5.4cm,xmin=0,xmax=0.92,ymin=0,ymax=0.94,
 xlabel={Error $e$},ylabel={Upper bound on $Q(X|B)$},
 title={(a) Sharp error curves},title style={font=\small},
 tick label style={font=\scriptsize},label style={font=\small},
 grid=major,grid style={gray!15},legend pos=north west,
 legend style={font=\scriptsize,draw=none,fill=white}]
\addplot[black,dashed,domain=0:0.92] {x};\addlegendentry{$e$}
\addplot[blue!70!black,thick] coordinates {(0,0) (0.0125,0.01187855018) (0.025,0.02290757006) (0.0375,0.03328764218) (0.05,0.04311429722) (0.0625,0.05244901131) (0.075,0.06133598955) (0.0875,0.06980906563) (0.1,0.07789520844) (0.1125,0.08561654004) (0.125,0.09299159785) (0.1375,0.1000361734) (0.15,0.1067638956) (0.1625,0.1131866516) (0.175,0.1193148992) (0.1875,0.1251579047) (0.2,0.1307239272) (0.2125,0.1360203649) (0.225,0.1410538714) (0.2375,0.145830451) (0.25,0.1503555364) (0.2625,0.1546340534) (0.275,0.1586704749) (0.2875,0.1624688667) (0.3,0.1660329254) (0.3125,0.1693660118) (0.325,0.1724711793) (0.3375,0.1753511972) (0.35,0.1780085727) (0.3625,0.1804455678) (0.375,0.1826642159) (0.3875,0.1846663345) (0.4,0.1864535373) (0.4125,0.1880272438) (0.425,0.1893886883) (0.4375,0.1905389266) (0.45,0.1914788424) (0.4625,0.1922091524) (0.475,0.1927304097) (0.4875,0.1930430074) (0.5,0.1931471806)};\addlegendentry{$\Phi_2(e)$}
\addplot[orange!90!black,thick] coordinates {(0,0) (0.01666666667,0.01585606212) (0.03333333333,0.03061132803) (0.05,0.04452905681) (0.06666666667,0.05773357703) (0.08333333333,0.07030439314) (0.1,0.08229834759) (0.1166666667,0.09375870658) (0.1333333333,0.1047197667) (0.15,0.1152094998) (0.1666666667,0.1252512043) (0.1833333333,0.1348645992) (0.2,0.1440665841) (0.2166666667,0.1528717859) (0.2333333333,0.1612929635) (0.25,0.1693413149) (0.2666666667,0.1770267154) (0.2833333333,0.1843579039) (0.3,0.1913426327) (0.3166666667,0.1979877874) (0.3333333333,0.2042994857) (0.35,0.2102831574) (0.3666666667,0.2159436116) (0.3833333333,0.221285092) (0.4,0.2263113227) (0.4166666667,0.2310255472) (0.4333333333,0.2354305595) (0.45,0.2395287312) (0.4666666667,0.2433220327) (0.4833333333,0.2468120503) (0.5,0.25) (0.5166666667,0.2528867373) (0.5333333333,0.255472764) (0.55,0.2577582318) (0.5666666667,0.259742943) (0.5833333333,0.2614263482) (0.6,0.2628075406) (0.6166666667,0.2638852475) (0.6333333333,0.2646578179) (0.65,0.2651232069) (0.6666666667,0.2652789553)};\addlegendentry{$\Phi_3(e)$}
\addplot[green!50!black,thick] coordinates {(0,0) (0.0225,0.02143778987) (0.045,0.04144657581) (0.0675,0.06037458863) (0.09,0.07838455585) (0.1125,0.09557972575) (0.135,0.1120333951) (0.1575,0.1278009705) (0.18,0.1429260642) (0.2025,0.1574439835) (0.225,0.1713839021) (0.2475,0.1847702958) (0.27,0.1976239348) (0.2925,0.2099625958) (0.315,0.2218015868) (0.3375,0.2331541433) (0.36,0.2440317324) (0.3825,0.2544442906) (0.405,0.2644004108) (0.4275,0.2739074918) (0.45,0.282971857) (0.4725,0.2915988499) (0.495,0.2997929102) (0.5175,0.3075576333) (0.54,0.3148958166) (0.5625,0.321809494) (0.585,0.3282999592) (0.6075,0.3343677794) (0.63,0.3400127998) (0.6525,0.345234138) (0.675,0.3500301691) (0.6975,0.3543985009) (0.72,0.3583359365) (0.7425,0.3618384248) (0.765,0.3649009947) (0.7875,0.3675176717) (0.81,0.3696813694) (0.8325,0.3713837532) (0.855,0.3726150626) (0.8775,0.3733638824) (0.9,0.373616839)};\addlegendentry{$\Phi_{10}(e)$}
\end{axis}
\end{tikzpicture}
\end{minipage}\hfill
\begin{minipage}[t]{0.48\textwidth}
\centering
\begin{tikzpicture}
\begin{axis}[width=\linewidth,height=5.4cm,xmin=0,xmax=1,ymin=0,ymax=0.52,
 xlabel={Overlap $s$},ylabel={Discrimination error},
 title={(b) Binary pure signals},title style={font=\small},
 tick label style={font=\scriptsize},label style={font=\small},
 grid=major,grid style={gray!15},legend pos=north west,
 legend style={font=\scriptsize,draw=none,fill=white}]
\addplot[black,thick] coordinates {(0,0) (0.0125,3.9064026e-05) (0.025,0.0001562744217) (0.0375,0.0003516861832) (0.05,0.000625391114) (0.0625,0.0009775180415) (0.075,0.001408233121) (0.0875,0.001917740228) (0.1,0.002506281447) (0.1125,0.00317413765) (0.125,0.003921629175) (0.1375,0.004749116609) (0.15,0.005657001668) (0.1625,0.006645728203) (0.175,0.007715783312) (0.1875,0.008867698578) (0.2,0.01010205144) (0.2125,0.01141946672) (0.225,0.01282061825) (0.2375,0.01430623074) (0.25,0.01587708172) (0.2625,0.01753400379) (0.275,0.01927788692) (0.2875,0.02110968114) (0.3,0.02303039929) (0.3125,0.0250411202) (0.325,0.02714299202) (0.3375,0.02933723591) (0.35,0.03162515012) (0.3625,0.03400811434) (0.375,0.03648759456) (0.3875,0.03906514831) (0.4,0.0417424305) (0.4125,0.04452119972) (0.425,0.04740332524) (0.4375,0.05039079469) (0.45,0.05348572251) (0.4625,0.05669035934) (0.475,0.06000710233) (0.4875,0.06343850662) (0.5,0.06698729811) (0.5125,0.07065638761) (0.525,0.07444888674) (0.5375,0.07836812561) (0.55,0.08241767279) (0.5625,0.08660135765) (0.575,0.0909232957) (0.5875,0.09538791726) (0.6,0.1) (0.6125,0.1047647062) (0.625,0.1096876251) (0.6375,0.1147748223) (0.65,0.1200328962) (0.6625,0.1254690433) (0.675,0.1310911359) (0.6875,0.1369078113) (0.7,0.1429285786) (0.7125,0.149163945) (0.725,0.1556255672) (0.7375,0.1623264335) (0.75,0.1692810861) (0.7625,0.1765058926) (0.775,0.1840193835) (0.7875,0.1918426741) (0.8,0.2) (0.8125,0.2085194046) (0.825,0.2174336361) (0.8375,0.2267813376) (0.85,0.2366086562) (0.8625,0.246971469) (0.875,0.2579385409) (0.8875,0.2695961426) (0.9,0.2820550528) (0.9125,0.2954616479) (0.925,0.3100164481) (0.9375,0.3260073637) (0.95,0.34387505) (0.9625,0.3643591599) (0.975,0.3888975698) (0.9875,0.4211904987) (0.99,0.4294663201) (0.999,0.4776449111) (0.9999,0.492929109) (0.99999,0.4977639376) (0.999999,0.4992928934) (0.9999999,0.4997763932) (0.99999999,0.4999292893) (1,0.5)};\addlegendentry{Actual = quadratic}
\addplot[blue!70!black,thick] coordinates {(0,0) (0.0125,2.604889687e-05) (0.025,0.0001042674394) (0.0375,0.0002348414664) (0.05,0.0004180441938) (0.0625,0.0006542241863) (0.075,0.0009437980661) (0.0875,0.001287245651) (0.1,0.001685106617) (0.1125,0.002137978221) (0.125,0.002646513803) (0.1375,0.003211421903) (0.15,0.003833465883) (0.1625,0.004513463976) (0.175,0.005252289704) (0.1875,0.006050872636) (0.2,0.006910199462) (0.2125,0.007831315359) (0.225,0.008815325649) (0.2375,0.009863397746) (0.25,0.01097676339) (0.2625,0.01215672116) (0.275,0.01340463933) (0.2875,0.01472195902) (0.3,0.01611019767) (0.3125,0.01757095292) (0.325,0.01910590683) (0.3375,0.02071683054) (0.35,0.02240558942) (0.3625,0.02417414863) (0.375,0.02602457932) (0.3875,0.02795906536) (0.4,0.02997991076) (0.4125,0.03208954781) (0.425,0.03429054604) (0.4375,0.03658562212) (0.45,0.03897765068) (0.4625,0.04146967636) (0.475,0.04406492711) (0.4875,0.04676682895) (0.5,0.04957902229) (0.5125,0.05250538022) (0.525,0.05555002886) (0.5375,0.05871737012) (0.55,0.06201210724) (0.5625,0.06543927361) (0.575,0.06900426518) (0.5875,0.07271287724) (0.6,0.07657134621) (0.6125,0.0805863973) (0.625,0.0847652991) (0.6375,0.08911592631) (0.65,0.09364683231) (0.6625,0.09836733327) (0.675,0.1032876064) (0.6875,0.1084188052) (0.7,0.1137731953) (0.7125,0.1193643161) (0.725,0.1252071742) (0.7375,0.1313184757) (0.75,0.1377169095) (0.7625,0.1444234942) (0.775,0.1514620074) (0.7875,0.1588595232) (0.8,0.1666470931) (0.8125,0.1748606203) (0.825,0.1835419981) (0.8375,0.1927406205) (0.85,0.202515421) (0.8625,0.2129376906) (0.875,0.2240950703) (0.8875,0.2360973826) (0.9,0.2490854594) (0.9125,0.2632451032) (0.925,0.2788303943) (0.9375,0.2962053911) (0.95,0.3159259325) (0.9625,0.3389222541) (0.975,0.3669960823) (0.9875,0.4047613597) (0.99,0.414578507) (0.999,0.4726666669) (0.9999,0.4913419121) (0.99999,0.4972614715) (0.999999,0.4991339778) (0.9999999,0.4997261388) (0.99999999,0.4999133975) (1,0.5)};\addlegendentry{Subentropy bound}
\addplot[orange!90!black,dashed,thick] coordinates {(0,0) (0.0125,3.90625e-05) (0.025,0.00015625) (0.0375,0.0003515625) (0.05,0.000625) (0.0625,0.0009765625) (0.075,0.00140625) (0.0875,0.0019140625) (0.1,0.0025) (0.1125,0.0031640625) (0.125,0.00390625) (0.1375,0.0047265625) (0.15,0.005625) (0.1625,0.0066015625) (0.175,0.00765625) (0.1875,0.0087890625) (0.2,0.01) (0.2125,0.0112890625) (0.225,0.01265625) (0.2375,0.0141015625) (0.25,0.015625) (0.2625,0.0172265625) (0.275,0.01890625) (0.2875,0.0206640625) (0.3,0.0225) (0.3125,0.0244140625) (0.325,0.02640625) (0.3375,0.0284765625) (0.35,0.030625) (0.3625,0.0328515625) (0.375,0.03515625) (0.3875,0.0375390625) (0.4,0.04) (0.4125,0.0425390625) (0.425,0.04515625) (0.4375,0.0478515625) (0.45,0.050625) (0.4625,0.0534765625) (0.475,0.05640625) (0.4875,0.0594140625) (0.5,0.0625) (0.5125,0.0656640625) (0.525,0.06890625) (0.5375,0.0722265625) (0.55,0.075625) (0.5625,0.0791015625) (0.575,0.08265625) (0.5875,0.0862890625) (0.6,0.09) (0.6125,0.0937890625) (0.625,0.09765625) (0.6375,0.1016015625) (0.65,0.105625) (0.6625,0.1097265625) (0.675,0.11390625) (0.6875,0.1181640625) (0.7,0.1225) (0.7125,0.1269140625) (0.725,0.13140625) (0.7375,0.1359765625) (0.75,0.140625) (0.7625,0.1453515625) (0.775,0.15015625) (0.7875,0.1550390625) (0.8,0.16) (0.8125,0.1650390625) (0.825,0.17015625) (0.8375,0.1753515625) (0.85,0.180625) (0.8625,0.1859765625) (0.875,0.19140625) (0.8875,0.1969140625) (0.9,0.2025) (0.9125,0.2081640625) (0.925,0.21390625) (0.9375,0.2197265625) (0.95,0.225625) (0.9625,0.2316015625) (0.975,0.23765625) (0.9875,0.2437890625) (0.99,0.245025) (0.999,0.24950025) (0.9999,0.2499500025) (0.99999,0.249995) (0.999999,0.2499995) (0.9999999,0.24999995) (0.99999999,0.249999995) (1,0.25)};\addlegendentry{Montanaro}
\draw[gray,densely dotted] (axis cs:0.7678757114,0) -- (axis cs:0.7678757114,0.1474082770);
\fill[black] (axis cs:0.7678757114,0.1474082770) circle[radius=1.5pt];
\node[font=\scriptsize,anchor=north,fill=white,inner sep=1pt]
 at (axis cs:0.65,0.066) {$s_*\approx0.767876$};
\draw[gray,thin] (axis cs:0.65,0.069) -- (axis cs:0.7678757114,0.142);
\end{axis}
\end{tikzpicture}
\end{minipage}
\caption{Sharp curves and their operational inversion.
(a) Each curve stops at its largest admissible optimal error, $1-1/n$.
Identical pure signals attain every point of the corresponding curve.
(b) For the pure signals with equal priors in \eqref{eq:binary_example}, the quadratic bound equals the actual error; the other curves are the subentropy bound and Montanaro's bound $s^2/4$.
The marked crossover is $s_*\approx0.767876$: Montanaro's bound is stronger for $0<s<s_*$, and the subentropy bound is stronger for $s_*<s\le1$.
The subentropy bound is exact at the orthogonal and identical endpoints.
Plot coordinates are rounded to ten significant digits.}
\label{fig:guessing}
\end{figure}
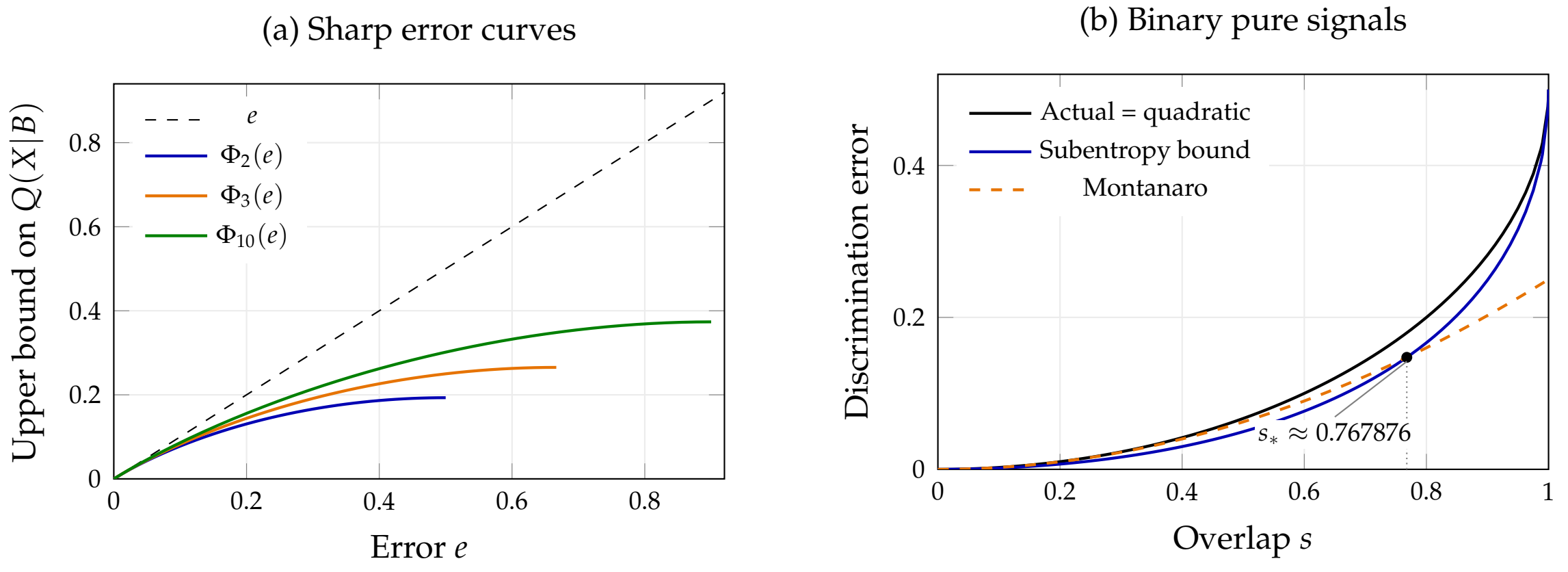

For equal priors, the overlap $s=|\langle\psi_0|\psi_1\rangle|$ also determines the two spectra: the joint nonzero spectrum is $(1/2,1/2)$ and the average spectrum is $((1+s)/2,(1-s)/2)$.
Hence
\begin{equation}\label{eq:binary_example}
 Q(X|B)=\log2-\frac12-\Phi_2\Bigl(\frac{1-s}{2}\Bigr),\qquad
 e=\frac{1-\sqrt{1-s^2}}2.
\end{equation}
The error formula is \eqref{eq:binary-quadratic-exact} specialized to equal priors.
For $0<u<1/2$ the subentropy curve has the expression
\[
 \Phi_2(u)=-\frac{(1-u)^2\log(1-u)-u^2\log u}{1-2u},
\]
with continuous values at the endpoints.
For this binary pure family, Montanaro's bound is $s^2/4=c$.
Table~\ref{tab:binary} and Figure~\ref{fig:guessing} compare it with the subentropy bound and the exact error.
\phantomsection\label{par:binary-crossover} For small overlap, $Q(X|B)=s^2/6+O(s^4)$ whereas $e=s^2/4+O(s^4)$.
Since $\Phi_2(u)\sim u$ at zero, its inverse captures two thirds of the leading error.
Montanaro's bound is strictly stronger than the subentropy bound for $0<s<s_*$, where $s_*\approx0.767876$ is the unique crossover in $(0,1)$, determined by
\[
 \Phi_2(s_*^2/4)=\log2-\frac12-\Phi_2\Bigl(\frac{1-s_*}{2}\Bigr).
\]
The subentropy bound is strictly stronger for $s_*<s\le1$; at $s=0$ both bounds vanish.
Appendix~\ref{app:binary-crossover} proves uniqueness and certifies the numerical value.
The quadratic bound equals the optimal error throughout this family.
\begin{table}[tbp]
\caption{Binary pure signals with equal priors.
Montanaro's bound is $s^2/4$; the subentropy bound is $\Phi_2^{-1}(Q(X|B))$.
The quadratic bound equals the actual error.
Actual errors are rounded to nearest; lower bounds are rounded downward to five places using rational enclosures.
Endpoints are evaluated analytically.}\label{tab:binary}
\centering
\begin{tabular}{rrrr}
\toprule
Overlap $s$ & Actual error & Montanaro bound & Subentropy bound\\
\midrule
0.00 & 0.00000 & 0.00000 & 0.00000 \\
0.25 & 0.01588 & 0.01562 & 0.01097 \\
0.50 & 0.06699 & 0.06250 & 0.04957 \\
0.80 & 0.20000 & 0.16000 & 0.16664 \\
0.95 & 0.34388 & 0.22562 & 0.31592 \\
1.00 & 0.50000 & 0.25000 & 0.50000 \\
\bottomrule
\end{tabular}
\end{table}

The subentropy bound can be stronger for three signals.
To see this, take pure signals with priors $p=(2/3,1/6,1/6)$ and weighted Gram matrix
\begin{equation}\label{eq:equal-tail-gram}
 G_\varepsilon=\begin{pmatrix}2/3&0&0\\0&1/6&\varepsilon\\0&\varepsilon&1/6\end{pmatrix},
 \qquad 0<\varepsilon<1/6.
\end{equation}
One realization is $\psi_1=e_1$, $\psi_2=e_2$, $\psi_3=6\varepsilon e_2+\sqrt{1-36\varepsilon^2}e_3$ in an orthonormal basis.
The average spectrum is $(2/3,1/6+\varepsilon,1/6-\varepsilon)$ and the joint nonzero spectrum is $p$, giving $c=\varepsilon^2$.
Writing $a=2/3$ and $b=1/6$, the integral representation gives the exact identity
\[
 Q(X|B)=\varepsilon^2\int_0^\infty
 \frac{\dd t}{(1+at)(1+bt)^2((1+bt)^2-t^2\varepsilon^2)}.
\]
Setting the perturbation to zero in the integrand gives the leading coefficient.
To justify the expansion and control the remainder, suppose $|\varepsilon|\le b/2$.
After removing the prefactor, the integrand is bounded by $4/[3(1+at)(1+bt)^4]$.
Its difference from the $\varepsilon=0$ integrand is bounded by $O(\varepsilon^2)$ times this integrable function, since $t/(1+bt)\le1/b$.
Therefore
\begin{equation}\label{eq:non-dominance-coefficient}
 Q(X|B)=\alpha\varepsilon^2+O(\varepsilon^4),\qquad
 \alpha=\int_0^\infty\frac{\dd t}{(1+2t/3)(1+t/6)^4}
 =\frac{256\log2-150}{27}>1.
\end{equation}
For an explicit evaluation of the coefficient, substitute $t=6x$ and use
\[
 \frac1{(1+4x)(1+x)^4}
 =\frac{256/81}{1+4x}-\frac{64/81}{1+x}
 -\frac{16/27}{(1+x)^2}-\frac{4/9}{(1+x)^3}-\frac{1/3}{(1+x)^4}.
\]
Integrating the first two terms together gives the stated value.
The first three positive terms in the power series for the inverse hyperbolic tangent give $\log2>842/1215>177/256$, proving $\alpha>1$.
Both inverse curves have right derivative one at zero, by Lemma~\ref{lem:purebound} and the quadratic formula.
Consequently $\Phi_3^{-1}(Q(X|B))=\alpha\varepsilon^2+o(\varepsilon^2)$, while $g_3^{-1}(c)=\varepsilon^2+O(\varepsilon^4)$.
The subentropy bound is therefore stronger for every sufficiently small positive perturbation.
Together with the binary example, this proves that neither bound dominates the other over all ensembles.
Appendix~\ref{app:cyclic-example} gives a further family of three signals with an explicit optimal measurement and certified values.

\subsection{Appending a specified independent ancillary state}\label{sec:ancilla-calibration}
Appending an independent ancillary state leaves the discrimination error unchanged, although conditional subentropy may change.
This also illustrates the distinction between conditional subentropy and the posterior average in Proposition~\ref{prop:posterior-comparison}.
Take a uniform binary label with identical signals $\rho_x=I_d/d$.
Every measurement leaves the posterior uniform, so its average subentropy is $\log2-1/2$.
The conditional subentropy instead equals
\begin{equation}\label{eq:mixed-ancilla}
 q_d=Q(I_{2d}/(2d))-Q(I_d/d)=\log2-H_{2d}+H_d.
\end{equation}
For $d=2$ this is $\log2-7/12$, already strictly smaller than the posterior average.
The same family can be obtained by appending an independent maximally mixed ancilla to identical pure signals.
Its guessing error stays $1/2$ while $q_d=1/(4d)+O(d^{-2})$ tends to zero.
Consequently, the conditional subentropy bound can decrease when independent mixed side information is added.
The bound remains pointwise sharp over all ensembles at each prescribed value of conditional subentropy.

To bound the unchanged error using the conditional subentropy of the extended state, we use a curve that depends on the ancillary state.
For an independent ancillary register $C$ in a specified state $\nu$, define
\begin{equation}\label{eq:ancilla-curve}
 \begin{aligned}
 C_\nu(\rho)&=Q(X|BC)_{\rho_{XB}\otimes\nu},\\
 f_\nu(p)&=Q(\diag(p)\otimes\nu)-Q(\nu),\\
 \Phi_{n,\nu}(u)&=f_\nu\Bigl(1-u,\frac{u}{n-1},\ldots,\frac{u}{n-1}\Bigr).
 \end{aligned}
\end{equation}
The product form $\rho_{XBC}=\rho_{XB}\otimes\nu$ is essential to the following comparison.

\begin{proposition}[Sharp curve for a specified ancilla]\label{prop:ancilla-calibration}
Fix a density operator $\nu$ and $n\ge2$.
For every CQ ensemble on the original system $B$, and every finite POVM on $B$ with posteriors $b^y$ and MAP error $e_M$,
\begin{equation}\label{eq:ancilla-posterior}
 0\le C_\nu(\rho)\le\sum_{y:q_y>0}q_yf_\nu(b^y)
 \le\Phi_{n,\nu}(e_M).
\end{equation}
Appending $\nu$ leaves the optimal guessing error $e$ unchanged, and $C_\nu(\rho)\le\Phi_{n,\nu}(e)$ is pointwise sharp at fixed $n$, $\nu$, and dimension of the original system $B$.
The curve is strictly increasing on $[0,1-1/n]$, and therefore
\begin{equation}\label{eq:ancilla-inverse}
 e\ge\Phi_{n,\nu}^{-1}(C_\nu(\rho)).
\end{equation}
Identical pure original signals with the priors in \eqref{eq:phi} attain the curve at every admissible error.
\end{proposition}
\begin{proof}
Use the Gram matrix $G$ from Proposition~\ref{prop:posterior-comparison} and the pure decomposition obtained there by refining the measurement.
For the pinching $\Delta\otimes\mathrm{id}$, the increase in subentropy $C_{Q,\nu}(G)=Q(\Delta(G)\otimes\nu)-Q(G\otimes\nu)$ is convex in $G$: Lemma~\ref{lem:convex} applies to the affine map $G\mapsto G\otimes\nu$.
For a pure state in the Gram representation with block weights $p$, its value is $f_\nu(p)$: the two spectra are $p\otimes\operatorname{spec}(\nu)$ and $\operatorname{spec}(\nu)$, up to zeros.
Moreover, $C_{Q,\nu}(G)=C_\nu(\rho)$.
Convexity therefore bounds conditional subentropy by the average over refined posteriors.
The function $f_\nu$ is symmetric and strictly concave, since $p\mapsto\diag(p)\otimes\nu$ is affine and injective.
Concavity bounds the refined average by the posterior average for the original measurement.
The permutation average and Jensen argument in Lemma~\ref{lem:classical-fano-q} then give \eqref{eq:ancilla-posterior}.
Nonnegativity follows from the increase in subentropy under pinching.

To compare the optimal errors before and after appending the ancilla, a guessing effect $E_x$ on the joint system induces the effect $\Tr_C[(I\otimes\nu^{1/2})E_x(I\otimes\nu^{1/2})]$ on the original system, with the same success probability.
The induced effects form a POVM.
Conversely, effects on $B$ lift to $E_x\otimes I_C$.
Thus the optimal error is unchanged, even for singular $\nu$.
An optimal measurement on the original system gives the error bound.
Identical pure original signals give $C_\nu=f_\nu(p)$ and error $1-\max p_x$, proving pointwise sharpness.
Strict concavity and permutation symmetry give a unique uniform maximum and strict increase up to it, justifying \eqref{eq:ancilla-inverse}.
\end{proof}

For $\nu=u_k$, write $C_k=C_{u_k}$ and $\Phi_{n,k}=\Phi_{n,u_k}$.
These bounds require only the original spectra and the specified ancillary state; no physical ancillary system needs to be prepared.
Although conditional subentropy tends to zero as the dimension of a maximally mixed ancillary state grows, inverting the corresponding curve gives a nontrivial limit.
\begin{corollary}[Bounds with a maximally mixed ancillary state]\label{cor:uniform-calibration}
For every original CQ ensemble with $n\ge2$ labels and error $e$,
\begin{equation}\label{eq:ancilla-dilution}
 0\le C_k(\rho)\le\Phi_{n,k}(e)\le\frac{e}{k}.
\end{equation}
The coefficient $1/k$ is sharp as $e\downarrow0$.
For the statistic $c(\rho)$ in \eqref{eq:quadratic-statistic}, one has
\begin{equation}\label{eq:ancilla-limit}
 \begin{aligned}
 kC_k(\rho)&\to c(\rho),\\
 \Phi_{n,k}^{-1}(C_k(\rho))&\to
 \frac{n-1}{n}\biggl(1-\sqrt{1-\frac{2n}{n-1}c(\rho)}\biggr).
 \end{aligned}
\end{equation}
For binary pure signals with arbitrary priors, the latter limit is the exact Helstrom error in \eqref{eq:binary-quadratic-exact}.
\end{corollary}
\begin{proof}
The preceding proposition bounds conditional subentropy by the curve.
Concavity and $\Phi_{n,k}(0)=0$ bound the curve by its tangent at zero, $\Phi_{n,k}(e)\le e\Phi_{n,k}'(0+)$.
To compute the slope, put $T=g^*u_kg$ and let $U$ be the average of the independent quadratic forms for the other $n-1$ blocks.
Then $\E U=1$ and the Gaussian formula gives
\[
 \Phi_{n,k}'(0+)=\E[T\log T]-\E\log T
 =\Tr(u_kK(u_k))=\frac1k.
\]
The integration by parts identity in Appendix~\ref{app:conditional-gradient} gives the middle equality.
The integrable logarithm bounds used there also justify differentiation at the boundary.
Identical pure original signals attain the curve, so the tangent coefficient is sharp.
Appendix~\ref{app:ancilla-rate} proves \eqref{eq:ancilla-limit}.
For binary pure signals, Equation~\eqref{eq:binary-quadratic-exact} identifies this limit as the optimal error.
\end{proof}
The choice of curve matters: applying the original inverse $\Phi_n^{-1}(C_k)$ gives a bound tending to zero.
Inverting the curve for the specified ancillary state instead converges to the quadratic bound in Proposition~\ref{prop:quadratic-certificate}.
This limiting bound is already computable from the original spectra.
The convergence does not imply exactness for general ensembles or monotonic improvement as $k$ increases.

\subsection{Numerical evaluation and error control}
State uncertainty and numerical error both affect the lower bounds obtained by inversion.
Let $\rho,\sigma$ be CQ states on the same alphabet of $n\ge2$ labels and quantum system $B$ of dimension $m$, with $D(\rho,\sigma)\le\delta\le1$.
Theorem~\ref{thm:cq-lipschitz}, the modulus of continuity of subentropy, contraction under partial trace, and the range in Theorem~\ref{thm:guess} give the bound
\begin{equation}\label{eq:conditional-continuity}
 \begin{aligned}
 |Q(X|B)_\rho-Q(X|B)_\sigma|&\le\mathcal R_{n,m}(\delta),\\
 \mathcal R_{n,m}(\delta)&=\min\{\delta,\,
 \omega_{nm}(\delta)+\omega_m(\delta),\,\log n-H_n+1\}.
 \end{aligned}
\end{equation}
The three terms come from the direct CQ continuity bound, the separate moduli of continuity, and the range of conditional subentropy.
The first improves the bound $2\delta$ obtained by bounding the two subentropies separately.
Their minimum need not be the optimal modulus of continuity for conditional subentropy.
If the marginal states on $B$ both equal $\tau$, the first term can be replaced by $\kappa(\tau)\delta$.
Suppose $\sigma$ approximates the true state $\rho$ and its computed conditional subentropy $\widehat q$ has a certified error $|\widehat q-Q(X|B)_\sigma|\le\eta$.
Then
\begin{equation}\label{eq:robust-certificate}
 e_{\mathrm{true}}\ge
 \Phi_n^{-1}\!\bigl(\max\{0,\widehat q-\eta-\mathcal R_{n,m}(\delta)\}\bigr).
\end{equation}
The simpler choice $\mathcal R_{n,m}(\delta)=\delta$ is valid in every dimension.
The argument of the inverse is at most the true conditional subentropy, so it lies in the domain of the inverse.
This guarantee requires a justified bound $\eta$ on the numerical error.

For a specified maximally mixed ancilla, Corollary~\ref{cor:ancilla} gives the smaller bound $\delta/k$ on the effect of state error.
If $|\widehat C_k-C_k(\sigma)|\le\eta$, the corresponding lower bound is
\begin{equation}\label{eq:calibrated-robust-certificate}
 e_{\mathrm{true}}\ge\Phi_{n,k}^{-1}
 \bigl(\max\{0,\widehat C_k-\eta-\delta/k\}\bigr).
\end{equation}
Both extended states must have the specified product form.
The lower bounds converge to the limiting value as $k\to\infty$ whenever the absolute evaluation errors satisfy $k\eta_k\to0$ and the state uncertainties satisfy $\delta_k\to0$.
With fixed tolerances, the bounds remain valid, but convergence to the limit is not guaranteed.

Near the maximum of either subentropy curve, uncertainty in the inverse scales as the square root of uncertainty in subentropy.
A small tolerance alone therefore cannot certify that the error equals its endpoint value.
Appendix~\ref{app:numerical-certification} gives the local expansions and the rational interval method used to certify the displayed numerical lower bounds.

\section{Discussion}\label{sec:discussion}
The comparison with measurement posteriors underlies both subadditivity of quantum subentropy and the sharp discrimination bound.
A classical Fano argument completes the latter proof: no stronger lower bound on the error is possible given only conditional subentropy and the number of labels.
Near equality, strong concavity quantifies proximity of the posteriors to the maximizing distributions.
Under perturbations of the CQ state, conditional subentropy satisfies optimal Lipschitz bounds, including when the marginal state of $B$ is fixed.

Although neither spectral bound dominates the other, appending an independent ancillary state relates them without changing the error.
With a maximally mixed ancilla of increasing dimension, inversion of the corresponding sharp curve converges to the quadratic bound computed from the original spectra.
For binary pure signals with arbitrary priors, this limiting bound is exact.
The error analysis gives sufficient conditions on the evaluation and state uncertainties for the computed bounds to converge to this limit.

The determinant comparison holds at every integration parameter, strengthening subadditivity and giving a proof of classical strong subadditivity.
In the quantum setting, strong subadditivity can fail even when one register is classical.
Subentropy need not be additive even for product states: $Q(I_2/2)=\log2-1/2$ and $Q(I_4/4)=\log4-13/12$ give a deficit of $1/12$.
The subadditivity deficit therefore cannot serve as a correlation measure that vanishes on product states.
Nor does a telescoping chain rule imply data processing for conditional subentropy, as the example with an optimal measurement shows.

The optimal modulus of continuity for $Q(X|B)$ on CQ states at a prescribed nonzero trace distance remains open; the best linear bound does not determine it.
For the signal ensemble itself, a further question is which assumptions yield an explicit stability rate beyond the posterior estimate and the qualitative compactness argument away from zero error.
Section~\ref{sec:ensemble-stability} shows why the absolute deficit alone cannot control stability uniformly down to zero error.

\section*{Acknowledgments}
The author thanks OpenAI for providing complimentary access to ChatGPT, which assisted with numerical experiments, exact computational checks, verification and simplification of proofs, the Lean formalization, literature searches, and proofreading.
The author assumes full responsibility for all content.

\bibliography{references_v12}

\appendixtitles{yes}
\appendixstart
\appendix
\renewcommand{\theHequation}{A\arabic{equation}}
\section{A polynomial proof of the inequality for four cells}
\label{app:four_cell_certificate}
After rescaling as in Lemma~\ref{lem:four}, assume $a+b+c+d=1$.
We clear the positive denominators and show that the resulting polynomial has nonnegative coefficients.
Set
\[
 u=(a+b)(c+d),\qquad v=(a+c)(b+d),\qquad
 w=ab+ac+ad+bc+bd+cd,
\]
\[
 x=abc+abd+acd+bcd,\qquad y=abcd,\qquad e=ad+bc,
 \qquad z=1+t.
\]
The denominators associated with the row sums, column sums, and cells are
\[
 P_A=z+ut^2,\qquad P_B=z+vt^2,\qquad
 P=z+wt^2+xt^3+yt^4.
\]
The left side minus the right side is
\[
 \frac1P+\frac1z-\frac1{P_A}-\frac1{P_B}
 =\frac{N(t)}{zP_AP_BP},
\]
where expansion gives
\begin{align}\label{eq:certificate}
 N(t)={}&et^2+(2e-x)t^3+(e-2x-y+2uv)t^4\notag\\
 &+(-x-2y+2uv)t^5+(-y+uvw)t^6+uvxt^7+uvyt^8.
\end{align}
To check the signs of these coefficients, group the entries along the two diagonals and put
\[
 m=ad,\qquad n=bc,\qquad s=a+d,\qquad k=b+c,
 \qquad \delta=m-n.
\]
The normalization and the definitions above imply
\[
 s+k=1,\qquad e=m+n,\qquad x=mk+ns,\qquad
 y=mn,\qquad w=e+sk,\qquad uv=x+\delta^2.
\]
In these variables, the coefficients of $t^2,\ldots,t^8$ are, in order,
\begin{align*}
 &e,\qquad e+ms+nk,\qquad e-y+2\delta^2,\\
 &m(k-n)+n(s-m)+2\delta^2,\\
 &xw-y+\delta^2w,\qquad uvx,\qquad uvy.
\end{align*}
Every entry lies between zero and one, so $m\le s$, $n\le k$, and $e\ge y$.
This proves nonnegativity except at degree six, where $m\le s^2/4$ and $n\le k^2/4$ give
\[
 xw\ge xsk=msk^2+ns^2k\ge4mn(s+k)=4y.
\]
The coefficient of degree six is therefore at least $3y+\delta^2w$ and hence nonnegative.
Since the denominator is positive for $t\ge0$, this proves the normalized inequality.

\section{Gaussian formulas for continuity of conditional subentropy}
\label{app:conditional-gradient}
The spatial sign formula \eqref{eq:K-matrix} is established in \cite{raninen2022bias}.
The Gaussian derivation below covers singular matrices and yields \eqref{eq:block-gradient-order} and sharpness at a fixed marginal.
For a nonzero positive semidefinite matrix $M$, choose an eigenbasis with eigenvalues $\lambda_i\ge0$.
Gaussian phase invariance makes the expectation in \eqref{eq:K-matrix} diagonal, with $i$th entry $\E[\lambda_i Z_i/T]$.
Here $T=\sum_i\lambda_iZ_i$, and the $Z_i$ are independent exponential variables of mean one.
Using $1/T=\int_0^\infty e^{-tT}\dd t$ and Tonelli's theorem gives
\[
 \E[\lambda_i Z_i/T]
 =\int_0^\infty\frac{\lambda_i}{1+t\lambda_i}
                  \prod_j(1+t\lambda_j)^{-1}\dd t.
\]
The normalized outer product has norm one and converges almost surely as matrices approach a nonzero positive semidefinite limit.
Dominated convergence gives continuity of $K$, including changes of rank.
Unitary covariance follows from Gaussian invariance, while positive $\lambda_i$ give positive eigenvalues of $K(M)$.

For positive definite $M$, the moment bound below justifies differentiating $F$ under the expectation.
Integration by parts in the exponential variables then gives
\begin{align*}
 \E[Z_i\log T]&=\E\log T+\lambda_i\E[Z_i/T],\\
 \nabla F(M)&=[1+\E\log(g^*Mg)]I+K(M).
\end{align*}
Phase invariance removes the off-diagonal entries.
To justify these operations, note that $T$ lies between positive multiples of $\sum_iZ_i$, with
\[
 \E\biggl[\Bigl(\sum_iZ_i\Bigr)\Bigl|\log\sum_iZ_i\Bigr|\biggr]<\infty.
\]
For positive definite CQ blocks, set $A=\sum_xA_x$ and $R=\bigoplus_xA_x$.
Subtracting their gradients gives
\begin{align*}
 \nabla_{A_x}[F(A)-F(R)]&=cI+K(A)-K_x(R),\\
 c&=\E\log(g^*Ag)-\E\log(h^*Rh).
\end{align*}
Here $g$ and $h$ have the dimensions of $A$ and $R$, respectively.
The scalar $c$ is common to all blocks, so it contributes zero along every perturbation of total trace zero.

To compare the block gradients, first consider the determinant factor.
For positive semidefinite $X,Y$, factorization and Sylvester's identity give
\[
 \det(I+X+Y)
 =\det(I+X)\det\!\bigl(I+Y^{1/2}(I+X)^{-1}Y^{1/2}\bigr)
 \le\det(I+X)\det(I+Y).
\]
The inequality follows from $Y^{1/2}(I+X)^{-1}Y^{1/2}\le Y$ and determinant monotonicity in the positive semidefinite order.
Iteration gives $\det(I+tA)^{-1}\ge\det(I+tR)^{-1}$.
For the matrix factor, use $A\ge A_x$ and the fact that, for $t>0$,
\[
 M(I+tM)^{-1}=t^{-1}[I-(I+tM)^{-1}]
\]
is operator monotone by order reversal under inversion.
The scalar determinant factor and positive matrix factor for $K(A)$ thus dominate their counterparts for $K_x(R)$.
Multiplication by positive scalars and integration preserve the matrix order, giving $0\le K_x(R)\le K(A)$ and hence \eqref{eq:block-gradient-order}.

For sharpness, start with a single nonzero CQ block $\tau$ and work on its support.
Choose $P$ as the projection onto a common eigenvector of the marginal and its spatial sign covariance matrix; let $\lambda>0$ be its eigenvalue for the marginal.
Splitting off $\varepsilon P$ into a new CQ block changes the joint quadratic form from $T$ to $T+\varepsilon(Z'-Z_i)$, where $Z'$ is a new independent exponential random variable with mean one.
The Gaussian functional then has right derivative
\[
 \frac{\dd}{\dd\varepsilon}F(R_\varepsilon)\Bigr|_{0+}
 =\E[(Z'-Z_i)(1+\log T)]
 =-\lambda\E[Z_i/T].
\]
For positive $\varepsilon<\lambda/2$, the perturbed quadratic form lies between $T/2$ and a fixed multiple of the full Gaussian squared norm.
Its logarithm is bounded by a constant plus absolute logarithms of the Gaussian norms on the support and full space.
Multiplying by $Z'+Z_i$ gives an integrable bound for the difference quotients.
Since the average remains $\tau$, the conditional right derivative is the corresponding eigenvalue of $K(\tau)$.
Choosing the eigenvector for the largest such eigenvalue proves sharpness in Corollary~\ref{cor:fixed-marginal}, also for singular $\tau$.

\section{Limit for maximally mixed ancillary states}\label{app:ancilla-rate}
After multiplication by $k$, the limit of conditional subentropy is determined by the purity difference between the average and joint states.
Their equal traces cancel the leading terms in the determinant expansion below.
Write $A=\rho_B=\sum_xA_x$ and $R=\bigoplus_xA_x$.
For a density operator $M$, set
\[
 L_{M,k}(t)=\det(I+(t/k)M)^{-k}.
\]
The integral representation gives $C_k=\int_0^\infty[L_{A,k}(t)-L_{R,k}(t)]\dd t/t^2$.
At each fixed $t$, expansion of $\log(1+u)$ gives
\[
 \log L_{M,k}(t)=-t+\frac{t^2}{2k}\Tr M^2+O(k^{-2}),\qquad
 \frac{k(L_{A,k}(t)-L_{R,k}(t))}{t^2}
 \to c(\rho)e^{-t},
\]
where $c(\rho)=(\Tr A^2-\Tr R^2)/2=\sum_{x<z}\Tr(A_xA_z)$.
For an integrable bound, the determinant factorization in Appendix~\ref{app:conditional-gradient} gives $D_k=\log L_{A,k}-\log L_{R,k}\ge0$.
The inequalities $u-u^2/2\le\log(1+u)\le u$ for $u\ge0$ give $D_k\le t^2\Tr A^2/(2k)$.
Hence, for $k\ge2$,
\[
 0\le\frac{k(L_{A,k}-L_{R,k})}{t^2}
 =\frac{kL_{A,k}(1-e^{-D_k})}{t^2}
 \le\frac{\Tr A^2}{2}L_{A,k}
 \le\frac{\Tr A^2}{2}(1+t/2)^{-2}.
\]
For the last step, $\det(I+(t/k)A)\ge1+t/k$ and $(1+t/k)^k\ge(1+t/2)^2$.
Since the final bound is integrable, dominated convergence proves $kC_k\to c(\rho)$.

Applying this limit to identical pure signals with the priors in \eqref{eq:phi} gives, for $0\le u\le e_n=1-1/n$,
\[
 k\Phi_{n,k}(u)\to
 g_n(u)=u-\frac{n}{2(n-1)}u^2.
\]
The limiting quadratic is continuous and strictly increasing on $[0,e_n]$.
To prove convergence of the inverse values, first note that the sharp bound gives $0\le c(\rho)\le g_n(e_n)$.
If $u_*=g_n^{-1}(c(\rho))$ lies in the interior, choose any two points on opposite sides of $u_*$.
By the preceding limits and strict increase, $\Phi_{n,k}^{-1}(C_k)$ eventually lies between them.
At either endpoint, use the interval boundary on one side and the same argument on the other.
Solving the quadratic gives \eqref{eq:ancilla-limit}.
For binary pure signals with priors $p,1-p$, $\Tr(A_0A_1)=p(1-p)s^2$, so \eqref{eq:binary-quadratic-exact} identifies the limit with the Helstrom error.

\section{Equality in the modulus of continuity}\label{app:ordinary-continuity-equality}
The equality cases in Theorem~\ref{thm:modulus} follow from its coupling and trace distance bounds.
\begin{proof}
Equality in the interior range forces $e=\tau=\delta>0$ by strict increase of the modulus.
The classical coupling must also attain the guessing bound.
By Theorem~\ref{thm:guess}, its $Y$ marginal is a point mass and its $X$ marginal is the distribution in \eqref{eq:phi}.
Thus the state with smaller subentropy is pure, say $|v\rangle\langle v|$, and the other has eigenvalues $(1-\delta,\delta/(d-1),\ldots)$.
Measuring the projection onto the pure state gives
\[
 D(|v\rangle\langle v|,\sigma)
 \ge 1-\langle v|\sigma|v\rangle
 \ge 1-\lambda_{\max}(\sigma)=\delta.
\]
The distance is at most $\delta$, so equality holds throughout.
Equality in the Rayleigh quotient makes $v$ an eigenvector of $\sigma$ for its largest eigenvalue, which has multiplicity one, giving \eqref{eq:modulus-attainers}.

On the plateau, equality requires the two states to attain the minimum and maximum of $Q$: pure states are exactly the minimizers, while strict concavity gives the uniform spectrum as the unique maximum.
Thus equality holds exactly for a pure state and the maximally mixed state, at distance $1-1/d$.
Zero distance forces identical states; dimension one is trivial.
\end{proof}

\section{A cyclic ensemble of three pure states}\label{app:cyclic-example}
To compare the bounds, consider three equally likely pure signals in dimension three,
\begin{equation}\label{eq:cyclic-signals}
 |\psi_k\rangle=\sum_{j=0}^{2}\sqrt{\lambda_j}\,\omega^{jk}|j\rangle,
 \qquad \omega=e^{2\pi i/3},\quad k=0,1,2,
\end{equation}
where $\lambda_j\ge0$ and $\sum_j\lambda_j=1$.
Averaging over the phases gives a diagonal state with spectrum $\lambda$.
The joint classical--quantum state has three nonzero eigenvalues $1/3$, so
\begin{equation}\label{eq:cyclic-conditional}
 q=Q(X|B)=\log3-H_3+1-Q(\lambda).
\end{equation}
The Fourier measurement projects onto $|\chi_k\rangle=3^{-1/2}\sum_j\omega^{jk}|j\rangle$ and achieves
\begin{equation}\label{eq:cyclic-success}
 \pg=\frac{1}{3}\Bigl(\sum_j\sqrt{\lambda_j}\Bigr)^2.
\end{equation}
For a matching upper bound, set $s=\sum_j\sqrt{\lambda_j}$ and let $Y$ be the diagonal matrix with entries $s\sqrt{\lambda_j}/3$ in the basis $\{|j\rangle\}$.
Weighted Cauchy--Schwarz gives $|\langle\psi_k|v\rangle|^2\le s\sum_j\sqrt{\lambda_j}|v_j|^2$ for every $v$, even when some $\lambda_j$ vanish.
Thus $Y\ge|\psi_k\rangle\langle\psi_k|/3$ for every $k$.
Any POVM $\{M_k\}$ therefore has success probability at most $\sum_k\Tr(M_kY)=\Tr Y=s^2/3$, matching the Fourier measurement and proving optimality.

Table~\ref{tab:cyclic} uses $\lambda=(1-2r,r,r)$ with $0\le r\le1/3$, from identical to orthogonal signals.
At the displayed interior points, the subentropy bound improves on the linear bound but remains below the optimal error.
Sharpness at fixed $q$ and number of labels concerns the optimum over all ensembles; individual ensembles need not attain the bound.
\begin{table}[tbp]
\caption{Three cyclic pure states with average state spectrum $\lambda=(1-2r,r,r)$, $0\le r\le 1/3$.
The optimal error follows from \eqref{eq:cyclic-success}, and \eqref{eq:cyclic-conditional} determines both lower bounds.
Numerical entries use five decimal places: errors are rounded to nearest, and lower bounds are rounded downward using rational enclosures.
Endpoints are analytic.}\label{tab:cyclic}
\centering
\begin{tabular}{rrrr}
\toprule
$r$ & Optimal error & Linear bound $q$ & Subentropy bound $\Phi_3^{-1}(q)$\\
\midrule
0 & $2/3$ & 0.26527 & $2/3$ \\
0.01 & 0.52801 & 0.24639 & 0.48126 \\
0.05 & 0.35049 & 0.18298 & 0.28014 \\
0.10 & 0.22288 & 0.12121 & 0.15987 \\
0.20 & 0.07145 & 0.03896 & 0.04323 \\
$1/3$ & 0.00000 & 0.00000 & 0.00000 \\
\bottomrule
\end{tabular}
\end{table}

\section{Numerical evaluation of the inverse bounds}\label{app:numerical-certification}
\subsection{Behavior near the uniform endpoint}
The curve $\Phi_n$ has zero slope at the uniform endpoint $e_n=1-1/n$.
The Gaussian Hessian there is $\partial_i\partial_jF=n(1+\mathbf1_{i=j})/(n+1)$.
The tangent direction has zero coordinate sum and squared norm $n/(n-1)$.
Symmetry cancels the linear term, giving
\begin{equation}\label{eq:endpoint-curvature}
 \Phi_n(e_n)-\Phi_n(e_n-h)
 =\frac{n^2}{2(n^2-1)}h^2+O(h^3).
\end{equation}
For a maximally mixed ancilla, the uniform joint spectrum has dimension $nk$ and the tangent vector has squared norm $n/[k(n-1)]$.
The marginal is fixed, so the same expansion gives
\begin{equation}\label{eq:calibrated-endpoint-curvature}
 \Phi_{n,k}(e_n)-\Phi_{n,k}(e_n-h)
 =\frac{n^2}{2(n-1)(nk+1)}h^2+O(h^3).
\end{equation}
Consequently, uncertainty in either inverse value scales as the square root of subentropy uncertainty near the maximum.
A tolerance alone cannot certify that the error equals $e_n$.

\subsection{Rational bounds for logarithms and inverse values}
The displayed numerical lower bounds are certified by rational interval arithmetic.
For a positive rational $x$, write $x=2^\ell y$ with $1\le y<2$ and set $z=(y-1)/(y+1)$.
For every integer $N\ge1$,
\[
 0\le\log y-2\sum_{j=0}^{N-1}\frac{z^{2j+1}}{2j+1}
 \le\frac{2z^{2N+1}}{(2N+1)(1-z^2)}.
\]
At the upper endpoint, the bound also encloses $\log2$; scaling then gives an enclosure of $\log x$.
Subentropy is evaluated using confluent divided differences in \eqref{eq:definition}, with derivative values at repeated eigenvalues.
On the increasing branch, a candidate is a certified lower bound on the inverse if an upper enclosure of $\Phi_n$ at that candidate does not exceed a lower enclosure of $Q(X|B)$.
When a comparison is inconclusive, we increase precision or test nearby rational candidates.
If it remains unresolved, we report no certified bound.
Table bounds are rounded downward; endpoints are evaluated analytically.
Plot coordinates are rounded to ten significant digits.

\subsection{Binary pure signals: comparison with Montanaro's bound}\label{app:binary-crossover}
Write $q(s)$ for the conditional subentropy in \eqref{eq:binary_example}.
Since $\Phi_2$ is strictly increasing on $[0,1/2]$, the subentropy bound is larger than $s^2/4$ exactly when $q(s)>\Phi_2(s^2/4)$.
To determine the number of crossings, set $t=s^2$ and define
\[
 f(t)=q(\sqrt t)-\Phi_2(t/4),\qquad 0\le t\le1.
\]
Expanding the logarithms in the binary subentropy formula gives
\[
 q(\sqrt t)=\sum_{k=1}^{\infty}\frac{t^k}{2k(4k^2-1)},\qquad 0\le t<1.
\]
The positive coefficients, including that of $t^2$, make the function defined by this series strictly convex.
Concavity of $\Phi_2$ therefore makes $f$ strictly convex, including at the endpoints by continuity.
Also $f(0)=0$, and Lemma~\ref{lem:purebound} gives $f'(0+)=1/6-1/4=-1/12$, whereas $f(1)=\Phi_2(1/2)-\Phi_2(1/4)>0$.
Strict convexity and $f(0)=0$ make $f(t)/t$ strictly increasing for $t>0$.
Thus $f$ has exactly one zero $t_*$ in $(0,1)$, is negative before it and positive after it.
This proves the ordering in Section~\ref{sec:comparison-bounds} with $s_*=\sqrt{t_*}$.

The rational logarithm bounds above, with $N=50$, certify a negative value of $f$ at $t=0.767875711400^2$ and a positive value at $t=0.767875711401^2$.
Consequently
\[
 0.767875711400<s_*<0.767875711401.
\]
The value in Figure~\ref{fig:guessing}(b) is rounded to six decimal places.
\end{document}